\documentclass[12pt]{article}
\usepackage{amssymb}
\begin{document}
\newcommand{\rf}[1]{eq.(\ref{#1})}
\newcommand{\oh}{\frac{1}{2}}
\newcommand{\oq}{\frac{1}{4}}
\newcommand{\beq}{\begin{equation}}
\newcommand{\eeq}{\end{equation}}
\newcommand{\bea}{\begin{eqnarray}}
\newcommand{\eea}{\end{eqnarray}}
\newcommand{\beas}{\begin{eqnarray*}}
\newcommand{\eeas}{\end{eqnarray*}}
\newcommand{\beqs}{\begin{displaymath}}
\newcommand{\eeqs}{\end{displaymath}}
\newcommand{\ra}{\rightarrow}
\newcommand{\Ra}{\Rightarrow}
\newcommand{\fslu}{\not\mbox{\hspace{-1.5mm}}}
\newcommand{\fsll}{\not\mbox{\hspace{-.5mm}}}
\newcommand{\br}{\langle}
\newcommand{\kt}{\rangle}
\newcommand{\bra}[1]{\langle {#1}|}
\newcommand{\ket}[1]{|{#1}\rangle}
\newcommand{\ul}{\underline}
\newcommand{\dg}{\dagger}
\newcommand{\5}{5 \! \times \! 5}
\newcommand{\nn}{\nonumber \\}
\newcommand{\ab}{abelian}
\newcommand{\ch}{chirality}
\newcommand{\Cou}{Coulomb}
\newcommand{\Gf}{Green functions}
\newcommand{\Fm}{Feynman}
\newcommand{\ham}{hamiltonian}
\newcommand{\lag}{lagrangian}
\newcommand{\Lo}{Lorentz}
\newcommand{\Mi}{Minkowski}
\newcommand{\mi}{minkowskian}
\newcommand{\Eu}{Euclid}
\newcommand{\eu}{euclidean}
\newcommand{\WT}{Ward-Takahashi}
\newcommand{\sm}{standard model}
\newcommand{\AdS}{AdS_{n+1}}
\renewcommand{\a}{\alpha}
\renewcommand{\b}{\beta}
\newcommand{\g}{\gamma}
\newcommand{\G}{\Gamma}
\newcommand{\del}{\delta}
\newcommand{\Del}{\Delta}
\newcommand{\ep}{\varepsilon}
\newcommand{\e}{\epsilon}
\newcommand{\ove}{\overline{\epsilon}}
\newcommand{\ka}{\kappa}
\newcommand{\kp}{\kappa}
\newcommand{\kpb}{\bar{\kappa}}
\renewcommand{\l}{\lambda}
\renewcommand{\L}{\Lambda}
\newcommand{\m}{\mu}
\newcommand{\mb}{\bar{\mu}}
\newcommand{\vmu}{\vec{\mu}}
\newcommand{\n}{\nu}
\newcommand{\nb}{\bar{\nu}}
\newcommand{\mn}{\mu\nu}
\newcommand{\om}{\omega}
\newcommand{\Om}{\Omega}
\newcommand{\vom}{\vec{\omega}}
\newcommand{\tphi}{\tilde{\phi}}
\newcommand{\phid}{\phi^{\dagger}}
\newcommand{\fT}{\phi^3}
\newcommand{\fF}{\phi^4}
\newcommand{\fn}{\phi^n}
\newcommand{\bpsi}{\bar{\psi}}
\newcommand{\rh}{\rho}
\newcommand{\rhb}{\bar{\rho}}
\newcommand{\sg}{\sigma}
\newcommand{\sgb}{\bar{\sigma}}
\newcommand{\vsig}{\vec{\sigma}}
\newcommand{\Sg}{\Sigma}
\newcommand{\Th}{\Theta}
\newcommand{\vp}{\varphi}
\newcommand{\dz}{\frac{dz}{2\pi i}}
\newcommand{\vB}{\vec{B}}
\newcommand{\cD}{{\cal D}}
\newcommand{\be}{\bar{e}}
\newcommand{\bl}{\bar{\l}}
\newcommand{\cL}{{\cal L}}
\newcommand{\vL}{\vec{L}}
\newcommand{\cM}{{\cal M}}
\newcommand{\cO}{{\cal O}}
\newcommand{\vS}{\vec{S}}
\newcommand{\cV}{{\cal V}}
\newcommand{\hk}{\hat{k}}
\newcommand{\vk}{\vec{k}}
\newcommand{\vek}{\vec{k}\mbox{\hspace{1mm}}}
\newcommand{\hq}{\hat{q}}
\newcommand{\vx}{\vec{x}}
\newcommand{\vex}{\vec{x}\mbox{\hspace{1mm}}}
\newcommand{\hr}{\hat{r}}
\newcommand{\Tr}{{\rm Tr}\;}
\newcommand{\ben}{\begin{equation}}
\newcommand{\een}{\end{equation}}
\newcommand{\bdm}{\begin{displaymath}}
\newcommand{\edm}{\end{displaymath}}
\newcommand{\lp}{\ell_P}
\newcommand{\hf}{\frac{1}{2}}
\newcommand{\vn}{\vec{\nabla}}
\newcommand{\ve}{\vec{e}}
\newcommand{\vep}{\vec{e}\mbox{\hspace{1mm}}'}
\newcommand{\La}{{\cal L}}
\newcommand{\Ha}{{\cal H}}
\newcommand{\pa}{\partial}
\newcommand{\opa}{\overline{\partial}}
\newcommand{\eps}{\epsilon}
\newcommand{\al}{\alpha}
\newcommand{\D}{{\cal D}}
\newcommand{\N}{{\cal N}}
\newcommand{\tb}{\bar{t}}
\newcommand{\rb}{\bar{r}}
\newcommand{\oz}{\overline{z}}
\newcommand{\of}{\overline{f}}
\newcommand{\ovh}{\overline{h}}
\newcommand{\ow}{\overline{w}}
\newcommand{\qh}{\hat{q}}
\newcommand{\phih}{\hat{\phi}}
\newcommand{\gb}{\overline{\gamma}}
\newcommand{\zt}{\zeta}
\newcommand{\ztb}{\overline{\zeta}}
\newcommand{\tq}{\tilde{q}}
\newcommand{\Asf}{{\sf A}}
\newcommand{\Bsf}{{\sf B}}
\newcommand{\Csf}{{\sf C}}
\newcommand{\ps}{\psi}
\newcommand{\psb}{\overline{\psi}}
\newcommand{\etb}{\overline{\eta}}
\newcommand{\Phc}{\Phi^{Cl}}
\renewcommand{\d}{\delta}
\newcommand{\cJ}{{\cal J}}
\newcommand{\vph}{\varphi}
\renewcommand{\dag}{\dagger}
\newcommand{\tr}{{\rm tr}\,}
\newcommand{\lla}{\left\langle}
\newcommand{\rra}{\right\rangle}
\newcommand{\DS}{\!\!\!\! /}
\newcommand{\ds}{\!\!\!\! \hspace{1mm} /}
\newcommand{\R}{\mathbb{R}}
\newcommand{\C}{\mathbb{C}}
\newcommand{\Q}{\mathbb{Q}}
\newcommand{\Z}{\mathbb{Z}}
\newcommand{\Nat}{\mathbb{N}}

\begin{titlepage}
\begin{flushright}
NBI-HE-99-05
\end{flushright}
\vskip 2.0cm
\begin{center}
{\Large\bf Introduction to the Maldacena Conjecture on AdS/CFT}\\
\vskip 1.cm
{\bf Jens~Lyng~Petersen}\\ 
\vskip 3mm
The Niels Bohr Institute, Blegdamsvej 17,\\
DK-2100 Copenhagen $\emptyset$, Denmark.
\end{center}
\vskip 1.0cm
\begin{center}
{\bf ABSTRACT}
\end{center}
\begin{quote}
These lectures do not at all provide a general review of this rapidly growing 
field. Instead a rather detailed account is presented of a number of 
the most elementary aspects.
\vskip 5mm
\hrule width 5.cm
\vskip 2.mm
{\small\small
\noindent
Based on Lectures presented at ``Quantum aspects of gauge theories, 
supersymmetry and unification'', School in Leuven, January 18-23, 1999; and on
Lectures presented at ``Nordic Course on Duality in String and Field 
Theories'', 4-14 August, 1998, Nordita, Copenhagen. 
Lecture notes prepared with the help of Troels Harmark. 
} \normalsize
\end{quote}
\vfill
\end{titlepage}

\section{Introduction}
The Maldacena conjecture \cite{maldacena} is a conjecture concerning 
string theory or M theory on certain backgrounds of the form 
$AdS_d\times M_{D-d}$. Here $AdS_d$ is an anti de Sitter space of space-time 
dimension $d$, and $M_{D-d}$ is a certain compactification space of dimension
$D-d$ with $D=10$ for string theory and $D=11$ for M theory. In addition, the
background is specified by a statement about the flux of a certain field 
strength differential form. The conjecture asserts that the quantum 
string- or M-theory on this background is mathematically equivalent - or dual
as the word goes - to an
{\em ordinary} but conformally invariant quantum field theory in a space-time 
of dimension $d-1$,
which in fact has the interpretation of ``the boundary'' of 
$AdS_d$. This seems to put the formulation of string/M-theory on a novel and
rather unexpected footing. Also the relation between quantum and classical 
theory is illuminated in a surprising way by the conjecture. Several details
in Maldacena's original formulation were left unspecified. Most of those
were subsequently given a precise formulation by independent works of 
Gubser, Klebanov and Polyakov \cite{gubser} and by Witten \cite{witten1}.
A priori it might seem very strange that quantum theories in different 
space-time dimensions could be equivalent. This possibility is related to the 
fact that the theory in the larger dimension is (among other things) a quantum
theory of gravity. For such theories the concept of {\em holography} has been
introduced as a generic property, and the Maldacena conjecture is an example of
the realization of that (for discussion, see for example \cite{susskindhol}).

In the meantime a large number of checks have been performed which we shall
not attempt to review in these notes (for some recent reviews with
many additional references, see for example 
\cite{klebanov99,schwarz98,sen,douglas99}).
Supposing the conjecture is true, it remains somewhat unclear what the most 
significant consequence will be. On the one hand the conjecture allows one to
obtain non perturbative information on ordinary, but mostly {\em conformally 
invariant} quantum field theories, especially at large $N$ (of a gauge group
$U(N)$), from classical string/M-theory or even classical supergravity. This
is a remarkable unexpected development, and the one that has mostly been
pursued until now. On the other hand it is conceivable that the conjecture 
will play an important role in the eventual non-perturbative formulation of
M-theory, for which the matrix model of BFSS \cite{bfss96} was a first proposal.
In a somewhat different line of development, Witten \cite{witten2} showed how to
apparently overcome the original restriction to conformally invariant 
(and mostly supersymmetric) quantum field theories, providing in fact an 
entirely new framework for studying large $N$ ``ordinary'' QCD and similar
theories. A rather new idea about how to achieve the same end in perhaps a more
efficient way has recently appeared \cite{0string}. That approach, however
will not be covered here at all (see also \cite{klebanov99}).
In any case, the AdS/CFT development attracts an enormous interest.

In these lectures we shall attempt a very elementary introduction to a 
somewhat restricted number of basic aspects. In sect. 2 we begin by reviewing
properties of anti de Sitter spaces, their isometries, the fact that
they may be associated with a ``boundary'' and the fact that the isometry group
of anti de Sitter space becomes the conformal group on the boundary. 
It follows that
{\em if} a quantum theory on anti de Sitter space is dual to another quantum 
theory on the boundary, then that second theory must necessarily be conformal.

In sect. 3 we expand on the discussion in \cite{maldacena} and provide a short
review of classical supergravity solutions in the presence of branes. 
Both so called extremal (BPS) and non-extremal solutions will be considered for
later reference. This 
subject has already been reviewed on numerous occasions (see for example
\cite{dkl95,stelle,fre98,argurio}). We describe how the so called near 
horizon approximation in some cases lead to geometries of the form 
$AdS_d\times S^{D-d}$. This fact has been known for several years by the 
experts, but its full significance was only realized by Maldacena.

In sec. 4 we follow rather closely \cite{witten1} and describe in detail several
instructive albeit rather trivial examples of how the duality between the bulk
theory and the boundary theory works in the case of free theories. An important
object which has a general significance is the generalized propagator 
describing propagation of certain modes from a space-time point in the bulk 
of anti de Sitter space to a ``point'' on the boundary. This propagator was
the key object in the discussions in \cite{gubser,witten1} and will be 
constructed in a few of the simplest cases. At the same time the Maldacena 
conjecture will be made more precise.

In sect. 5 we follow \cite{witten2} and describe how certain finite temperature
scenarios may be used to provide a mechanism for breaking conformal invariance 
and supersymmetry, and thereby obtain a framework for studying large $N$ QCD.  

\section{Elementary properties of anti de Sitter spaces}
We begin by considering the Einstein-Hilbert action with a cosmological term.
\ben
S=-s\frac{1}{16\pi G_D}\int d^Dx\sqrt{|g|}(R+\L)
\een
We consider (first) Minkowski metric with $s=-1$, and we take it to be 
``mostly plus''. We shall also consider Euclidean signature, $s=+1$.
Notice that the sign of the action flips if we go from a ``mostly plus'' 
to a ``mostly minus'' metric. 
Anti de Sitter space (AdS) as well as de Sitter space are solutions of the 
empty space Einstein equation:
\bea
R_{\mu\nu}-\hf g_{\mu\nu}R&=&\hf\L g_{\mu\nu} \Ra\nn
R&=&\frac{D}{2-D}\L\Ra\nn
R_{\mu\nu}&=&\frac{\L}{2-D}g_{\mu\nu}
\label{EinsteinLambda}
\eea
So these spaces have the property that the Ricci tensor is proportional to the 
metric tensor: They are Einstein spaces. We shall be interested in various 
examples of such spaces, in particular in ones with maximal symmetry, for which
in addition we have
\ben
R_{\mu\nu\rho\sg}=\frac{R}{D(D-1)}(g_{\nu\sg}g_{\mu\rh}-g_{\nu\rh}g_{\mu\sg})
\label{maxsym}
\een
Such spaces are (for $R\neq 0$): spheres, $S^D$, de Sitter spaces, $dS_D$, 
and anti de Sitter spaces, $AdS_D$. 
The difference between de Sitter space and anti de Sitter space 
is the sign of the cosmological constant. With the above conventions, AdS 
spaces have $\L >0$ (see below).

\subsection{$AdS_{n+1}$ by embedding}
It is useful to consider an $(n+1)$-dim $AdS_{n+1}$ as a submanifold of a
pseudo-Euclidean $(n+2)$-dimensional embedding 
space with coordinates $(y^a)=(y^0,y^1,...,y^n,y^{n+1})$ and metric
$$\eta_{ab}=\mbox{diag}(+,-,-,...,-,+)$$
with ``length squared''
$$y^2\equiv (y^0)^2+(y^{n+1})^2-\sum_{i=1}^{n}(y^i)^2$$
preserved by the ``Lorentz-like'' group $SO(2,n)$ (with ``two times'') acting as
\ben
y^a\ra {y'}^a={\L^a}_by^b,\ \ \ {\L^a}_b\in SO(2,n)
\label{so(2,n)}
\een
A possible definition of $AdS_{n+1}$ is then as the locus of 
\ben
y^2=b^2=\mbox{const.}
\label{submanifold}
\een
For de Sitter spaces we would use a ``mostly plus'' metric and the same 
definition (or equivalently, $b^2\ra -b^2$) and similarly for the spherical
spaces, for which of course the metric is positive definite.
We shall demonstrate below that this implies \rf{maxsym}. 

If instead of $AdS_{n+1}$ we consider the $n+1$ dimensional Minkowski space,
we know that our theory should be invariant under the Poincar\'e group, which in
$n+1$ dimensions has dimension $n+1$ (for the translations) plus $\hf n(n+1)$
(for the Lorenz transformations) in total $\hf (n+1)(n+2)$. In fact the 
Poincar\'e group is exactly the isometry group of flat space: invariant 
intervals, squared, are preserved by the Poincar\'e group. With the 
definition of $AdS_{n+1}$ just given, it is obvious that the isometry group of 
that space instead is 
$SO(2,n)$. In fact, let $y^a_0, y^a_0+dy^a_{(1)}$ and $y^a_0+dy^a_{(2)}$ 
be  3 points lying in the submanifold given by \rf{submanifold}, and let
${y'}^a_0, {y'}^a_0+{dy'}^a_{(1)}$ and ${y'}^a_0+{dy'}^a_{(2)}$ be the 
corresponding images under the $SO(2,n)$ transformation \rf{so(2,n)}. Clearly
these also lie in $AdS_{n+1}$, and in particular we have
\ben
dy_{(1)}\cdot dy_{(2)}=dy'_{(1)}\cdot dy'_{(2)}
\een
where for any vectors in $(n+2)$-dimensional pseudo-Euclidean space
$$x\cdot y\equiv\eta_{ab}x^ay^b$$   
Since the vectors $dy^a_{(1)},dy^a_{(2)},{dy'}^a_{(1)},{dy'}^a_{(2)}$ are all
vectors in $AdS_{n+1}$,
this proves that the metric on $AdS_{n+1}$ inherited from that of 
the embedding space, is $SO(2,n)$ invariant. It follows that quantum theories on
$AdS_{n+1}$ should have an $SO(2,n)$ invariance. Clearly the dimension of that
group is (the same as the dimension of $SO(n+2)$) $\hf (n+1)(n+2)$. So the
invariance group for theories on $AdS$ is just a large as for theories on a flat
space of the same dimensionality.

\subsubsection{Polar/Stereographic coordinates}
Introduce coordinates $(x^\mu)=(x^1,...,x^{n+1})$ on $AdS_{n+1}$ by
\bea
y^0&=&\rho\frac{1+x^2}{1-x^2}\nn
y^\mu&=&\rho\frac{2x^\mu}{1-x^2}, \ \ \mu=1,...,n+1
\label{adsstereo}
\eea 
where 
$$x^2\equiv (x^1)^2+...+(x^n)^2-(x^{n+1})^2$$
We may think of the set 
$(\rho,x^\mu)$ as a possible set of coordinates on the $(n+2)$-dimensional
embedding space. Clearly $y^2=\rho^2$ and AdS is $\rho=b$.
The metric in the embedding space is (convention: ``mostly minus'')
\ben
ds^2=(dy^0)^2+(dy^{n+1})^2-d\vec{y}^2
\label{embeddingmetric}
\een
where $d\vec{y}=(dy^1,...,dy^n)$.
From this we may work out the metric in $x$ coordinates. We get
(Notation: $\mu =1,...,n+1$ and we raise and lower $\mu$ by the flat Minkowski 
metric $\mbox{diag}(+,+,...,+,-)$)
\bea
dy^0&=&d\rho\frac{1+x^2}{1-x^2}+4\rho\frac{x_\mu dx^\mu}{(1-x^2)^2}\nn
dy^\mu&=&d\rho\frac{2x^\mu}{1-x^2}+\frac{2\rho}{(1-x^2)^2}
\{(1-x^2)\delta^\mu_\nu+2x^\mu x_\nu\}dx^\nu
\eea
Then work out
\ben
ds^2=d\rho^2-\frac{4\rho^2}{(1-x^2)^2}dx^2
\een
We see that in these coordinates the metric factorizes into a trivial
``radial'' part and an interesting (``angular'') AdS part:
\ben
g_{\mu\nu}=+\frac{4b^2}{(1-x^2)^2}\eta_{\mu\nu}
\label{polarmetric}
\een
(convention: ``mostly plus''). 

We now want to verify that this metric indeed satisfies the Einstein equation
in vacuum \rf{EinsteinLambda} with a cosmological term:
$$R_{\mu\nu}\propto g_{\mu\nu}$$
and determine the constant in terms of the dimension $D=n+1$ and the 
AdS-scale, $b$. We shall do even more, and verify that the spaces also satisfy
the maximal symmetry condition, \rf{maxsym}.

Let us in fact consider a general ``conformally flat'' metric of the form
$$g_{\mu\nu}(x)=e^{\phi(x)}\eta_{\mu\nu}$$
In our case
$$\phi(x)=\log 4b^2 - 2\log(1-x^2)$$
Then work out
\bea
\G^\mu_{\nu\rho}&=&\hf g^{\mu\l}(\pa_\nu g_{\rho\l}+\pa_\rho g_{\nu\l}-
\pa_\l g_{\nu\rho})\nn
&=&\hf(\pa_\nu\phi\del^\mu_\rho+\pa_\rho\phi\del^\mu_\nu-
\pa^\mu\phi\eta_{\nu\rho})\nn
{R^\mu}_{\nu\rho\sigma}&=&\pa_\rho\G^\mu_{\nu\sigma}-\pa_\sigma\G^\mu_{\nu\rho}
+\G^\mu_{\l\rho}\G^\l_{\nu\sigma}-\G^\mu_{\l\sigma}\G^\l_{\nu\rho}\nn
\pa_\rh\G^\mu_{\nu\sg}-(\rh\leftrightarrow\sg )&=&
\hf \left (\del^\mu_\sg\pa_\rh\pa_\nu\phi +
\del^\mu_\nu\pa_\rh\pa_\sg\phi \right.\nn
&&\left. -\eta_{\nu\sg}\pa_\rh\pa^\mu\phi\right ) -
(\rh\leftrightarrow\sg )\nn
&=&\hf\left (\del^\mu_\sg\pa_\rh\pa_\nu\phi-\del^\mu_\rh\pa_\sg\pa_\nu\phi
-\eta_{\nu\sg}\pa_\rh\pa^\mu\phi+\eta_{\nu\rh}\pa_\sg\pa^\mu\phi\right )\nn
\G^\mu_{\l\rho}\G^\l_{\nu\sigma}-(\rh\leftrightarrow\sg)&=&
\frac{1}{4}\left (\del^\mu_\rh\pa_\nu\phi\pa_\sg\phi+
\eta_{\sg\nu}\pa^\mu\phi\pa_\rh\phi+\eta_{\nu\rh}\del^\mu_\sg(\pa\phi)^2
\right )-(\rh\leftrightarrow\sg)\nn
\pa_\mu\phi\pa_\nu\phi&=&\frac{16x_\mu x_\nu}{(1-x^2)^2}\nn
\pa_\mu\pa_\nu\phi&=&\frac{4}{1-x^2}\eta_{\mu\nu}+\frac{8x_\mu x_\nu}{(1-x^2)^2}
\nn
{R^\mu}_{\nu\rh\sg}&=&-\frac{4}{(1-x^2)^2}\left (\eta_{\nu\sg}\del^\mu_\rh
-\eta_{\nu\rh}\del^\mu_\sg\right )\nn
&=&-\frac{1}{b^2}(-g_{\nu\rh}\del^\mu_\sg + g_{\nu\sg}\del^\mu_\rh)
\eea
the last equality being the statement of maximal symmetry. Then further
\bea
R_{\nu\sigma}&=&-\frac{D-1}{b^2}g_{\nu\sigma},
\ \ \ (D\equiv n+1)
\label{adscosmological}
\eea
And we see that indeed \rf{EinsteinLambda} is satisfied with
\ben
\L=\frac{n(n-1)}{b^2}=\frac{(D-1)(D-2)}{b^2}
\een
(Notice that under a shift in convention $g_{\mu\nu}\ra -g_{\mu\nu}$, i.e. 
``mostly plus'' $\ra$ ``mostly minus'', $R_{\mu\nu}$ is unchanged, so the 
cosmological term will appear with the opposite sign.)\\[.2cm]
{\bf Exercise:} Show for the conformally flat metric, that in general
\ben
R_{\mu\nu}=\Big(1-\frac{D}{2}\Big)(\pa_\nu\pa_\mu -\hf\pa_\nu\phi\pa_\mu\phi)
+\hf\eta_{\mu\nu}\Big(\left[1-\frac{D}{2}\right](\pa\phi)^2-\pa^2\phi\Big)
\een
Use this to provide yet another derivation of \ref{adscosmological}.\\[.2cm]
We shall also need to consider the case of Euclidean signature or 
imaginary times. Then
$$dx^2=\sum_{\mu=1}^{n+1} (dx_\mu)^2$$
In such coordinates the ``Euclidean version'' of $AdS_{n+1}$ is topologically
the ball $B_{n+1}$
$$\sum_{\mu=1}^{n+1} (x_\mu)^2<1$$
The ``boundary'' of the ball lies infinitely far away as measured in the 
AdS metric. We shall come back to that.\\[.3cm]
{\bf Exercise:} In the Euclidean case, $AdS_{n+1}$ may be viewed as the
hyperbola
$$(y^0)^2-r^2=b^2$$
where
$$r^2\equiv \sum_{\mu=1}^{n+1}(y^\mu)^2$$
Denoting the point $(y^0,r)=(-b,0)$ as the ``South Pole'', show that the 
coordinates $x^\mu$ \rf{adsstereo} (Euclidean version) are the
``stereographic'' projections of $AdS_{n+1}$ from the South Pole to the
``equatorial plane'' $y^0=0$ (in units of $b$).\\[.3cm]
Now define ``light cone coordinates''
\ben
u=y^0+iy^{n+1},\ \ v=y^0-iy^{n+1}
\een
or for Euclidean signature where the Euclidean $AdS_{n+1}$ is the locus of
$$y_E^2\equiv (y^0)^2-(y^{n+1})^2-\vec{y}^2=b^2$$
with isometry group $SO(1,n+1)$,
\ben
u=y^0+y^{n+1},\ \ v=y^0-y^{n+1}
\een 
so that in both cases
$$y^2=uv-\vec{y}^2=b^2$$

We consider in turn various coordinates on $AdS_{n+1}$ and the corresponding 
metrics.
 
The first set is the one used for example by Maldacena in \cite{maldacena}.
Define
\bea 
\xi^\alpha&\equiv&\frac{y^\alpha}{u}, \ \ \alpha=1,...,n\nn
\vec{\xi}^2&\equiv&\sum_{\alpha=1}^n(\xi^\alpha)^2
\eea
then 
\bea
y^2&=&uv-\vec{y}^2=uv-u^2\vec{\xi}^2=b^2\nn
\Ra v&=&\xi^2u+\frac{b^2}{u}
\eea
Use the set $(u,\xi^\alpha)$ on $AdS_{n+1}$:
\bea
dv&=&2\xi\cdot d\xi u + \xi^2 du-\frac{b^2du}{u^2}\nn
dy^\alpha&=&ud\xi^\alpha+\xi^\alpha du\nn
(ds^2)_{\mbox{embedding}}&=&dudv-\vec{dy}^2\nn 
&=&-\frac{b^2du^2}{u^2}-u^2d\vec{\xi}^2\ \ (y^2\equiv b^2)\nn
\Ra (ds^2)_{AdS_{n+1}}&=&+\frac{b^2du^2}{u^2}+u^2d\vec{\xi}^2
\ \ (\mbox{``mostly plus''})\nn
\label{maldacenametric}
\eea

The second set is similar to Poincar\'e coordinates on the projective plane.
For simplicity put $b\equiv 1$ and use the set 
$$(\xi^0,\vec{\xi})\equiv (u^{-1},\vec{\xi})$$
Then $\log u=-\log \xi^0$ and $\frac{du}{u}=-\frac{d\xi^0}{\xi^0}$.
Hence
\ben
ds^2=\frac{(d\xi^0)^2}{(\xi^0)^2}+\frac{d\vec{\xi}^2}{(\xi^0)^2}=
\frac{1}{(\xi^0)^2}\left ((d\xi^0)^2+d\vec{\xi}^2\right )
\label{poincaremetric}
\een
This is one of the forms used by Witten, \cite{witten1}.

\subsection{The ``boundary'' of $AdS_{n+1}$}
Anti de Sitter space has a kind of ``projective boundary''. The idea is in
embedding space to consider $(y^0,y^\mu)$ very large with $y\in \AdS$. Hence 
define new variables
\ben
y^a=R\tilde{y}^a,\ \ u=R\tilde{u}, \ \ v=R\tilde{v}
\een
and take $R\ra\infty$. Then 
\ben
y^2=b^2\Ra \tilde{u}\tilde{v}-\vec{\tilde{y}}^2=b^2/R^2\ra 0
\een
So the boundary is somehow the manifold
\ben
\tilde{u}\tilde{v}-\vec{\tilde{y}}^2=0
\een
But since $tR$ is just as good as $R$ for any $t\in\R$, we have to consider 
the boundary to be the projective equivalence classes 
\bea
uv-\vec{y}^2&=&0\nn
(u,v,\vec{y})&\sim&t(u,v,\vec{y})
\eea
so the boundary is $n$-dimensional -- as it should be. Using the 
equivalence scaling, the boundary may be considered to be represented by 
(Minkowski signature) 
\ben
(y^0)^2+(y^{n+1})^2=1=\vec{y}^2
\een
so that topologically the boundary is $S^1\times S^{n-1}$. In another use of 
scaling, for points with $v\neq 0$  we may scale to $v=1$. 
Then $u=\vec{y}^2$ and we may use $\vec{y}$ as coordinates on the boundary.
Equivalently, if also $u\neq 0$ we may instead scale $u$ to 1 and use 
coordinates $\vec{\tilde{y}}$ and have $v=\vec{\tilde{y}}^2$. Clearly the 
connection between the two sets is
\ben
\vec{\tilde{y}}=\frac{\vec{y}}{y^2}
\label{boundaryinversion}
\een
When either $v=0$ or $u=0$, only one of the two sets may be used. For $v=0$,
$\vec{\tilde{y}}=\vec{0}$ whereas for $u=0$, $\vec{y}=\vec{0}$. We may think of
the (one) point $v=0$ as ``the point at infinity'' in the $\vec{y}$ 
coordinates, and similarly for $u=0$. So the boundary is automatically 
compactified. The situation is analogous to compactifying the Riemann sphere
including the point $z=\infty$ with z a good coordinate in a neighbourhood of 
$z=0$, and $\zeta =1/z$ a good coordinate in a neighbourhood of $z=\infty$.

The above definition of $\AdS$ and its boundary in terms of the embedding 
space, implies that the isometry group $SO(2,n)$ ($SO(1,n+1)$ for Euclidean 
signature) acts in an obvious way on points of the boundary. The crucial result
on which we would like to elaborate, is that {\em the isometry group 
$SO(2,n)$ ($SO(1,n+1)$) acts on 
the boundary as the conformal group acting on Minkowski (Euclidean) space.}

\subsection{The conformal group}
For definiteness, consider $n$-dimensional Euclidean space $E^n$. We first 
want to understand that the conformal group is $SO(1,n+1)$. Begin by counting
the number of generators $=$ number of generators in $SO(n+2)$ $=$ number of
linearly independent antisymmetric $(n+2)\times (n+2)$ matrices 
\ben
\mbox{dim} SO(1,n+1)=\hf (n+2)(n+1)
\een
By comparison, the Poincar\'e group in $n$ dimensions has $n$ translation
generators and $\hf n(n-1)$ rotation generators
\ben
\mbox{dim Poincar\'e}(E^n)=\hf n(n+1)
\een
so the difference is $n+1$. This just fits with the following ``extra'' 
possible conformal transformations:\\
\underline{Dilations}
\ben
\vec{x}\ra \l\vec{x}, \ \ \l\in\R
\een 
gives one generator and \\
\underline{The Special Conformal Transformations}
\bea
\vec{x}&\ra&\vec{x}'\ \ \mbox{such that}\nn
\frac{{x'}^\mu}{{x'}^2}&=&\frac{x^\mu}{x^2}+\al^\mu
\label{specconf1}
\eea
involve the $n$ parameters $\al^\mu, \mu=1,...,n$ and give rise to the 
additional $n$ generators. Equivalently we may write
\ben
{x'}^\mu=\frac{x^\mu+\al^\mu x^2}{1+2\vec{\al}\cdot\vec{x}+\al^2x^2}
\label{specconf2}
\een
The equivalence between \rf{specconf1} and \rf{specconf2} follows after noting
(from \rf{specconf2}) that
\ben
{x'}^2=\frac{x^2}{1+2\vec{\al}\cdot\vec{x}+\al^2x^2}
\een
To verify that these are really {\em conformal} transformations, consider
3 neighbouring points
$$\vec{x}, \vec{x}+d\vec{x}_1, \vec{x}+d\vec{x}_2$$
and their images
$$\vec{x}', \vec{x}'+d\vec{x}'_1, \vec{x}'+d\vec{x}'_2$$
The statement that the transformation is conformal, is the statement that 
the angles are preserved, or
\ben
\frac{d\vec{x}_1\cdot d\vec{x}_2}{\sqrt{dx_1^2 dx_2^2}}=
\frac{d\vec{x}'_1\cdot d\vec{x}'_2}{\sqrt{(dx'_1)^2 (dx'_2)^2}}
\een
But \rf{specconf1} implies
\bea
\frac{{x'}^2d{x'}^\mu-2\vec{x}'\cdot d\vec{x}'{x'}^\mu}{({x'}^2)^2}&=&
\frac{x^2dx^\mu-2\vec{x}\cdot d\vec{x}x^\mu}{(x^2)^2}\nn
\Ra\frac{d\vec{x}'_i\cdot d\vec{x}'_j}{{x'}^4}&=& 
\frac{d\vec{x}_i\cdot d\vec{x}_j}{x^4}, \ \ i,j =1,2
\eea
and the claim follows.

Now we want to show that the action of $SO(1,n+1)$ on boundary points give
conformal transformations. A point in $\AdS$: $(u,v,\vec{y})$ with
$uv-\vec{y}^2=b^2$ is mapped by $SO(1,n+1)$ to $(u',v',\vec{y}')$ as
\ben
\L\left (\begin{array}{c}u\\v\\\vec{y}\end{array}\right )=
\left (\begin{array}{c}u'\\v'\\\vec{y}'\end{array}\right ), 
\een
where $\L\in SO(1,n+1)$ i.e. $\L$ preserves the norm $uv-\vec{y}^2$.

Similarly, a point on the boundary has coordinates $(u,v,\vec{y})$ subject to
\bea
(i)&&uv-\vec{y}^2=0\nn
(ii)&&(u,v,\vec{y})\sim\l (u,v,\vec{y})
\eea
and is mapped by $\L$ to $(u',v',\vec{y}')$ as before. (Notice that of course
$(u_2,v_2,\vec{y}_2)=\l (u_1,v_1,\vec{y}_1)\Ra 
(u'_2,v'_2,\vec{y}'_2)=\l (u'_1,v'_1,\vec{y}'_1)$).

Now consider the infinitesimal transformation $\L={\sf 1}_{n+2} +\omega$ with
$\omega$ infinitesimal. In order for
the relevant norm to be preserved, the $(n+2)\times (n+2)$ dimensional matrix, 
$\omega$ must be of the form
\ben
\omega=\left (\begin{array}{ccc}
a&0&\vec{\al}^T\\
0&-a&\vec{\beta}^T\\
\hf\vec{\beta}&\hf\vec{\al}&\omega_n\end{array}\right )
\een
where $\vec{\al},\vec{\beta}$ are $n$-vectors represented as columns and 
$\omega_n$ is an $n\times n$ antisymmetric matrix. (The strange looking 
factors $\hf$ are due to the fact that we have a non-trivial metric on 
$E^{n+2}$ in the coordinates $(u,v,\vec{y})$, and that our 
matrices have indices like ${\L^a}_b$ rather than two lower indices, say.)

Indeed
\ben
({\sf 1}_{n+2}+\omega)\left (\begin{array}{c}u\\v\\ \vec{y}\end{array}\right )
=\left (\begin{array}{c}u'\\v'\\ \vec{y}'\end{array}\right )=
\left (\begin{array}{c}u(1+a)+\vec{\al}\cdot\vec{y}\\
v(1-a)+\vec{\beta}\cdot\vec{y}\\ 
\left (\vec{y}+\frac{u}{2}\vec{\beta}+\frac{v}{2}\vec{\al}\right )
+\omega_n\vec{y}\end{array}
\right )
\label{boundarytrans}
\een
and one checks that $u'v'-{\vec{y'}}^2=uv-\vec{y}^2$ to first order in the 
infinitesimal quantities, $a,\vec{\al},\vec{\beta},\omega_n$.

Now, choose a representative $(u,v,\vec{y})$ for a boundary point with 
$v=1,u=\vec{y}^2$. (This can always be done except for $v=0$ corresponding
to ``a point at infinity'' on the boundary.) We have seen that with this 
representation, $\vec{y}$ is a convenient representation of the boundary point.
Now, the image point according to \rf{boundarytrans} is not in the same 
convention: $v'\neq 1$ in general. But the image point is equivalent to
$(u'/v',1,\vec{y}'/v')$, which is in the same convention. Thus, the effect of
the mapping is
\ben
\vec{y}\ra \vec{y}'/v'=\vec{y}(1+a-\vec{\beta}\cdot\vec{y})+\frac{y^2}{2}
\vec{\beta}+\hf\vec{\al}+\omega_n\vec{y}
\een
Let us verify that this transformation is in fact a combination of infinitesimal
(i) translations (ii) (Lorentz-) rotations (iii) dilations and (iv) special
conformal transformations:\\
\underline{(i) Only $\vec{\al}\neq 0 \Ra$}
\ben
\vec{y}\ra\vec{y}+\hf\vec{\al}
\een
i.e. translations.\\
\underline{(ii) Only $\omega_n \neq 0 \Ra$}
\ben
\vec{y}\ra\vec{y}+\omega_n\vec{y}
\een
i.e. rotations.\\
\underline{(iii) Only $a\neq 0 \Ra$}
\ben
\vec{y}\ra \vec{y}(1+a)
\een
i.e. dilation.\\
\underline{(iv) Only $\vec{\beta}\neq 0 \Ra$}\\
\ben
\vec{y}\ra \vec{y}(1-\vec{\beta}\cdot\vec{y})+\hf y^2\vec{\beta}
\een
If we compare with \rf{specconf2} and put $\vec{\al}$ in that equation equal to 
$\vec{\beta}/2$ we find
\ben
\vec{y}\ra\frac{\vec{y}+\hf\vec{\beta}y^2}{1+\vec{\beta}\cdot\vec{y}+
\frac{1}{4}\beta^2y^2}=\vec{y}(1-\vec{\beta}\cdot\vec{y})++\hf y^2\vec{\beta}
+O(\beta^2)
\een
in agreement with the above, i.e. indeed we find in this case the 
(infinitesimal) special conformal transformations.

This completes the main result in this section that $SO(1,n+1)$ (and $SO(2,n)$ 
in the Minkowski case) acts (i) as an isometry on $\AdS$ and (ii) as the 
conformal group on the boundary of $\AdS$.

\subsection{The conformal algebra}
For completeness let us work out the Lie algebra of the conformal group.
We choose the simplest possible representation, which is in terms of scalar 
fields $\phi(x)$ with $\vec{x}$ an $n$-tuple of Cartesian coordinates. It is
trivial to check that the generators are represented as follows:\\
\underline{Translations} $P_\mu=-i\pa_\mu$\\
\underline{(Lorentz-)rotations} $M_{\mu\nu}=i(x_\mu\pa_\nu-x_\nu\pa_\mu)=
-(x_\mu P_\nu-x_\nu P_\mu)$\\
\underline{Dilations} $D=-ix^\mu\pa_\mu$\\
\underline{Special Conformal Transformations}
$K_\mu=i(2x_\mu x\cdot\pa-x^2\pa_\mu)=-2x_\mu D+x^2 P_\mu$

One then easily finds:
\bea
{[}M_{\mu\nu},P_\rho{]}&=&i(g_{\nu\rho}P_\mu-g_{\mu\rho} P_\nu)\nn
{[}M_{\mu\nu},M_{\rho\tau}{]}&=&i\left (g_{\mu\tau}M_{\nu\rho}+
g_{\nu\rho}M_{\mu\tau}-g_{\mu\rho}M_{\nu\tau}-g_{\nu\tau}M_{\mu\rho}\right )\nn
{[}M_{\mu\nu},K_\rho{]}&=&i(g_{\nu\rho}K_\mu-g_{\mu\rho}K_\nu)\nn
{[}D,P_\mu{]}&=&+iP_\mu\nn
{[}D,K_\mu{]}&=&-iK_\mu\nn
{[}P_\mu,K_\nu{]}&=&2i(g_{\mu\nu}D+M_{\mu\nu})
\label{confalgebra}
\eea
all others zero.
\section{The Maldacena Conjecture}
\subsection{On $Dp$-branes and other $p$-branes.}
In this section we first briefly review the construction of solitonic $p$-branes
in low energy effective supergravity. There are many excellent reviews 
available, for example. refs. \cite{dkl95,stelle,alwis,fre98} 
to which we refer for 
more details and references to the extensive original literature.
We begin by writing down the effective low energy string action for type II 
(A or B) strings in the string frame:
\ben
S_s=-s\frac{1}{16\pi G_{10}}\int d^{10}x\sqrt{|g|}\left ( e^{-2\phi}
(R+4g^{\mu\nu}\pa_\mu\phi\pa_\nu\phi)-\hf\sum_n\frac{1}{n!}F^2_n + ...\right )
\label{stringframe}
\een
($s=-1 (+1)$ for Minkowski (Euclidean) signature, flipping to ``mostly minus'' 
signature introduces an additional sign of $(-)^n$ in front of $F_n^2$)).
Here the dots represent fermionic terms as well as the NS-NS 
3-form field strength term. $\phi$ is the dilaton, and the
$n$-form field strengths $F_n$ belong to the RR sector. 
For the Newton constant in $D$ dimensions we write
$$16\pi G_D=2\kappa_D^2$$
We shall only
be concerned with the terms given. For IIA strings (IIB strings) we only have
even (odd) values of $n$. For the IIB string the $n=5$ field strength tensor 
is self-dual (In Minkowski space, see below), and it is not strictly speaking
possible to describe the theory by the simple action above. A more complicated
formulation nonetheless exists \cite{pst}. 
However, it turns out to be suffcient to adopt the above action for deriving 
the equations of motions, and imposing self-duality a posteriori (making sure 
that the normalization of $F_5^2$ is unchanged). We shall therefore employ 
that procedure. It is convenient for various reasons to
also represent the action for fields in the Einstein frame, obtained by a
certain Weyl rescaling. In fact the following identity in $D$ space-time 
dimensions may be verified \cite{hull95}
\bea
g_{\mu\nu}&\ra&e^{2\sg\phi}g_{\mu\nu} \Ra\nn
\sqrt{|g|}e^{-2\phi}R&\ra&\sqrt{|g|}e^{-\phi(\sg(D-2)+2)}
\{R+2\sg(D-1)\frac{1}{\sqrt{|g|}}\pa_\mu(\sqrt{|g|}\pa^\mu\phi)\nn
&&-\sg^2(D-1)(D-2)(\pa\phi)^2\} 
\eea
We may therefore choose 
$$\sg=-\frac{2}{D-2}$$
and get rid of a total derivative, specifically in 10 dimensions:
\ben
g_{\mu\nu}(\mbox{Einstein})= e^{-\hf\phi}g_{\mu\nu}(\mbox{string})
\label{einstein2string}
\een
so we obtain in the Einstein frame
\ben
S_E=-s\frac{1}{16\pi G_{10}}\int d^{10}x\sqrt{|g|}\left (
R-\hf g^{\mu\nu}\pa_\mu\phi\pa_\nu\phi -\hf\sum_n\frac{1}{n!}e^{a_n\phi}
F^2_n + ...\right )
\label{einsteinframe}
\een
with 
$$a_n=-\hf(n-5)$$
We shall also be concerned with low-energy M-theory in the form of 
11-dimensional super gravity. The bosonic fields of that theory are just the 
metric and a 3-form gauge potential $C$ with a $4$-form field strength tensor
$$K=dC$$
The bosonic part of the action is
\ben
S_{\mbox{bosonic}}(\mbox{11-dim SUGRA})=-s\frac{1}{2\kappa_{11}^2}\left (
\int d^{11}x\sqrt{|g|}\{R-\frac{1}{48}K^2\}-\frac{1}{6}\int C\wedge K\wedge K
\right )
\label{sugra11}
\een
which only makes sense in the Einstein frame - there is no dilaton.
 
We shall be interested in classical solutions of the above theories, 
specifically in the ones describing Dp-branes. We shall consider static 
solutions corresponding to flat translationally invariant $p$-branes, isotropic
in transverse directions. For such solutions the last term in \rf{sugra11}
will vanish. Hence one is able to cover all cases by considering the
generic action
\ben
S=-s\frac{1}{2\kappa_D^2}\int d^Dx\sqrt{g}\{R-\hf 
g^{\mu\nu}\pa_\mu\phi\pa_\nu\phi -\hf\sum_n\frac{1}{n!}e^{a_n\phi}
F^2_n + ..\}
\label{sgeneric}
\een
in particular with $a=0$ and $\phi\equiv 0$ for 11-dimensional supergravity.
The $p$-brane is a source of charge for the $p+1$ form (RR-) gauge field and 
the $n=p+2$ form field strength. We write
\ben
D=(p+1)+d
\label{Dpd}
\een
where $d$ is the number of dimensions transverse to the $p$-brane.
\subsection{Summary on differential forms}
An $n$-form $F$ has components related to it by
\ben
F=\frac{1}{n!}F_{\mu_1...\mu_n} dx^{\mu_1}\wedge \cdots\wedge dx^{\mu_n}
\een
We define the Levi-Civita symbol (unconventionally) as a non-tensor, so that 
simply
\ben
\e_{01...(D-1)}\equiv \e^{01...(D-1)}=1
\een
From that we define the proper $D$-form, the volume form $\om$ with tensor 
components
\ben
\om_{\mu_1...\mu_D}=\sqrt{g}\e_{\mu_1...\mu_D}, \ \ \ 
\om^{\mu_1...\mu_D}=\frac{s}{\sqrt{g}}\e^{\mu_1...\mu_D}
\een
Here $s=1$ ($s=-1$) for Euclidean (Minkowski) metric (both mostly plus).
We have dropped from now on the numerical signs around the determinant of the 
metric but they are always to be understood.

Integration of an $n$-form over a flat $n$-dimensional domain $M$ becomes
\bea
\int_M F_n&=&\int_MF_{01...(n-1)}dx^0\wedge ...\wedge dx^{n-1}
=\int d^nx F_{01...(n-1)}\nn
&=&\frac{1}{n!}\int d^nx \e^{\mu_1 ...\mu_n}
F_{\mu_1...\mu_n}
\eea
Generally
\ben
V(M)=\int_M\om=\int_M d^nx \sqrt{g_{\mbox{ind}}}
\een
with $(g_{\mbox{ind}})_{ab}$ the induced metric on the sub manifold $M$.
The Hodge dual satisfies
\bea
(*F)^{\mu_{n+1}...\mu_D}&=&\frac{1}{n!}\om^{\mu_1...\mu_D}F_{\mu_1...\mu_n}
=\frac{1}{n!}\frac{s}{\sqrt{g}}\e^{\mu_1...\mu_D}F_{\mu_1...\mu_n}\nn
(*F)_{\mu_{n+1}...\mu_D}&=&\frac{1}{n!}\om_{\mu_1...\mu_D}F^{\mu_1...\mu_n}
=\frac{1}{n!}\sqrt{g}\e_{\mu_1...\mu_D}F^{\mu_1...\mu_n}\nn
(F\wedge*F)_{01...(D-1)}&=&\sqrt{g}\frac{1}{n!}F_{\mu_1...\mu_n}
F^{\mu_1...\mu_n}\nn
{**F}&=&s(-1)^{n(D-n)}F\nn
F\wedge *F&=&s*F\wedge*(*F)\nn
F^2&\equiv&F_{\mu_1...\mu_n}F^{\mu_1...\mu_n}\nn
\frac{1}{n!}F^2&=&s\frac{1}{(D-n)!}(*F)^2
\eea
A self-dual tensor satisfies
$$*F=F\Ra F=**F$$
which for a given dimension and rank is obviously impossible for both 
Minkowski signature and Euclidean 
signature. In particular, the 5-form field strength in 10 dimensional IIB string
theory is self dual only for Minkowski signature. Even in Euclidean signature
it continues to be true for the D3-brane solution considered below, 
that it will be of the form
\ben
F_5=A_5+*A_5
\een
Also
\bea
*A&\equiv&d*F \Ra\nn
A^{\mu_1...\mu_{n-1}}&=&\frac{1}{\sqrt{g}}\pa_\mu(\sqrt{g}
F^{\mu_1...\mu_{n-1}\mu})
\eea
\subsection{Equations of motion}
The equations of motion for the generic problem (Einstein frame) \rf{sgeneric}
are (often we shall write $a$ for $a_n$):
\bea
{R^\mu}_\nu&=&\hf\pa^\mu\phi\pa_\nu\phi+\frac{1}{2n!}e^{a\phi}\left (
nF^{\mu\xi_2 ...\xi_n}F_{\nu\xi_2 ...\xi_n}-\frac{n-1}{D-2}\del^\mu_\nu F_n^2
\right )\nn
\nabla^2\phi&=&\frac{1}{\sqrt{g}}\pa_\mu(\sqrt{g}\pa_\nu\phi g^{\mu\nu})=
\frac{a}{2n!}F_n^2\nn
\pa_\mu(\sqrt{g}e^{a\phi}F^{\mu\nu_2 ...\nu_n})&=&0
\label{eqmotgen}
\eea
where for simplicity we have considered the case with $F_n\neq 0$ only for one
value of $n$, as will be the case.
The Bianchi identity for $F_n$ is
\ben
\pa_{{[}\mu_1} F_{\mu_2 ...\mu_{n+1}{]}}=0
\een
Our $p$-brane ansatz makes use of coordinates
\ben
z^\mu=(t,x^i,y^a),\ \ \mu=0,...,D-1;\ \ i=1,2,...,p;\ \ a=1,2, ...,d;
\ \ D=p+1+d
\een
A metric respecting the symmetries is
\ben
ds^2=g_{\mu\nu}dz^\mu dz^\nu=sB^2dt^2+C^2\sum_{i=1}^p(dx^i)^2
+F^2 dr^2+G^2r^2d\Om_{d-1}^2
\een
which is a diagonal metric, the components of which are all functions of 
the transverse ``distance'' coordinate,
$$r^2=\sum_{a=1}^d (y^a)^2$$
only.
Also $d\Om_{d-1}^2$ is the metric on the unit sphere $S^{d-1}$ in the 
transverse space. There is a gauge freedom which may be disposed of by putting 
$F=G$ or $G=1$ or something else. We shall leave it for the time being in 
order to find a convenient form of the solutions in which we shall be 
interested. It is furthermore part of the $p$-brane ansatz to require, that 
the metric should tend to a flat value at $r\ra\infty$, i.e. that all the
coefficients $B,C,F,G$ should tend to $1$ in that limit.

A $p+1$ form gauge potential couples naturally to the world volume of the 
$Dp$ brane. The resulting $p+2$ form field strength tensor is termed
electric. However, we shall also need the magnetic possibility,
which is a consequence of an electric/magnetic duality in the problem. In fact,
defining
\ben
\tilde{F}_{D-n}=e^{a\phi}*F_n
\label{dualf}
\een
and  using the Hodge duality relations of the previous
subsection, it is possible to verify that the equations of motions are 
invariant under the ``duality transformations'':
\ben
a\phi\ra -a\phi, \ \ n\ra D-n,\ \ F_n\ra \tilde{F}_{D-n}
\label{emduality}
\een
It is helpful first to establish
\bea
\sqrt{g}e^{a\phi}F^{\mu_1...\mu_n}&=&\frac{1}{(D-n)!}e^{\mu_1...\mu_D}
\tilde{F}_{\mu_{n+1}...\mu_D}\nn
\tilde{F}_{\mu_{n+1}...\mu_D}&=&\frac{1}{n!}e^{a\phi}F^{\mu_1...\mu_n}\sqrt{g}
\e_{\mu_1...\mu_D}\nn
\frac{1}{n!}e^{a\phi}F_n^2&=&-\frac{1}{(D-n)}e^{-a\phi}\tilde{F}_{D-n}^2\nn
\frac{n}{n!}e^{a\phi}F^{\mu\xi_2...\xi_n}F_{\nu\xi_2...\xi_n}&=&
\frac{1}{(D-n)!}e^{-a\phi}\left ((D-n)\tilde{F}^{\mu\xi_2...\xi_{D-n}}
\tilde{F}_{\nu\xi_2...\xi_{D-n}}-\del^\mu_\nu\tilde{F}^2_{D-n}\right )
\eea
The {\em electric ansatz} for the field strength is
\ben
F_{ti_1...i_pr}(r)=\e_{i_1...i_p}k(r)
\een
Since
\ben
\sqrt{g}=BC^pF(Gr)^{d-1}\sqrt{\g_{d-1}}
\een
with $\g_{\a\b}$ the metric on $S^{d-1}$, we find
\ben
F^{ti_1...i_pr}=\frac{s}{B^2C^{2p}F^2}\e_{i_1...i_p}k(r)\nn
\een
and the equation of motion for $F_n$ becomes
\ben
\left (\frac{1}{BC^pF}(Gr)^{d-1}e^{a\phi}k(r)\right )'=0
\een
with the result
\bea
k(r)&=&e^{-a\phi}BC^pF\frac{Q}{(Gr)^{d-1}}\nn
F_{ti_1...i_pr}&=&\e_{i_1...i_p}e^{-a\phi}BC^pF\frac{Q}{(Gr)^{d-1}}
\eea
and $Q$ a constant of integration.
\bea
\tilde{F}_{\a_1...\a_{d-1}}&=&\sqrt{\g_{d-1}}\e_{\a_1...\a_{d-1}}Q\nn
\mu_{p}&=&\frac{1}{\sqrt{16\pi G_D}}\int_{S^{d-1}}\tilde{F}_{d-1}
=\frac{\Om_{d-1}Q}{\sqrt{16\pi G_D}}
\label{electricflux}
\eea
where $\mu_p$ is the density of electric charge on the $p$-brane. The 
$\a_i=1,...,d-1 $ are indices on the unit sphere in the transverse space, and
$\Om_{d-1}$ is the volume of $S^{d-1}$:
\bea
\Om_d&=&\frac{2\pi^{\frac{d+1}{2}}}{\G(\frac{d+1}{2})}\nn
\Om_{2n-1}&=&\frac{2\pi^n}{(n-1)!}\nn
\Om_{2n}&=&\frac{2(2\pi)^n}{(2n-1)!!}
\label{volumesd}
\eea
For future reference we work out
\bea
\frac{1}{n!}F_n^2&=&F_{t12...pr}F^{t12...pr}=se^{-2a\phi}
\frac{Q^2}{(Gr)^{2(d-1)}}\nn
\frac{1}{(n-1)!}F^{\mu\xi_2...\xi_n}F_{\nu\xi_2...\xi_n}&=&
\del^\mu_\nu F_{t1...pr}F^{t1...pr}=s\del^\mu_\nu e^{-2a\phi}
\frac{Q^2}{(Gr)^{2(d-1)}}
\eea
with $\mu,\nu\in\{t,1,...,p,r\}$, else $0$.

The {\em magnetic ansatz} has $n=D-(p+2)=d-1$ with the non zero components 
of the field strength tensor given by
\ben
F_{\a_1...\a_{d-1}}=\sqrt{\g_{d-1}}\e_{\a_1...\a_{d-1}}Q
\label{magnetic}
\een
and $Q=Q(r)$. But the equation of motion for $F_n$ is trivially 
satisfied, whereas the Bianchi identity requires $Q$ to be a constant. The 
magnetic charge (density) is
\ben
g_p=\frac{1}{\sqrt{16\pi G_D}}\int_{S^{d-1}}F_{d-1}=
\frac{\Om_{d-1}}{\sqrt{16\pi G_D}}Q
\een
Now
\ben
F^{\a_1...\a_{d-1}}=\frac{1}{\sqrt{\g_{d-1}}}\e_{\a_1...\a_{d-1}}
\frac{Q}{(Gr)^{2(d-1)}}
\een
from which
\bea
\frac{1}{n!}F_n^2&=&\frac{Q^2}{(Gr)^{2(d-1)}}\nn
\frac{1}{(n-1)!}F^{\mu\xi_2...\xi_n}F_{\nu\xi_2...\xi_n}&=&\del^\mu_\nu
\frac{Q^2}{(Gr)^{2(d-1)}}
\eea
with $\del^\mu_\nu$ non-vanishing only for $\mu\nu$ indices belonging 
to the sphere
$S^{d-1}$. The similarity to the electric case will allow us to cover 
both possibilities at the same time.

To find the form of the equations of motion for our ansatz, we must work out
the Riemann tensor for the metric. We choose to work via the spin connection,
expressed in terms of the vielbein $e_\mu^a$ as (flat indices are either small
latin letters from the beginning of the alphabet, or bar'ed Greek letters)
\bea
\om_{\mu ab}&=&\hf e_\mu^c(\Om_{cab}+\Om_{bac}+\Om_{bca})\nn
\Om_{abc}&=&e^\mu_ae^\nu_b(\pa_\mu e_{\nu c}-\pa_\nu e_{\mu c})\nn
\Om_{bac}&=&-\Om_{abc},\ \ \om_{\mu ba}= -\om_{\mu ab}\nn
R_{\mu\nu ab}&=&S_{\mu\nu ab}+K_{\mu\nu ab}\nn
S_{\mu\nu ab}&=&\pa_\mu\om_{\nu ab}-\pa_\nu\om_{\mu ab}\nn
K_{\mu\nu ab}&=&{\om_{\mu a}}^c\om_{\nu cb}-{\om_{\nu a}}^c\om_{\mu cb}
\eea
Our ansatz is of the ``diagonal'' type:
\ben
ds^2=s(A_0)^2(dz^0)^2+\sum_{\mu =1}^{D-1}(A_\mu)^2(dz^\mu)^2
\een
Then we may employ a diagonal vielbein
$$e_\mu^{\bar{\mu}}=A_\mu$$
and work out (no sums over $\mu$ and $\nu$):
\bea
\Om_{\nb\mb\mb}&=&-\Om_{\mb\nb\mb}=\eta_{\mb\mb}\frac{1}{A_\mu A_\nu}
\pa_\nu A_\mu\nn
\om_{\mu\mb\nb}&=&\eta_{\mb\mb}\frac{1}{A_\nu}\pa_\nu A_\mu\nn
S_{\mb\nb\mb\nb}&=&\eta_{\nb\nb}\frac{1}{A_\mu^2}\left (
-\pa_\mu\pa_\mu\log A_\nu
-(\pa_\mu \log A_\nu)^2+(\pa_\mu \log A_\mu)(\pa_\mu\log A_\nu)\right )\nn
&&+\eta_{\mb\mb}\frac{1}{A_\nu^2}\left (
-\pa_\nu\pa_\nu\log A_\mu
-(\pa_\nu \log A_\mu)^2+(\pa_\nu \log A_\nu)(\pa_\nu\log A_\mu)\right )\nn
K_{\mb\nb\mb\nb}&=&-\sum_{\kp\neq \mu,\nu}\eta_{\mb\mb}\eta_{\nb\nb}
\eta_{\kpb\kpb}\frac{1}{A_\kp^2}(\pa_\kp\log A_\mu)(\pa_\kp\log A_\nu)
\eea
When the metric only depends on one coordinate, $r$, these are the only 
non-vanishing components (up to symmetries). Also notice that 
$S_{\mb\nb\mb\nb}$ and $K_{\mb\nb\mb\nb}$ are never simultaneously 
non-vanishing. 

We define $f(r)$ by
\ben
f(r)r^{d-1}\equiv BC^pF^{-1}(Gr)^{d-1}
\een
Using the above formulas, we may then work out
\bea
{R^{\tb}}_{\tb}&=&-\frac{1}{F^2}\left ((\log B)''+
(\log B)'(\log(fr^{d-1}))'\right )\nn
{R^{\bar{i}}}_{\bar{i}}&=&-\frac{1}{F^2}\left ((\log C)''+
(\log C)'(\log(fr^{d-1}))'\right )\nn 
{R^{\rb}}_{\rb}&=&-\frac{1}{F^2}\left ((\log (Ffr^{d-1}))''-
(\log F)'(\log (Ffr^{d-1}))' +((\log B)')^2 \right.\nn
&& \left.+p((\log C)')^2 +(d-1)((\log Gr)')^2\right )\nn
{R^{\bar{\a}}}_{\bar{\a}}&=&-\frac{1}{F^2}\left ((\log Gr)''+
(\log Gr)'(\log(fr^{d-1}))' -(d-2)\frac{F^2}{G^2r^2}\right )
\eea
The equations of motion for the metric and the dilaton for the electric case 
then take the forms (no summation over indices):
\bea
{R^{\tb}}_{\tb}&=&-\frac{(d-2)e^{-a_n\phi}n!}{2(D-2)}F_n^2\equiv 
-(d-2)\frac{K^2}{F^2}\nn
{R^{\bar{i}}}_{\bar{i}}&=&-(d-2)\frac{K^2}{F^2}\nn
{R^{\rb}}_{\rb}&=&-(d-2)\frac{K^2}{F^2}+\frac{1}{2F^2}(\phi')^2\nn
{R^{\bar{\a}}}_{\bar{\a}}&=&(p+1)\frac{K^2}{F^2}\nn
\phi''+\phi'(\log(fr^{d-1}))'&=&a_n(D-2)K^2\nn
K^2&\equiv&\frac{1}{2(D-2)}e^{-a\phi}F^2\frac{Q^2}{(Gr)^{2(d-1)}}
\label{equmot}
\eea
We see already here, that in order for the metric to reduce to a form
similar to $AdS_q\times S^{D-q}$, the Riemann tensor in each sub space 
has to be proportional to the metric tensor, and  
a necessary condition therefore is that the dilaton decouples and becomes a 
constant (in particular, zero). This requires either $n=5$ i.e. IIB string 
theory and $AdS_5\times S^5$, or  M-theory (or 11-dimensional supergravity)
where 2-branes and 5-branes are possible corresponding to $AdS_4\times S^7$
or $AdS_7\times S^4$.

\subsection{The extremal and non-extremal $p$-brane solutions}
We begin by providing the final solution (in the Einstein frame)
\bea
B&=&f^{\hf}H^{-\frac{d-2}{\Delta}},\ C=H^{-\frac{d-2}{\Delta}},\ 
F=f^{-\hf}H^{\frac{p+1}{\Delta}}, \ G=H^{\frac{p+1}{\Delta}},\ 
e^\phi=H^{a\frac{D-2}{\Delta}}\nn
&&\mbox{i.e.}\nn
ds^2&=&H^{-2\frac{d-2}{\Delta}}\left (sfdt^2+\sum_{i=1}^p(dx^i)^2\right )
+H^{2\frac{p+1}{\Delta}}\left (f^{-1}dr^2+r^2(d\Omega_{d-1})^2\right )\nn
H&=&1+\left (\frac{h}{r}\right )^{d-2},\ f=1-\left (\frac{r_0}{r}\right )^{d-2}
\nn
\Delta&=&(p+1)(d-2)+\hf a^2(D-2)\nn
h^{2(d-2)}+r_0^{d-2}h^{d-2}&=&\frac{\Delta Q^2}{2(d-2)(D-2)}
\label{branesolution}
\eea
Notice that indeed the diagonal metric tensor components 
tend to 1 as $r\ra\infty$. In the electric case
\ben
F_{ti_1...i_pr}=\e_{i_1...i_p}H^{-2}\frac{Q}{r^{d-1}}
\een
In the magnetic case the solutions are obtained by the above duality relations
\rf{emduality}. For the 5-form in IIB string theory we replace 
$F_5\ra F_5+*F_5$.

The above solutions are not the most general ones, but represent a 2-parameter 
sub-family of solutions. For $r_0=0$ we have $f\equiv 1$ and we obtain the 
extremal solution, depending only on a single parameter, $Q$ related to 
the common mass and charge density of the BPS D-brane. For $r_0\neq 0$ a 
horizon develops at $r=r_0$.

Thus we are seeking a 2-parameter solution to be represented in a suitable form 
by some convenient gauge choice. Let us try the ansatz
\ben
\log \left (\frac{B}{C}\right ) =c_B\log f,\ \log \left (\frac{F}{G}\right )
=c_F\log f
\een
with $c_B$ and $c_F$ constants to be sought for. Further define
\ben
g\equiv C^{p+1}G^{d-2}=f\frac{CG}{BF}=f^{1-(c_B-c_F)}
\een
From the Einstein equations we derive
\ben
(\log B-\log C)''+(\log B-\log C)'\left [(\log f)'+\frac{d-1}{r}\right  ]=0
\een
or
\ben
(\log f)''+((\log f)')^2+\frac{d-1}{r}(\log f)'=0
\een
giving
\bea
f''+\frac{d-1}{r}f'&=&0\nn
f&=&1-\left (\frac{r_0}{r}\right )^{d-2}
\eea
since we demand $f\ra 1$ for $r\ra\infty$, which is our first result for 
$f(r)$. From the equations of motion we further obtain
\ben
(\log g)''+(\log g)'\left [(\log fr^{d-1})'\right ]+\frac{d-1}{r}\left [
(\log f)'+\frac{d-2}{r}\Big(1-\left (\frac{F}{G}\right )^2\Big)\right ]=0
\een
This suggests trying to solve with $g\equiv 1$, hence $c_B-c_F=1$. Then
\ben
(\log f)'+\frac{d-2}{r}(1-f^{2c_F})=0
\een
or with the above solution for $f$
$$c_F=-\hf,\ B=f^{\hf}C,\  F=f^{-\hf}G$$
With this the remaining equations of motion become
\bea
(\log G)''+(\log G)'(\log fr^{d-1})'&=&-(p+1)K^2\nn
(a\log G-\frac{p+1}{D-2}\phi)''+(a\log G-\frac{p+1}{D-2}\phi)'(\log fr^{d-1})'
&=&0\ \Ra\nn
a\log G&=&\frac{p+1}{D-2}\phi
\eea
{\bf Exercise:} Complete the calculation and derive the solution,
 \rf{branesolution}.

\subsection{The extremal, non-dilatonic solutions. The near horizon 
approximation}
The extremal solution is obtained by putting $r_0=0$. It corresponds to the
brane being in the ground state in a quantum description. The non-extremal 
solution (presumably) represents excitations, corresponding to a definite 
temperature. In the extremal case we further consider the case with no 
dilaton coupling. We have already identified the relevant cases as 
$$(D,n)=(10,5),\ (11,4),\ (11,7)$$
For these values, fortuitously one has the ``accidental'' identity
\bea
\Delta &=&(p+1)(d-2)=2(D-2)\Ra\nn
h^{d-2}&=&\frac{Q}{d-2}
\eea
The solution then simplifies as follows:
\bea
f(r)&\equiv&1\nn
H&=&1+\frac{Q}{(d-2)r^{d-2}}\nn
ds^2&=&H^{-\frac{2}{p+1}}(sdt^2+\sum_{i=1}^p(dx^i)^2)+
H^{\frac{2}{d-2}}\sum_{a=1}^d(dy^a)^2\nn 
\sum_{a=1}^d(dy^a)^2&\equiv& dr^2+r^2(d\Om_{d-1})^2
\eea
We now want to consider $N$ coincident branes. For a single extremal $Dp$
brane, the flux as normalized in \rf{electricflux} is given by 
\cite{polchinski,alwis}
\ben
\mu_p=T_p\sqrt{16\pi G_{10}}
\label{fluxunit}
\een  
where the $p$-brane tension and Newton's constant are given by
\bea
T_p&=&\frac{2\pi}{(2\pi\ell_s)^{p+1}g_s}\nn
16\pi G_{10}&=&\frac{(2\pi\ell_s)^8}{2\pi}g^2_s
\eea
In fact in $11$-dimensional supergravity, almost the same formulas apply, with
an obvious change in dimensionality, but with the understanding that $g_s$ is
absent (say $g_s\equiv 1$ in $11$ dimensions). With this somewhat vulgar 
notation, we may write
\bea
T_p&=&\frac{2\pi}{(2\pi\ell)^{p+1}g_s}\nn
16\pi G_D&=&\frac{(2\pi\ell)^{D-2}}{2\pi}g^2_s
\eea
The length $\ell$ is the string length $\ell_s^2=\a'$ for $D=10$ and the 
$11$-dimensional Planck length for $11$-dim. sugra.
We therefore insist that we should choose $Q$ so that
\ben
\frac{\mu_p}{T_p\sqrt{16\pi G_D}}=N
\label{fluxnormalization}
\een
or from \rf{electricflux}
\ben
\frac{Q\Om_{d-1}}{15\pi G_D}\frac{(2\pi\ell)^{p+1}g_s}{2\pi}=N
\een
or
\ben
Q=Ng_s\frac{(2\pi\ell)^{d-2}}{\Om_{d-1}},\ \ 
h_d^{d-2}=Ng_s\frac{(2\pi\ell)^{d-2}}{(d-2)\Om_{d-1}}
\een
The Maldacena-conjecture arises by considering the so-called near horizon limit 
in which we consider the region very close to $r=0$ and subsequently scale 
this region up in a singular way to be described. In this limit we simply have
\ben
H\simeq \frac{h^{d-2}_d}{r^{d-2}}
\een
We see in particular that the $r^2$ in front of $d\Om_{d-1}^2$ will get 
cancelled so that the metric becomes a direct product with an $S^{d-1}$.
We now consider the various cases in turn.

\underline{$D=10,\ p+1=4,\ d=6\ra AdS_5\times S^5$}

This is the case of $N$ $D3$ branes in IIB string theory. So according to the
above prescription we take (cf. also \rf{volumesd})
\ben
H=1+\frac{4\pi g_s N\ell_s^4}{r^4}
\een
Also define the scaled variable
\ben
U=r/\ell_s^2
\een
We consider the limit $\a'=\ell_s^2\ra 0$ and again also $r\ra 0$ in such a 
way that $U$ becomes the meaningful variable:
\bea
H&\simeq &\frac{4\pi g_sN}{U^4\ell_s^4}\nn
ds^2&=&\ell_s^2\left \{\frac{U^2}{\sqrt{4\pi g_sN}}dx_4^2+\sqrt{4\pi g_sN}
\left (\frac{dU^2}{U^2}+d\Om_5^2\right )\right \}\nn
&=&\frac{U^2}{L^2}d\tilde{x}_4^2+L^2\frac{dU^2}{U^2}+L^2d\Om_5^2
\eea
with 
$$dx_4^2\equiv s dt^2+\sum_{i=1}^3 (dx^i)^2$$
and $\tilde{x}$ a suitably scaled version of the coordinate $x$.
Comparing with \rf{maldacenametric} we see that we exactly have the metric of 
$AdS_5(L)\times S^5(L)$ where we have indicated the length parameter, $L=b$, of 
$AdS_5$ and and the radius $L$ of $S^5$. When $\ell_s\ra 0$ the metric has to 
be 
rescaled to get a finite result - by removing the $\ell_s^2$ overall factor. 
This is the singular blowing up alluded to above. The radius parameter is given 
by
\ben
L^4=b^4=4\pi g_sN\ell_s^4
\label{ads5radius}
\een
Also, compared with \rf{maldacenametric} $\xi^i=\frac{\a'}{b^2}x^i$.

\underline{$D=11,\ p+1= 6,\ d=5\ra AdS_7\times S^4$}

This is the case of $N$ $M5$-branes. We find
\bea
ds^2&=&H^{-\frac{1}{3}}dx_6^2+H^{\frac{2}{3}}dy_5^2=H^{-\frac{1}{3}}dx_6^2+
H^{\frac{2}{3}}(dr^2+r^2d\Om_4^2)\nn
H&=&1+\frac{\pi N\ell_{11}^3}{r^3}\sim \frac{\pi N\ell_{11}^3}{r^3}
\eea
This time define
\bea
U^2&\equiv& \frac{r}{\ell_{11}^3}\nn
H&\simeq&\frac{\pi N}{\ell_{11}^6}\frac{1}{U^6}
\eea
Then
\bea
dr&=&\ell_{11}2UdU,\ \ r^2=\ell_{11}^6U^4\nn
ds^2&=&\ell_{11}^2\left\{\frac{U^2}{(\pi N)^{1/3}}dx_6^2+
4(\pi N)^{2/3}\frac{dU^2}{U^2}+(\pi N)^{2/3}d\Om_4^2\right \}\nn
&=&\frac{U^2}{4L^2}d\tilde{x}_6^2+4L^2\frac{dU^2}{U^2}+L^2d\Om_4^2
\eea
which this time is the metric of 
$$AdS_7(2L)\times S^4(L)$$
with
\ben
L^2=(\pi N)^{2/3}\ell_{11}^2
\label{L74}
\een

\underline{$D=11,\ p+1= 3,\ d=8\ra AdS_4\times S^7$}

This is the case of $N$ $M2$-branes. Here
\bea
ds^2&=&H^{-2/3}dx_3^2+H^{1/3}(dr^2+r^2d\Om_7^2)\nn
H&=&1+\frac{N\ell_{11}^62^5\pi^2}{r^6}\simeq \frac{N\ell_{11}^62^5\pi^2}{r^6}
\eea
We introduce
\bea
U^{\hf}&=&\frac{r}{\ell_{11}^{3/2}}\nn
H&\simeq& \frac{2^5\pi^2 N}{\ell_{11}^3U^3}\nn
ds^2&=&\ell_{11}^2\left\{\frac{U^2}{(2^5\pi^2N)^{2/3}}dx_3^2+
\left ( \hf\pi^2 N\right )^{1/3}\frac{dU^2}{U^2}+4
\left ( \hf\pi^2 N\right )^{1/3}d\Om_7^2\right\}\nn
&=&\frac{4U^2}{L^2}d\tilde{x}_3^2+\frac{L^2}{4}\frac{dU^2}{U^2}+L^2d\Om_7^2
\eea
corresponding to
$$AdS_4(\hf L)\times S^7(L)$$
with
\ben
L^2=4\ell_{11}^2\left ( \hf\pi^2 N\right )^{1/3}
\een
{\bf Exercise} show that the spaces, $AdS_5(L)\times S^5(L)$, 
$AdS_7(2L)\times S^4(L)$ and $AdS_4(\hf L)\times S^7(L)$ are in fact {\em exact}
solutions to the equations of motion of the appropriate low energy effective
Lagrangians, in particular, that the last term in \rf{sugra11} does not cause
any modifications. Notice, however, that unlike the brane solutions, these 
solutions do not become asymptotically flat.

\subsubsection{Non-extremal branes}
For completeness and later reference we give her also the form of the 
non-extremal $p$-brane solutions \rf{branesolution} in the near horizon 
approximation and with the same scaled $U$-variables as in the extremal case. We
have used an independent scaling of the boundary coordinates $t,x^1,...,x^{n-1}$
in some cases.

\underline{$D=10,\ p+1=4,\ d=6\ra AdS_5\times S^5$}

\bea
ds^2&=&\frac{U^2}{L^2}(s f(U)dt^2+d\vex_3^2)+L^2
\frac{dU^2}{f(U)U^2}+L^2d\Om_5^2\nn
L^2&=&\ell_s^2\sqrt{4\pi g_sN}\nn
f(U)&=&1-\left (\frac{U_0}{U}\right )^4
\label{nonexn4}
\eea

\underline{$D=11,\ p+1= 6,\ d=5\ra AdS_7\times S^4$}

\bea
ds^2&=&\frac{U^2}{4L^2}(s f(U)dt^2+d\vex_5^2)+
4L^2\frac{dU^2}{f(U)U^2}+L^2d\Om_4^2\nn
L^2&=&\ell_{11}^2(\pi N)^{2/3}\nn
f(U)&=&1-\left (\frac{U_0}{U}\right )^6
\label{nonexn6}
\eea

\underline{$D=11,\ p+1= 3,\ d=8\ra AdS_4\times S^7$}

\bea
ds^2&=&\frac{4U^2}{L^2}(sf(U)dt^2+d\vex_2^2)+
\frac{L^2}{4}\frac{dU^2}{f(U)U^2}+L^2d\Om_7^2\nn
L^2&=&4\ell_{11}^2\left ( \hf\pi^2 N\right )^{1/3}\nn
f(U)&=&1-\left (\frac{U_0}{U}\right )^3
\label{nonexn3}
\eea

\subsection{The Brane Theory and the Maldacena Conjecture}
From the discussion in the previous section we know that IIB string theory on
$AdS_5$ compactified on $S^5$ or M-theory on $AdS_4$ compactified on $S^7$ 
or finally on $AdS_7$ compactified on $S^4$, are all quantum theories with
isometry groups (for Minkowski signature) $SO(2,4),\ SO(2,3)$ and $SO(2,6)$
respectively. The remarkable Maldacena conjecture \cite{maldacena} 
is that these 
various quantum theories are exactly mathematically equivalent to (dual to)
certain quantum theories on the boundary of the relevant  $AdS$ spaces, i.e. 
on the coincident branes. 

What could these brane theories be? They would have to be conformally 
invariant quantum field theories according to the discussion in the previous
section. And in particular for the case of $AdS_5\times S^5$ the argumentation
is perhaps not too far fetched. The important point is that we have seen that
what we are looking at, is a small portion of space-time very close to the 
branes, and then subsequently blown up by formally letting $\a'\ra 0$. But in
precisely that limit we think we know what the effective quantum theory on the 
$N$ coincident $D3$-branes should be \cite{witten95}: It should be $N=4$ super 
Yang Mills with gauge group $U(N)$. (For references on D-branes, see 
\cite{polchinski,johnson98}).

Let us try a heuristic argumentation: Excitations of $D$-branes may be thought 
of in terms of open strings with end-points on the $D$-branes. We are dealing 
with IIB string theory which is a theory of oriented strings, and there are
$N$ different branes for the strings to end on, so these open strings are 
automatically equipped with Chan-Paton labels relevant to $U(N)$. They 
interact with each other, but in the singular limit, $\a'\ra 0$ we consider, 
only the zero-mass modes need be considered, and in 10-dimensions they form the
$N=1$ multiplet of pure Super Yang-Mills (for gauge group $U(N)$). Also,
as has been known since the middle 1970'ies \cite{ss74}, 
they interact exactly according to
that theory, the gauge coupling being related to the open string coupling
$g^o_s$ and $\a'$. For a given Feynman-diagram, the corresponding 
open-string-diagram will have boundaries, which come with a natural orientation,
because the
string is oriented (the string connects two boundaries and is oriented, so the 
two boundaries are different: have opposite orientations). Further the
different boundaries have labels $i=1,...,N$ equal to the label of the 
$D3$-brane. Thus the open string-diagram has an appearance identical to
the (super) Yang-Mills Feynman-diagram in the 't Hooft double line 
representation for $U(N)$. In 10 dimensions such a string theory is anomalous,
but here we are considering end points restricted to the $D$-branes and there is
no such problem. 
  
In our case, however, the zero-mass particles are confined 
strictly to the 4-dimensional world volume of the coincident $D3$ branes. Hence
the theory is naturally 10-dim SYM dimensionally reduced to 4 dimensions, and
that is exactly the $N=4$ SYM theory mentioned above. The SYM-coupling $g_{YM}$
is essentially the open string coupling constant which is itself the square root
of the closed IIB string coupling constant. More precisely for $N$ $p$-branes:
\ben
\frac{g^2_{YM}}{4\pi}= g_s(2\pi\ell_s)^{p-3}
\label{YMcoupling}
\een 
(in a suitable normalization). The dimensionality is the well known one.\\[.3cm]
{\bf Exercise:} Derive \rf{YMcoupling} from the Born-Infeld-action
\ben
I_{BI}=T_pTr\left\{\int d^{p+1}x
\sqrt{\det\Big(G_{\mu\nu}+2\pi\ell_2^2F_{\mu\nu}\Big)}\right\}
\een
Here, for the gauge group $U(N)$, $F_{\mu\nu}$ is treated as an $N\times N$
matrix, but the ``det'' refers only to the $(p+1)\times(p+1)$ index structure of
indices $\mu\nu$. The interpretation of the determinant in this case is via the
symmetrized trace, denoted by $Tr$ above. However, in the calculation needed 
here, only an expansion to 2nd order is required, and no ambiguity exists. 
\vspace{.3cm}
So for a $D3$-brane
\ben
\frac{g^2_{YM}}{4\pi}= g_s,\ \ \ D3\mbox{-brane}
\een
From the perspective of the 4-dim. field theory, $N$ is the $N$ of the gauge 
group $U(N)$. From the perspective of IIB string theory on $AdS_5\times S^5$,
$N$ is the flux (normalized carefully as in \rf{fluxnormalization}) 
through $S^5$. 

So, the remarkable conjecture is that IIB quantum {\em string} 
theory on $AdS_5$ compactified on $S^5$ is identically equivalent to the 
quantum {\em field} theory in 4 dimensions. Let us collect a number of points
in favour of this hypothesis. We shall try to compare properties of the two
quantum theories: (i) IIB string theory on $AdS_5\times S^5$ with $N$ units of 
5-form flux through $S^5$ - to be referred simply as IIB; and (ii) $N=4$ 
super-Yang-Mills with gauge-group $U(N)$ in 4 dimensions - to be referred to
simply as SYM.

First compare global symmetries. The IIB-theory has an isometry group 
$SO(2,4)\times SO(6)$ with the last $SO(6)$ being the isometry group of the 
5-sphere. Actually, because spinors are involved the relevant groups for 
$AdS_5$ and the 
sphere, $S^5$ are the covering groups $SU(4)$ of $SO(6)$ and $SU(2,2)$ of 
$SO(2,4)$, so we have $SU(2,2)\times SU(4)$. But the 32 Majorana spinor
supercharges of the IIB theory (which are all preserved in this background)
transform under this symmetry in such a way that in fact the full invariance
is given by the Lie-supergroup $SU(2,2|4)$. 

Now we should try to understand that this is also the relevant 
invariance to consider for the SYM. We have already understood that the 
$SO(2,4)$ or $SU(2,2)$ part is realized as a conformal invariance. Indeed the 
SYM is known to have vanishing beta-function and be conformally invariant. How
about $SO(6)$ (or $SU(4)$)? does the SYM theory know about the 5-sphere? Yes,
indeed. That is the R-symmetry of SYM. In fact, consider briefly the field 
content of SYM:

In $10$ dimensions, $N=1$ pure Super Yang-Mills contains the gauge field 
potentials $A_\mu, \ \mu=0,1,...,9$ (the ``gluons'') 
giving $10-2=8$ bosonic physical 
degrees of freedom, all in the adjoint representation of the $U(N)$. 
Further we have the $8$-dimensional Majorana-Weyl ``gluinos'' 
$\l_\a, \ \a =1,...,8$, also
all in the adjoint representation. The theory has $16$ Majorana supercharges
$Q_\a, \ \a =1,...,16$. Under dimensional reduction, the gluon fields turn into
$4-2=2$ gauge fields and a remaining 6 scalar fields, $\phi_1,...,\phi_6$. The
gluino fields turn into $4\cdot 2$ Weyl spinors in 4 dimensions, 
$\l_\a^A, \ \a =1,2, \ A=1,2,3,4$. An $N=1$ description in 4 dimensions put
one of these spinors together with the gauge field in a gauge superfield, and
the remaining 3 spinors each combine with a pair of scalars to give 3 
scalar chiral superfields. The 16 supercharges turn into 4 sets of complex 
Majoranas $Q_\al^A, \bar{Q}^A_{\dot{\a}},\ \ \a =1,2, \ A=1,2,3,4$ 
transforming as a $\{\mbox{\bf 4}\}$ and a $\{\bar{\mbox{\bf 4}}\}$
of the R-symmetry group $SU(4)$, and the $\phi_i$ transform as a 
$\{\mbox{\bf 6}\}$ (the fundamental rep. of $SO(6)$ or the antisymmetric rank 2
tensor under $SU(4)$), so we see that $SO(6)$ (or $SU(4)$) is indeed present. 

But the IIB theory had 32 fermionic supercharges, the SYM only 16. 
Indeed from the perspective of the $N$ coincident BPS $D3$-branes, half the IIB 
supersymmetries are broken. In any case, where are the
remaining 16 fermionic generators? The answer is that they arise as part of an 
extension of the conformal group that takes place when supersymmetry is 
present as described in the famous paper by  
Haag, Lopuzanski and Sohnius \cite{hls75}.
For completeness let us give here their form of the superconformal algebra:

Define first
\ben
P_{\a\dot{\a}}\equiv P_\mu\sg^\mu_{\a\dot{\a}}, \ \ 
K_{\a\dot{\a}}\equiv K_\mu\sg^\mu_{\a\dot{\a}}
\een
and
\ben
\sg^\mu_{\a_1\dot{\b_1}}\sg^\nu_{\a_2\dot{\b_2}}M_{\mu\nu}\equiv
M_{\a_1\a_2}\e_{\dot{\b_1}\dot{\b_2}}+\bar{M}_{\dot{\b_1}\dot{\b_2}}
\e_{\a_1\a_2}
\een
In addition to the generators of the conformal algebra with commutation 
relations, \rf{confalgebra}, one now has the {\em 16 new fermionic generators} 
$$Q_\b^{(1)A}, \bar{Q}_{\dot{\b}}^{(1)A}$$
obtained as
\ben
{[}K_{\a\dot{\b}},Q^A_\g{]}=2i\e_{\a\g}\bar{Q}_{\dot{\b}}^{(1)A}
\een
The $R$-symmetry $SU(4)$ generators, $T^{AB}$, commute with the supersymmetry
generators in a way dictated by the fact that these transform as a fundamental 
and and anti-fundamental 4-dimensional multiplet under $R$.
Some of the remaining commutation relations are
\bea
{[}Q_\a^A,D{]}&=&\frac{i}{2}Q_\a^A\nn
{[}P_{\a\dot{\b}},\bar{Q}_{\dot{\g}}^{(1)A}{]}&=&2i\e_{\dot{\b}\dot{\g}}
Q_\a^A\nn
\{Q_\a^A,\bar{Q}_{\dot{\b}}^{(1)B}\}&=&\del^{AB}K_{\a\dot{\b}}\nn
\{Q_\a^A,Q_\b^{(1)B}\}&=&\del^{AB}(\e_{\a\b}D+M_{\a\b})+i\e_{\a\b}T^{AB}\nn
{[}\bar{Q}_{\dot{\a}}^{(1)A},T^{BC}{]}&=&2\del^{AB}\bar{Q}_{\dot{\a}}^{(1)C}
-\hf\del^{BC}\bar{Q}_{\dot{\a}}^{(1)A}
\label{superconfalgebra}
\eea
Indeed the combination of \rf{confalgebra} and 
\rf{superconfalgebra} together with the standard $N=4$ supersymmetry algebra 
constitute the Lie super-algebra $SU(2,2|4)$.

It is well known that the IIB theory (almost certainly) contains a 
(non-perturbative) $SL(2,\Z)$
invariance \cite{schwarz95}. It is best viewed as arising from compactification 
of M-theory on a 2-torus with modular parameter
$$\tau=\chi+ie^{-\phi}$$
with $\chi$ the RR-scalar of IIB (the ``axion''). In $N=4$ Super Yang-Mills 
there is a corresponding $SL(2,\Z)$ invariance of the theory with modular
parameter in this case
$$\tau=\frac{\theta}{2\pi}+\frac{4\pi i}{g_{YM}^2}$$
In this latter case the symmetry is represented by an $SL(2,\Z)$ transformation
of the lattice of electric and magnetic charges in that theory:
$q+ig=g_{YM}(n_e+\tau n_m)$ by treating $(n_e,n_m)$ as a doublet.

This finishes our very brief comparison of symmetries of the two theories.

\subsection{Implication of the Maldacena conjecture}
We continue for definiteness to focus on the case of $AdS_5\times S^5$. As we 
have seen in \rf{ads5radius} the common ``radius-'' or length-parameter is given
be
\ben
b^4=\ell_s^4 4\pi g_sN=\ell_s^4 g^2_{YM}N=\ell_s^4\l
\een
with 
$$\l\equiv g^2_{YM}N$$
the 't Hooft coupling relevant to large $N$ Yang-Mills theory. Thus,
it is tempting in particular to consider the limit $\l$ fixed while 
$N\ra\infty$. We see that in this limit the string coupling tends to zero, so 
that we may perform calculations on the string theory side, simply by 
restricting ourselves to string tree-diagrams, the {\em classical} limit of 
string theory! The full {\em quantum} non-perturbative description of $N=4$ 
Super Yang-Mills would be obtained from this classical theory - in the large 
$N$ limit. This seems like a program which might have some success eventually,
even though the NS-R formulation of IIB string theory on the $AdS_5\times S^5$
background with $N$ units of RR 5-form flux is unknown (for a preliminary 
attempt,
see \cite{dpolyakov98}). In the Green-Schwarz-formulation there is a proposal
\cite{kt98}, \cite{p98}, but non-trivial calculations remain to be performed.
In any case, this looks like a concrete proposal for the so called 
Master-field idea of Witten, that the large-$N$ $U(N)$ Yang-Mills
theory path integral could be described by a single field configuration,  
reminiscent indeed of classical theory.

The situation becomes even more astonishing if we furthermore consider the 
strong coupling limit, i.e. large $\l$ limit of SYM. If we keep the AdS/sphere
radii fixed, we are therefore dealing with the $\ell_s^2=\a'\ra 0$ limit of 
string theory, or in other words, with the limit in which classical string
theory simply becomes classical supergravity! This is the limit mostly 
considered in concrete calculations so far. Even in that extreme limit the 
conjecture has dramatic predictions: It predicts how the SYM theory at large 
$N$ behaves in the extreme non-perturbative, strong coupling regime. String 
excitations become infinitely heavy and decouple in that limit, but since we 
kept the radius of the sphere $S^5$ fixed, we cannot at all neglect the 
Kaluza-Klein states associated with the compactification on that sphere.
(Strictly speaking one cannot send the dimensionful string length 
to zero. What one means by this is to consider energy scales for which string
excitations may be neglected. If we consider the string length fixed and still 
take $\l$ large, the radius of $S^5$ tends to infinity. People often phrase the 
situation that way. Then all curvatures are ``small'' and quantum gravity
corrections may be neglected: 
classical supergravity is adequate. KK-masses now become very small. 
Obviously the situation is entirely equivalent to our formulation, but as usual
one properly has to consider dimensionless ratios for such arguments to make
sense.)

\section{A Detailed Specification of the Conjecture. Sample Calculations.}
\subsection{Presentation of the idea}
Maldacena conjectured an equivalence - a duality - between two theories:
(i) String/M-theory on a manifold
of the form $AdS_d\times{\cal M}$, with ${\cal M}$ being a compactification
manifold, and (ii) an appropriate conformal field theory on the boundary of 
$AdS_d$. But his conjecture did not specify the precise way in which these two 
theories should be mapped onto each other. Subsequently a detailed proposal was
made independently by Gubser, Klebanov and Polyakov \cite{gubser} and by Witten
\cite{witten1}. Here we shall describe that and carry out several sample 
calculations considered in ref. \cite{witten1} filling in a few details
(while leaving several aspects of \cite{gubser,witten2} untreated).
For definiteness of presentation we would think of the canonical 
$AdS_5\times S^5$ example, but in fact the discussion will be much more general,
ignoring mostly the details of the compactification manifold.

On the boundary theory of $n$-dimensional Minkowski (or Euclidean) space-time,
we should typically like to understand a general correlator of the the form
$$\br {\cal O}_1(x_1)...{\cal O}_q(x_q)\kt$$
One wants to know how to obtain the same object in terms of the quantum theory 
on $AdS_{n+1}$. The proposal of \cite{gubser} and \cite{witten1} 
is to identify this
object with the result of a path integral in the $AdS_{n+1}$ theory, 
with certain
fields attaining specific boundary properties. The fields in question would
be related to the corresponding boundary theory operators ${\cal O}_i(x_i)$
by the requirement that their boundary values - to be 
identified in a non-trivial 
way - should couple to the operators in a way consistent with the symmetries 
of the problem. In practice the resulting path integral might be evaluated by
means of generalized Feynman diagrams with $q$ ``external'' propagators 
``ending'' on the boundary at points $x_i$. In particular in the large $N$ 
limit, as we have seen, the Feynman diagrams would be tree diagrams only, albeit
string tree diagrams in general. If furthermore we consider the limit of large
't Hooft coupling, the string tree diagrams become tree diagrams of 
supergravity.  We shall need to understand how to evaluate these generalized
propagators. A particularly neat formulation is possible if we can construct
a standard generating functional for the correlators on the boundary:
\ben
Z(\{\phi_i\})=\sum_q\frac{1}{q!}\int\prod_{k=1}^q d^nx_k
\br {\cal O}_1(x_1)...{\cal O}_q(x_q)\kt\phi_1(x_1)...\phi_q(x_q)
=\br exp\{\int dx\sum_i\phi_i(x){\cal O}_i(x)\}\kt
\een
This requires short distance singularities of the correlators to be integrable,
or else the introduction of some device to render the expression meaningful.
In any case we might always go back to considering individual correlators. 
Obviously, if the operators ${\cal O}_i$ on the boundary CFT have conformal
dimension $\Del_i$ then the currents $\phi_i$ should have conformal dimension
$n-\Del_i$. Similarly any other quantum number the operators may have (say as
multiplets of $SU(2,2|4)$ in the case of $AdS_5\times S^5$) would have to be 
supplemented by conjugate quantum numbers of the currents so that singlets may
be formed.
Supposing the generating functional of the ``currents'' $\{\phi_i(x)\}$ 
($x\in\pa \Big(AdS_{n+1}\Big)\sim E^{n}$) makes
sense, the description of this object in $AdS_{n+1}$ should be by a path 
integral with fields $\phi_i(y)$ in that theory 
(i.e. with $y\in AdS_{n+1}$) tending to the boundary currents $\{\phi_i(x)\}$
in a certain prescribed way, that we shall have to infer. In the large $N$ 
limit we would just have to work out the classical action, and in the large 
$N$ strong coupling limit, just the classical supergravity action on fields
satisfying the equations of motion and tending to the prescribed boundary
``currents'' in some particular way.

\subsection{Free scalar fields on $AdS_{n+1}$}
By a scalar field we mean one which transforms as a scalar under the $AdS$ 
isometry group, hence it would tend to a boundary value with conformal 
dimension zero, and couple to operators with conformal dimension $n$. In that 
case it turns out to be meaningful to simply require the path integral on 
$AdS_{n+1}$ with the scalar field in question tending to a definite {\em value}
on the boundary
$$\phi(x)\ra\phi_0(x')$$
for $x\in AdS_{n+1}$ and $x'\in E^n$ and (somehow) $x\ra x'$. Let us study the 
case of free scalar fields. Here the path integral becomes trivial, and is 
simply  equal to the exponential of (minus or $i$ times) the classical action,
up to a normalization constant. Thus in this case the classical approximation 
is the exact result. The action on $AdS_{n+1}$ is
\ben
I(\phi)=\hf\int_{AdS_{n+1}}d^{n+1}x \sqrt{g}\pa_\mu\phi\pa^\mu\phi
\een
So we seek a classical field, satisfying the equation of motion
\ben
D_\mu D^\mu\phi(x)=\frac{1}{\sqrt{g}}\pa_\mu\Big (\sqrt{g}\pa^\mu\phi(x)\Big )=0
\label{scalareqmot}
\een
throughout $AdS_{n+1}$, but such that $\phi(x)\ra\phi_0(x')$ whenever the point 
$x$ in $AdS_{n+1}$ runs away to $\infty$ in the particular way that defines the 
boundary point $x'\in E^n$ of the boundary.

It is plausible that the classical solution to this problem is unique. Indeed, 
imagine $\phi_1(x)$ and $\phi_2(x)$  both being solutions of the equation of
motion with the same boundary value. Then $\d\phi(x)\equiv\phi_1(x)-\phi_2(x)$ 
has boundary value zero and also satisfies \rf{scalareqmot}. We should show 
that $\d\phi$ vanishes identically. Since it vanishes at infinity we take 
it to be square integrable. Then
\ben
0=-\int d^{n+1}x\sqrt{g}\d\phi D_\mu D^\mu\d\phi=\int d^{n+1}x\sqrt{g}\pa_\mu
\d\phi \pa^\mu\d\phi
\een
Since $\pa_\mu\d\phi \pa^\mu\d\phi$ is positive (semi) definite (in the 
Euclidean case), we find
\ben
\pa_\mu\d\phi\equiv 0
\een
or $\d\phi$ constant. But as it tends to zero, it must vanish everywhere.
 
We may solve the problem in terms of a Greens function, the generalized 
propagator, 
$$K(x,x')$$
with $x\in AdS_{n+1}$ and $x'\in E^n$ (or $E^{1,n-1}$ in the Minkowski case),
or since we have seen the boundary to be compactified, $S^n$, which we will
take to denote the boundary. Thus we seek a solution to the problem
\ben
\frac{1}{\sqrt{g}}\pa^x_\mu\Big (\sqrt{g} \pa^\mu_xK(x,x')\Big )=0
\label{laplaceK}
\een
and somehow $K(x,x')$ a delta function when $x$ is on the boundary. Then we
may construct the sought for classical solution as
\ben
\phi(x)=\int_{S^n}d^nx' K(x,x')\phi_0(x'),\ \ x\in AdS_{n+1}
\een
Following Witten \cite{witten1}, we construct this propagator in the 
coordinates \rf{poincaremetric}, taking for definiteness Euclidean signature:
\ben
ds^2=\frac{1}{(x^0)^2}\sum_{\mu=0}^n(dx^\mu)^2
\een
Now $AdS_{n+1}$ is described by the upper half space $x^0>0$, and 
$x^\mu\equiv x_\mu, \ \mu =1,...,n$ are coordinates in $E^n$, the coordinates
of the boundary. However, we know that we must be dealing with a compactified 
boundary $S^n$, and there is an extra ``point at infinity'' described at some 
length around \rf{boundaryinversion}. The boundary here is $x^0=0$; the
single point at infinity is $x^0=\infty$, a single point indeed, no matter what
the values of the remaining coordinates are, since the metric
tensor vanishes there.    

In these coordinates then
\ben
g_{\mu\nu} =\frac{1}{(x^0)^2}\del_{\mu\nu},\ \sqrt{g}=\frac{1}{(x^0)^{n+1}}, \ 
g^{\mu\nu}=(x^0)^2\del^{\mu\nu}
\een
{\bf Exercise:} Verify by explicit calculation, that
\ben
K(x^0,\vex;\vex')=c\frac{(x^0)^n}{((x^0)^2+(\vex-\vex')^2)^n}
\een
satisfies Laplace's equation \rf{laplaceK} for $x^0\neq 0$ and 
$\vex\neq \vex'$, and becomes
the desired delta function in the limit $x_0\ra 0$.\\[.3cm]
We may also infer the above more elegantly from Witten's trick 
\cite{witten1}, useful in the sequel. First notice, that just as the scalar 
propagator in flat space is Poincar\'e invariant so the propagator in our case
is invariant under the AdS isometry group. Then let the boundary point 
$\vex'$ represent the point, $P$ at infinity:
\ben
K(x;P)=K(x^0,\vex;P)
\een 
This cannot depend on $\vex$ due to translation invariance, so in that 
particular case, $K$ is a function of $x^0$ only, and Laplace's equation 
becomes
$$\pa_0\Big(\sqrt{g}\pa^0 K(x^0)\Big)=0$$
Here
$$\pa_0=\frac{\pa}{\pa x^0};\ \pa^0 =g^{00}\pa_0=(x^0)^2\pa_0$$
so the equation becomes
\ben
\frac{d}{dx^0}\Big((x^0)^{-n+1}\frac{d}{dx^0}K(x^0)\Big)=0
\een
If we try to solve with $K(x^0)=c (x^0)^p$ we find
$$p(-n+p)=0$$
The solution with $p=0$ cannot describe something with delta function support 
``at infinity'', $x^0=\infty$, since such a delta function should vanish
on the boundary $x^0=0$ for any $\vex$. Hence 
$$K(x^0,\vex;P)=c(x^0)^n$$
Next apply a transformation to map $P\ra \vex'=\vec{0}$:
\ben
x^\mu\ra z^\mu\equiv\frac{x^\mu}{(x^0)^2+\vex^2},\ \mu=0,...,n
\een 
This transformation is indeed an $SO(1,n+1)$ isometry of $AdS_{n+1}$ 
(in this case with Euclidean signature). In fact we find with
$x^2\equiv (x^0)^2+\vex^2$
\ben
dz^\mu=\frac{x^2dx^\mu-x^\mu 2x\cdot dx}{(x^2)^2},\ 
(x\cdot dx\equiv\sum_{\nu=0}^nx^\nu dx^\nu)
\een
It follows that
\ben
dz^2=\frac{dx^2}{(x^2)^2}\Ra \frac{dz^2}{(z^0)^2}=\frac{dx^2}{(x^0)^2}
\een
so that the transformation is an isometry and the parametrization space is 
preserved. Under this mapping 
\bea
x^0&\ra&\frac{x^0}{(x^0)^2+\vex^2}\nn
K(x^0,\vex;P)&\ra&K(x^0,\vex;\vec{0})=c\frac{(x^0)^n}{((x^0)^2+\vex^2)^n}
\eea
as claimed. Finally from translational invariance in the boundary we find
\ben
K(x^0,\vex;\vex')=c\frac{(x^0)^n}{((x^0)^2+(\vex-\vex')^2)^n}
\een
For $x^0\ra 0^+$ this becomes proportional to a $\del^n(\vex-\vex')$. 
Indeed, clearly for $\vex'\neq\vex$, $K\ra 0$ for $x^0\ra 0$. Also
$$\int d^nx\frac{(x^0)^n}{((x^0)^2+\vex^2)^n}$$
is independent of $x^0$ and convergent, as seen by scaling to the new 
integration variable $x^i/x^0$,
so that a limit is obtained for $x^0\ra 0$, and we may adjust $c$ to get 
proper normalization if desired.

So we have obtained the sought for classical solution
\ben
\phi(x^0,\vex)=c\int d^nx'\frac{(x^0)^n}{((x^0)^2+(\vex-\vex')^2)^n}\phi_0(\vex')
\een
We want to evaluate the classical action on that field, remembering that
\ben
\Delta\phi=\frac{1}{\sqrt{g}}\pa_\mu\Big(\sqrt{g}\pa^\mu\phi\Big) =0
\een
or (as we have seen, since $\pa^\mu=(x^0)^2\pa_\mu$)
\ben
\sum_{\mu=0}^n\pa_\mu\Big((x^0)^{-n+1}\pa_\mu\phi\Big) =0
\een
Now
\bea
I(\phi)&=&\hf\int d^{n+1}x\sqrt{g}\pa_\mu\phi\pa^\mu\phi=
\hf\int d^{n+1}x(x^0)^{-n+1}\pa_\mu\phi\pa_\mu\phi\nn
&=&\hf\int d^{n+1}x\pa_\mu\left ( (x^0)^{-n+1}\phi\pa_\mu\phi\right )
-\hf\int d^{n+1}x\phi\left \{\pa_\mu\Big((x^0)^{-n+1}\pa_\mu\phi\Big)\right\}
\eea
The last term vanishes by the equation of motion, and the total derivative 
term vanishes in all directions except in the $x^0$ direction where we have a 
boundary. To avoid the divergence for $x^0=0$ we put first $x^0=\e$. Then
\ben
I(\phi)=-\hf\int_{x^0=\e} d^nx(x^0)^{-n+1}\phi(x^0,\vex)\pa_0\phi (x_0,\vex)
\label{scalaraction}
\een
In the limit $x^0\ra 0$ we may put $\phi(x^0,\vex)=\phi_0(\vex)$. We then 
evaluate
\bea
\pa_0\phi(x^0,\vex)&=&c\frac{\pa}{\pa x^0}
\int d^nx'\frac{(x^0)^n}{((x^0)^2+(\vex-\vex')^2)^n}\phi_0(\vex')\nn
&=&cn(x^0)^{n-1}\int d^nx'\frac{1}{(\vex-\vex')^{2n}}\phi_0(\vex')
+{\cal O}((x^0)^{n+1}), \ x^0\ra 0
\eea
Inserting into \rf{scalaraction} we see that the singular $x^0$ behaviour 
drops out and we get
\ben
I(\phi)=-\frac{cn}{2}\int d^nx d^nx'\frac{\phi_0(\vex')\phi_0(\vex)}
{(\vex-\vex')^{2n}}
\een
In the classical (super)gravity limit the generating function for operators 
${\cal O}(\vex)$ in the boundary theory coupling to the ``source'' 
$\phi_0(\vex)$, is then given by the exponential of (minus) that. It follows 
that in this (trivial) example, there are only connected 2-point functions
\ben
\br{\cal O}(\vex){\cal O}(\vex')\kt\sim\frac{1}{(\vex-\vex')^{2n}}
\label{2pointscalar}
\een
This is the expected result: The $SO(1,n+1)$-scalar, $\phi(x)$  
will couple to conformal 
operators of dimension $n$ on the boundary in order for the coupling
$$\int d^nx {\cal O}(\vex)\phi_0(\vex)$$
to be invariant. And conformal invariance (indeed dilatation invariance) 
suffices then to fix the form of the 2-point function to \rf{2pointscalar}. 
In the case of $AdS_5\times S^5$ this operator will turn out to be the 
$Tr(F^2)$ of the YM field strength (sect. 4.5).

Notice that the 2-point function \rf{2pointscalar} is characteristic of a 
{\em quantum} theory with a non trivial short distance singularity. This is so
even though it was derived from a classical calculation in the bulk of 
$AdS_{n+1}$.

\subsection{Massless abelian gauge field on $AdS_{n+1}$}
We continue to follow ref.\cite{witten1}.
A gauge field $A_\mu(x)$ in $AdS_{n+1}$ with $\mu = 0,1,...,n$ gives rise to a
field strength
$$F_{\mu\nu}=\pa_\mu A_\nu-\pa_\nu A_\mu$$
satisfying the free equation of motion (no currents)
\ben
\frac{1}{\sqrt{g}}\pa_\mu(\sqrt{g}F^{\mu\nu})=0
\label{freemaxwell}
\een 
Now again we seek to construct, first a propagator with one delta function 
source on the boundary. With that we then build any gauge field 
$A_\mu(\vex)$ in 
the bulk with the property that the components $A_i(x_0,\vex),\ i\geq 1$ tend
to prescribed functions on the boundary, corresponding a certain 1-form on
the boundary
\ben
A_0(\vex)=a_i(\vex)dx^i
\label{gaugeboundary}
\een
As before, we use the trick of first working out the propagator when the point 
on the boundary is the point $P$ ``at infinity''. Again, in that case 
we expect the propagator to be independent of any $\vex$. Further, the 
propagator should be a 1-form, but in that case, one with no $0$-component. 
The remaining components are all treated the same way. We treat the components 
in the boundary one by one.

Thus we look for a 1-form in the bulk of $AdS_{n+1}$, only depending on $x^0$, 
and with a single component only, say the $i$'th
$$A^{(i)}(x)=f(x^0)dx^i$$  
It should satisfy the equation of motion, \rf{freemaxwell}. We have
\bea
A^{(i)}_i&=&f(x^0),\ \ A^{(i)}_\mu\equiv 0,\ \mu \neq i\nn
F_{0i}&=&f'(x^0)=-F_{i0},\ \mbox{all other} \ F_{\mu\nu}\equiv 0\nn
F^{0i}&=&(x^0)^4f'(x^0)\nn
\sqrt{g}F^{0i}&=&(x^0)^{-n+3}f'(x^0)
\eea
Then the equations of motion give
\bea
\frac{d}{dx^0}(\sqrt{g}F^{0i})&=&\frac{d}{dx^0}
\left ( (x^0)^{-n+3}f'(x^0)\right )=0\Ra\nn
f'(x^0)&\propto&(x^0)^{n-3}\Ra f(x^0)\propto (x^0)^{n-2}\Ra\nn
A^{(i)}&=&\frac{n-1}{n-2}(x^0)^{n-2}dx^i \ (\mbox{fixed}\ i)
\eea
(the normalization is for later convenience). We hope this 1-form will have a 
delta-function singularity at $P$. We exhibit this as before using the 
$SO(1,n+1)$ isometry (inversion)
\bea
x^\mu&\ra&\frac{x^\mu}{(x^0)^2+\vex^2}\Ra\nn
A^{(i)}&\ra&\frac{n-1}{n-2}\left (\frac{x^0}{(x^0)^2+\vex^2}\right )^{n-2}
d\left (\frac{x^i}{(x^0)^2+\vex^2}\right )
\eea
This new propagator represents propagation from $\vex'=\vec{0}$ on the 
boundary to $(x^0,\vex)$ in the bulk. When we work out the derivatives we see
that this new 1-form propagator will have components along all the different 
$dx^\mu$'s. We may simplify, using the fact that the propagator 
is unique only up to 
gauge transformation. We shall in fact get a simpler expression if we subtract 
the ``pure gauge''
$$\frac{1}{n-2}d\left (\frac{(x^0)^{n-2}x^i}{ ((x^0)^2+\vex^2)^{n-1}}\right )$$
Then
\bea
A^{(i)}&=&\frac{n-1}{n-2}\left (\frac{x^0}{(x^0)^2+\vex^2}\right )^{n-2}
d\left (\frac{x^i}{(x^0)^2+\vex^2}\right )-\frac{1}{n-2}d
\left (\frac{(x^0)^{n-2}x^i}{ ((x^0)^2+\vex^2)^{n-1}}\right )\nn
&=&\frac{1}{n-2}\left\{(n-1)(x^0)^{n-2}\frac{x^i}{ ((x^0)^2+\vex^2)^{n-2}}
d\left (\frac{1}{(x^0)^2+\vex^2}\right )+(n-1)
\frac{(x^0)^{n-2}}{ ((x^0)^2+\vex^2)^{n-1}}dx^i\right.\nn
&-&\left.d\left (\frac{(x^0)^{n-2}x^i}{ ((x^0)^2+\vex^2)^{n-1}}\right )\right\}
\nn
&=&\frac{1}{n-2}\left\{(x^0)^{n-2}x^id\left (\frac{1}{((x^0)^2+\vex^2)^{n-1}}
\right ) +(n-1)\frac{(x^0)^{n-2}}{((x^0)^2+\vex^2)^{n-1}}dx^i\right.\nn
&-&\left.d\left (\frac{(x^0)^{n-2}x^i}{ ((x^0)^2+\vex^2)^{n-1}}\right )\right\}
\nn
&=&\frac{1}{n-2}\left\{-\frac{1}{((x^0)^2+\vex^2)^{n-1}}
d\left ((x^0)^{n-2}x^i\right )+(n-1)\frac{(x^0)^{n-2}}
{((x^0)^2+\vex^2)^{n-1}}dx^i\right\}\nn
&=&\frac{1}{((x^0)^2+\vex^2)^{n-1}}\left\{-(x^0)^{n-3}dx^0x^i+(x^0)^{n-2}dx^i
\right\}
\eea
Or,
\ben
A^{(i)}_\mu=\frac{1}{((x^0)^2+\vex^2)^{n-1}}\left\{\begin{array}{ll}
-(x^0)^{n-3}x^i&\mu=0\\
+(x^0)^{n-2}&\mu=i\\
0&\mbox{otherwise}\end{array}\right.
\een
We may now collect results and obtain for the general classical solution
1-form field
\bea
A^{(i)}(x^0,\vex)&=&\int d^nx'\sum_{i=1}^nA^{(i)}(x^0,\vex;\vex')a_i(\vex')\nn
&=&\int d^nx'\left\{\frac{(x^0)^{n-2}}{((x^0)^2+(\vex-\vex')^2)^{n-1}}
a_i(\vex')dx^i\right.\nn
&&\left. -(x^0)^{n-3}dx^0\frac{(x'-x)^ia_i(\vex')}{((x^0)^2+
(\vex'-\vex)^2)^{n-1}}\right\}
\label{classicalvector}
\eea
Notice that only the first term acts as a delta-function for $x^0\ra 0$. In 
fact, a function of the form
\ben
\frac{\e^\b}{(\e^2+\vex^2)^\a}
\label{deltafunction}
\een
is a model of $\del^n(\vex)$ only if $0<\b=2\a-n$. Then, namely
\bdm
\int d^nx\frac{\e^\b}{(\e^2+\vex^2)^\a}=\e^{\b +n-2\a}\int
d^n\left(\frac{x}{\e}\right) 
\frac{1}{\left (1+\left (\frac{\vex}{\e}\right)^2\right)^\a}
\edm
is a constant, independent of $\e$. For $x^0=\e$, the first term in 
\ref{classicalvector} is of that 
form. The last term in \rf{classicalvector} has an extra power of $\e$ 
and vanishes for $\e\ra 0$. This is perhaps not completely obvious. In fact
it looks like there is a power {\em less} of $\e$. But first we must remember to
scale also the factor $(x'-x)^i$, and second we see by expanding $a_i(\vex')$
around $\vex$ that the leading term vanishes since the integrand is odd, and 
the following terms {\em do} have extra powers of $\e$. 
Hence, clearly \rf{classicalvector} will tend to 
(up to a constant normalization)
\ben
A_0(\vex)=\sum_{i=1}^na_i(\vex)dx^i, \ \mbox{for } \ x^0\ra 0
\een
We are now instructed to evaluate the classical action for the classical 
solution \rf{classicalvector}.

In form language
\ben
I(A)=\hf\int_{AdS_{n+1}}F\wedge *F
\een
where $F=dA$, and the equation of motion is $*d*F=0$ i.e. $d*F=0$. Then
\ben
I(A)=\hf\int_{AdS_{n+1}} dA\wedge *F=\hf\int_{AdS_{n+1}} d(A\wedge *F)
=\hf\int_{\mbox{boundary}(\e)}A\wedge *F
\een
where again we take the boundary to be $x^0=\e$ at first, and only let 
$\e\ra 0$ at the end. On the boundary, we only need the $i=1,...,n$
components of $A\wedge*F$. Thus a component of $A$, $A_i$ has $i=1,...,n$. 
Also $*F$ has components $j_1,...,j_{n-1}=1,...,n$, but never $0$ or $i$. 
Therefore we need exactly the components $F_{0i}$. In coordinates on the
boundary
\ben
I\sim\int d^nx\sqrt{h}A^\ell n^0F_{0\ell}
\een
Near the boundary the metric is 
$$h_{ij}=\frac{1}{(x^0)^2}\del_{ij},\ \ i,j=1,...,n$$
The fact that the metric {\em on the boundary} is not uniquely obtained (seems
singular) is related to the fact that the boundary theory as a conformal 
theory knows of no metric - only of a conformal class. We shall come back to 
that. $n^\mu$ is an outward pointing unit vector normal to the boundary. 
We may take
$$n_\mu=(-\frac{1}{x^0},0,...,0);\ \ n^\mu=(-x^0,0,...,0)$$
and $\sqrt{h}=(x^0)^{-n}$. We must find $F_{0\ell}$. $F=dA$ is obtained 
hitting $A$ with $d=dx^0\pa_0+dx^i\pa_i$, but in the result we only need bother
about terms with a $dx^0$. Introducing 
$$D\equiv (x^0)^2+(\vex-\vex')^2$$
we find
\bea
F&=&(n-2) (x^0)^{n-3}dx^0\wedge \int d^nx'\frac{a_i(x')dx^i}{D^{n-1}}\nn
&&-2(n-1)(x^0)^{n-1}dx^0\wedge \int d^nx'\frac{a_i(x')dx^i}{D^n}\nn
&&+2(n-1)(x^0)^{n-3}dx^\ell\wedge dx^0\int d^nx'(x^\ell-(x')^\ell)a_i(x')
(x^i-(x')^i)\frac{1}{D^n}\nn
&&-(x^0)^{n-3}dx^i\wedge dx^0\int d^nx'\frac{a_i(x')}{D^{n-1}}
+\ \mbox{terms with no }\ dx^0
\eea
Using $dx^0\wedge dx^i=-dx^i\wedge dx^0$ we get further
\bea
F&=&(n-1)(x^0)^{n-3}dx^0\wedge\int d^nx'\frac{a_i(x')dx^i}{D^{n-1}}\nn
&&-2(n-1)(x^0)^{n-1}dx^0\wedge\int d^nx'\frac{a_i(x')dx^i}{D^n}\nn
&&-2(n-1)(x^0)^{n-3}dx^0\wedge\int d^nx'
\frac{(\vex-\vex')\cdot d\vex a_i(x')(x^i-(x')^i)}{D^n}+...
\label{F2nd}
\eea
Now
\ben
I=\int d^nx' (x^0)^{-n+3}A_i(x^0,\vex')F_{0i}(x^0,\vex')
\een
Here, on the boundary $A_i\ra a_i(\vex')$ and from \rf{F2nd}
\bea
F_{0i}(x^0,\vex)&=&(x^0)^{n-3}\left\{(n-1)\int d^nx'\frac{a_i(x')}{D^{n-1}}
\right.\nn
&&\left.-2(n-1)\int d^nx'(x_i-x'_i)\frac{a_k(x')(x^k-(x')^k)}{D^n}\right\}\nn
&&+{\cal O}((x^0)^{n-1})
\eea
We use a notation with $x_i\equiv x^i$. Only the term with $(x^0)^{n-3}$
survives for $x^0\ra 0$, and we find
\ben
I(a)=\int d^nxd^nx'a_i(\vex)a_j(\vex')\left (
\frac{\del^{ij}}{(\vex-\vex')^{2n-2}}-
\frac{2(x-x')^i(x-x')^j}{( \vex-\vex')^{2n}}\right )
\label{vectoraction}
\een
This is the final result. To see that this is in accord with the conjecture, 
notice that a gauge-field 1-form in $AdS_{n+1}$ is a scalar under $SO(1,n+1)$. 
Hence, the {\em components} have conformal dimension $+1$ on the boundary, 
so they couple to operators, $J_i$, in the conformal field theory on the 
boundary with conformal dimension $n-1$, but these ``currents'' must be 
conserved by virtue of gauge invariance in the bulk. According to the conjecture
we have calculated the generating function for these operators
in \rf{vectoraction}. We see that they have only non vanishing 2-point 
functions:
\ben
\br J_i(\vex)J_j(\vex')\kt\sim\frac{1}{(\vex-\vex')^{2(n-1)}}
\left\{\del_{ij}-\frac{2(x_i-x_i')(x_j-x_j')}{(\vex-\vex')^2}\right\}
\een
The last term ensures current conservation:
\bea
&&\pa^x_i \br J_i(\vex)J_j(\vex')\kt\nn
&=&\pa_i\left\{\frac{\del_{ij}}{(\vex-\vex')^{2(n-1)}}
-\frac{2(x_i-x_i')(x_j-x_j')}{(\vex-\vex')^{2n}}\right\}\nn
&=&0
\eea
We conclude that the conjecture also works for free massless gauge fields.
Furthermore, we have constructed a propagator also in that case (in a 
particular gauge).

The case of massless gravitons in the bulk is a little more complicated
\cite{gubser,witten1}. 
They couple to the energy momentum tensor on the boundary.

\subsection{Free massive fields on $AdS_{n+1}$}
Following Witten again\cite{witten1} we shall argue that a massive scalar with
mass $m$ in $AdS_{n+1}$ must couple to operators ${\cal O}_\Del$ with 
conformal dimension $\Del$ in the boundary theory, given by the largest root of
\ben
\Del(\Del -n)=m^2
\een
Of course we have already checked the case of $m=0$. In the massive case it 
turns out that we have to reinterpret the idea that the field $\phi$ 
``should tend to'' a definite (current) field on the boundary.

We take the free massive theory in the bulk to be described by
\ben
I(\phi)=\hf\int d^{n+1}x\sqrt{g}(\pa_\mu\phi\pa^\mu\phi +m^2\phi^2)
\een
Consider coordinates $x^\mu,\ \mu=0,...,n$ with metric \rf{polarmetric}
\ben
ds^2=\frac{4}{(1-x^2)^2}\sum_0^n(dx^\mu)^2
\een
with
\ben
x^2\equiv\sum_0^n(x^\mu)^2\equiv r^2, \ 0\leq r <1
\een
Then change variable to
\bea
r&=&\tanh\frac{y}{2},\ \ 0\leq y <\infty\nn
dr&=&\frac{dy}{2\cosh^2 y/2},\  1-r^2=\frac{1}{\cosh^2 y/2}\nn
\frac{r^2}{(1-r^2)^2}&=&\sinh^2\frac{y}{2}\cosh^2\frac{y}{2}=\frac{1}{4}\sinh^2y
\eea
Next write
$$\sum_0^n(dx^\mu)^2=dr^2+r^2d\Om_n^2$$
with $d\Om_n$ the metric of the unit $S^n$. Thus the metric on $AdS_{n+1}$ may
be expressed as
\ben
ds^2=dy^2+\sinh^2y d\Om_n^2
\een 
In these coordinates
\bdm
\det g=\sinh^{2n}y \det \g
\edm
with $\g_{\al\b}$ the metric tensor on $S^n$. Now the Laplacian on scalars 
becomes
\bea
\Del&=&\frac{1}{\sqrt{g}}\pa_\mu\sqrt{g}\pa^\mu\nn
&=&\frac{1}{\sinh^ny}\frac{d}{dy}\sinh^ny\frac{d}{dy}+
\frac{1}{\sqrt{\g}}\pa_\a\sqrt{\g}\pa^\a\nn
&=&\frac{1}{\sinh^ny}\frac{d}{dy}\sinh^ny\frac{d}{dy}-\frac{L^2}{\sinh^2y}
\label{ylaplacian}
\eea
where
\ben
-L^2=\frac{1}{\sqrt{\g}}\pa_\a\sqrt{\g}\pa^\a
\label{spherelaplacian}
\een
is the Laplacian on the sphere, the ``angular momentum'' or centrifugal 
contribution. (The notation is perhaps slightly confusing here: In 
\ref{ylaplacian} 
$$\pa^\a=g^{\a\nu}\pa_\nu=g^{\a\b}\pa_\b=\frac{1}{\sinh^2y}\g^{\a\b}\pa_\b$$
In \ref{spherelaplacian} instead, $\pa^\a$ is just $\g^{\a\b}\pa_\b$).
One might imagine expanding $\phi$ in eigenmodes of $L^2$ 
which indeed as wee shall see in sect. 4.5, is the Casimir of $SO(n+1)$. 
We want to understand the behaviour 
of $\phi(y,\theta^\a)$ for $y\ra\infty$ ($\theta^\a$ are coordinates on $S^n$).
The Klein-Gordon equation in these coordinates becomes
\bdm
\frac{1}{\sinh^ny}\frac{d}{dy}\big(\sinh^ny\frac{d}{dy}\phi\big)
-\frac{L^2}{\sinh^2y}\phi =m^2\phi
\edm
For large $y$, the centrifugal term is negligible and we get approximately
\bdm
e^{-ny}\frac{d}{dy}\big(e^{ny}\frac{d}{dy}\phi\big)=m^2\phi
\edm
The solution is an exponential $e^{\l y}$ provided
\ben
\l(n+\l)=m^2
\een
Thus there are 2 linearly independent solutions which asymptotically behave as
\bdm
e^{\l_+ y}\ \mbox{and}\ e^{\l_- y}
\edm
with $\l_+$ and $\l_-$ the larger and the smaller solutions for $\l$ 
respectively. 
Only one particular linear combination is an allowed solution free of 
singularities in the interior of $AdS_{n+1}$, and hence it's asymptotic 
behaviour is dominated by 
$$e^{\l_+ y}$$

It follows that we cannot assume that $\phi(x)$ tends to a definite value on
the boundary!

We might try to assume (with $\vex$ a coordinate on the boundary)
\ben
\phi(y,\vex)\sim(e^y)^{\l_+}\phi_0(\vex)
\een
near the boundary. But the form $e^y$ is rather arbitrary. In fact, the function
$$e^{-y}\equiv f(y,\vex)$$
has a 1st order zero on the boundary (taking into account the metric on 
$AdS_{n+1}$). It is therefore just of the form needed to build a finite 
metric from the divergent AdS one. But this construction has a degree of 
arbitrariness about it. Thus, if we transform 
$$  f(y,\vex)=e^{-y}$$
into
$$e^{w(\vex)}e^{-y}\equiv \tilde{f}(y,\vex)$$
that new function has an equally good 1st order zero. Thus, demanding an 
asymptotic behaviour
$$\phi(y,\vex)\sim \Big(f(y,\vex)\Big)^{-\l_+}\phi_0(\vex)$$
this freedom in $f$ implies a freedom in $\phi_0$:
\ben
f\ra e^w f\Ra \phi_0\ra e^{w\l_+}\phi_0
\een
So, the 
arbitrariness, $f\ra e^w f$, shows that the metric on $AdS_{n+1}$ in a natural 
way only defines a {\em conformal class} of metrics on the boundary: if 
$h_{ij}$ is a metric on the boundary defined by means of $f$, then 
$e^{2w}h_{ij}$ is a conformally transformed metric obtained from $e^w f$.

This is all consistent with the field theory on the boundary being 
conformally invariant. Then namely, the field theory is insensitive to the
conformal rescalings of the metric. The field theory can only conceive of a 
conformal class.

But then the behaviour
\bdm
h_{ij}\ra e^{2w}h_{ij}\ \Ra\ \phi_0\ra e^{w\l_+}\phi_0
\edm
is the statement that $\phi_0$ is not a function, but rather 
a {\em density} with 
conformal weight $-\l_+$. Indeed, a density of weight $d$ would have the 
property that for a small length $\Del \ell$
$$(\Del\ell)^d\phi_0$$
should be invariant under conformal scaling. That works, since the above is
transformed into
$$(\Del\ell e^w)^de^{w\l_+}\phi_0$$
which is invariant for $d=-\l_+$.

So the only natural procedure is to require that massive fields should tend
to {\em densities} $\phi_0$, and therefore should couple to operators 
${\cal O}_\Del$ with conformal dimension $n+\l_+$ so that 
$${\cal O}_\Del\phi_0$$
is a density with weight $n$.    

We now verify that this state of affairs is in full accordance with the general
prescription.

As before we want to solve for a propagator. We use again coordinates with 
metric (in units where the ``radius'' $b$ of $AdS_{n+1}$ has been put equal 
to 1)
\bdm
ds^2=\frac{1}{(x^0)^2}\sum_0^n(dx^\mu)^2
\edm
Again we begin with a propagator vanishing on the boundary $x^0=0$, but 
developing a delta-function at $P$: $x^0=\infty$, and thus being independent of 
$\vex$. Denoting again the propagator as $K(x^0,\vex;\vex')$ and in particular
$K(x^0,\vex;P)=K(x^0)$, the equation of motion is
\bdm
\Big(-(x^0)^{n+1}\frac{d}{dx^0}(x^0)^{-n+1}\frac{d}{dx^0}+m^2\Big)K(x^0)=0
\edm
We may find a solution of the form $K(x^0)\propto (x^0)^{\l+n}$ provided
\ben
-(\l +n)\l+m^2=0
\een
with the solutions $\l=\l_\pm$
\bea
\l_+&=&\hf(-n+\sqrt{n^2+4m^2})\nn
\l_-&=&\hf(-n-\sqrt{n^2+4m^2})
\eea
Only the solution $K(x^0)=(x^0)^{n+\l_+}$ will vanish for $x^0=0$. As before 
the propagator $K(x^0,\vex;\vec{0})$ is found by the inversion
$$x^\mu\ra\frac{x^\mu}{(x^0)^2+\vex^2}$$
giving
\bea
K(x^0,\vex;\vec{0})&=&\frac{(x^0)^{n+\l_+}}{((x^0)^2+\vex^2)^{n+\l_+}}\nn
K(x^0,\vex;\vex')&=&\frac{(x^0)^{n+\l_+}}{((x^0)^2+(\vex-\vex')^2)^{n+\l_+}}
\eea
Notice that according to the rule \rf{deltafunction}, this does {\em not} tend
to a delta-function when $x^0\ra 0$, rather it is
\bdm
\frac{(x^0)^{n+2\l_+}}{((x^0)^2+(\vex-\vex')^2)^{n+\l_+}}
\edm
which tends to $\del^n(\vex-\vex')$ for $x^0\ra 0$. Hence, when we build the 
classical field as
\bea
\phi(x^0,\vex)&=&c\int d^nx'
\frac{(x^0)^{n+\l_+}}{((x^0)^2+(\vex-\vex')^2)^{n+\l_+}}\phi_0(\vex')\nn
&=&(x^0)^{-\l_+}c\int d^nx'
\frac{(x^0)^{n+2\l_+}}{((x^0)^2+(\vex-\vex')^2)^{n+\l_+}}\phi_0(\vex')
\eea
we see that for $x^0\ra 0$, this behaves as
\bdm
(x^0)^{-\l_+}\phi_0(\vex)\sim e^{\l_+ y}\phi_0(\vex)
\edm
as anticipated.

As explained, $\phi_0(\vex)$ has conformal dimension $-\l_+$ and couples on the
boundary to operators ${\cal O}_\Del$ with conformal dimension 
$\Del = n+\l_+$. Therefore we would expect to find a 2-point function
\ben
\br {\cal O}_\Del(\vex) {\cal O}_\Del(\vex')\kt=
\frac{1}{(\vex-\vex')^{2n+2\l_+}}
\label{massive2point}
\een
We now verify that this is indeed what is obtained, using the by now well 
established prescription.

Namely, we evaluate the classical free action on the classical field as follows:
\bea
I(\phi)&=&\hf\int d^{n+1}x\sqrt{g}(\pa_\mu\phi\pa^\mu\phi+m^2\phi^2)\nn
&=&\hf\int d^{n+1}x\sqrt{g}\left\{\frac{1}{\sqrt{g}}\pa_\mu
\Big(\sqrt{g}\phi\pa^\mu\phi\Big)-
\phi\big(\frac{1}{\sqrt{g}}\pa_\mu\sqrt{g}\pa^\mu\phi-m^2\phi\Big)\right\}
\eea
the last term vanishes, and the first term is evaluated as in the massless case
\ben
I(\phi)=-\hf\int_{x^0=\e}d^nx(x^0)^{-n+1}\phi(x^0,\vex)\pa_0\phi(x^0,\vex)
\een
Now, 
\bea
\pa_0\phi(x^0,\vex)&=&c(n+\l_+)(x^0)^{n+\l_+-1}\int d^nx'\frac{\phi_0(\vex')}
{((x^0)^2+(\vex-\vex')^2)^{n+\l_+}}\nn
&&+ \mbox{ non leading terms as } x^0\ra 0
\eea
and as we have seen, $\phi(x^0,\vex)\sim (x^0)^{-\l_+}\phi_0(\vex), \ x^0\ra 0$.
Hence
\ben
I_{Cl}(\phi_0)\propto\int d^nxd^nx'\frac{\phi_0(\vex)\phi_0(\vex')}
{(\vex-\vex')^{2(n+\l_+)}}
\een
in complete agreement with the expectation \rf{massive2point}. Since $\l_+$ 
is the larger root of
$$\l(\l+n)=m^2,$$
$\Del$ is the larger root of ($\Del =n+\l, \l=\Del-n$)
\bea
(\Del-n)\Del&=&m^2\nn
\Del&=&\hf\Big(n+\sqrt{n^2+4m^2}\Big)
\label{massivedimension}
\eea
It is intuitively plausible how to generalize to fields other than scalars. We 
saw that the massless 1-form gauge-field $A=A_\mu dx^\mu$ restricting to 
$A_i dx^i$ on the boundary, naturally had component fields of dimension 1. 
Likewise a massless $p$-form field $C_p$ has component fields with dimension 
$p$ and couples to operators on the boundary with dimension $n-p$.

A {\em massive} $p$-form field would couple to operators which would have 
dimensions shifted as in the scalar case to
\ben
\Del =n+\l_+ -p
\een
or
\bea
(\Del-n+p)(\Del+p)&=&m^2 \nn
\Del&=&\hf\Big(n+\sqrt{n^2+4m^2-4np}\Big)
\eea

\subsection{Comparison of multiplet data in the bulk and on the boundary}
In ref.\cite{witten1} Witten makes a check on the Maldacena conjecture in the 
case of $AdS_5\times S^5$. Similar checks may be performed in more complicated 
situations. The check is restricted to the case of the 
strong coupling (in the boundary theory) and large $N$ approximation, which may
be treated by classical supergravity. Even though we can neglect string 
excitations in that limit, the compactification on $S^5$ gives rise to 
Kaluza-Klein
excitations with massive modes. We may use the inverse radius of of $S^5$ as
our unit of mass (as in the preceding subsections), i.e. continue to put that
equal to 1. Thus we should at first analyze the spectrum of KK excitations
(for a general discussion, see \cite{dnp86}).
This was done some time ago, in \cite{krn85} by studying small fluctuations
of the supergravity fields around the $AdS_5\times S^5$ background, and in
\cite{gm85} by applying the powerful technique of (super) group representation
theory. These analyses lead to several infinite families of massive field modes
with definite masses and transformation properties under $SO(1,5)\times SO(6)$
or even better, under $SU(1,3|4)$ ($SU(2,2|4)$ for Minkowski signature). 
The representation theory of that
supergroup has been considered for example in \cite{gm85,gunaydin,ferrara}.
According to the Maldacena conjecture,
these give rise to predictions concerning the spectrum of conformal operators 
in the $N=4$ $U(\infty)$ boundary theory. For a given set of quantum numbers
(conjugate to the ones for the modes in the bulk theory) we may predict 
conformal dimensions using the result of the previous subsection.
We must ask whether in fact quantum corrections would upset the result of such a
simple analysis. However, it turns out that both in the bulk theory and in the
boundary theory, there exist large classes of {\em small} representations
with properties similar to properties of BPS states and for which such quantum 
corrections cannot occur. This makes it possible to perform meaningful checks.
In the boundary theory these are the so called chiral primaries (see for example
\cite{howewest}).

Here we shall not attempt an account of this which is anywhere near complete.
Instead we shall restrict ourselves to analyzing certain aspects of one family,
the one corresponding to KK-excitations of the dilaton field. We shall show,
that the masses of these excitations obey the rule
\ben
m^2 =k(k+4),\ \ k=0,1,2,...
\label{KKdilatonmass}
\een
In the boundary conformal field theory Witten pointed out that the 
corresponding operators are of the form
\ben
{\cal O}_{(i_1,...,i_k)}(x)=
Tr(\phi_{(i_1}\cdots\phi_{i_k)}F_{\mu\nu}F^{\mu\nu})(x)
\label{boundaryfamily}
\een
But only when the {\em symmetrized} tensor in $\{i_1,...,i_k\}$ 
is taken, do these fields belong to 
a multiplet of chiral primaries. Precisely then do the fields transform as an
irreducible representation of $SO(6)$ (see below).
The trace is over the adjoint of the $U(N)$ gauge group. The scalar 
($N\times N$ matrix valued) fields
$\phi_i(x), \ i=1,...,6$ have been mentioned before. They transform in the 
$\{{\bf 6}\}$ vector representation of $SO(6)$. The $F_{\mu\nu}$ is the $U(N)$
field strength matrix written as an $N\times N$ matrix. It is trivial to count
the conformal dimension in {\em the weak coupling limit} where free field 
dimensions apply. There the scalars have dimension 1, and the field strength 
tensor dimension 2, so ${\cal O}_{1_1,...,i_k}(x)$ has dimension $\Del_k$
\ben
\Del_k=k+4
\een 
This fits with the formula of the previous subsection \rf{massivedimension}
(for $k=\l$) in the case of the mass values \rf{KKdilatonmass}. In
general the check would not be convincing since we used a strong coupling 
argument in the bulk and a weak coupling one on the boundary. But because it may
be shown that we are dealing with the above mentioned small representations, the
result will survive quantum corrections. As emphasized we shall not go into 
these crucial matters, but here restrict ourselves to an elementary account
of KK-modes of the dilaton field.\\[.3cm]
{\bf Exercise:} Consider the case of $AdS_5\times S^5$ and consider the 
KK-mode of the dilaton field which is independent of the coordinates on $S^5$,
the ``s-wave''. Supposing it couples indeed to $Tr(F^2)$ as described,
work out the 2-point function of that operator in the supergravity picture, 
using the result of sect. 4.2. In particular work out the coefficient in the 
value of the classical action,
left out in the calculation there, taking into account the integration over
$S^5$ and the scale $b$. Verify that the coefficient is a numerical constant 
times $N^2$. Argue
that the form of the 2-point function is exactly the expected one for the 
operator $Tr\Big(F^2\Big)$ in the large $N$ limit (cf. \cite{gubser}).\\[.3cm]
As we have seen before, the dilaton decouples in the background 
$AdS_5\times S^5$ and satisfies the free $10-$dimensional equation of motion
\ben
\frac{1}{\sqrt{g}}\pa_\mu\Big(\sqrt{g}\pa_\nu\phi g^{\mu\nu}\Big)=0
\een
We may use a splitting of this into the 5 components of the 
$AdS_5$ and the 5 components of the $S^5$. Thus, if we expand the dilaton field
on $S^5$ in eigen modes of the Laplacian on $S^5$, we see that these eigenvalues
will play the role of (minus) $m^2$-values  in $AdS_5$.

Our first task will be to understand the connection between the Laplacian on 
$S^5$ (more generally $S^{n+1}$) and the quadratic Casimir of $SO(6)$ 
($SO(n+2)$).

We may think of $S^{n+1}$ as imbedded in $R^{n+2}$ in close analogy to the case
of $AdS_{n+1}$ (indeed many of the results below carry over to results for
$SO(2,n)$, but subtle important differences exist). Thus define $S^{n+1}$ by 
the condition
\ben
y_0^2+y_1^2+...+y_n^2+y_{n+1}^2=1
\een
We are interested in scalar fields defined on $S^{n+1}$, but we may trivially 
extend those to scalar fields on $\R^{n+2}$. In fact, define
$$\rho^2=\sum_{\mu=0}^{n+1} (y_\mu)^2$$  
Then a scalar field, $\phi$  on $S^{n+1}$ is only defined for $\rho =1$, but 
we may 
extend the definition to $\R^{n+2}$ by demanding $\phi$ independent of $\rho$
along fixed directions in  $\R^{n+2}$. More concretely, introduce coordinates
$(\rho, x_\mu)$ similar to \rf{adsstereo} by
\bea
y_0&=&\rho\frac{1-x^2}{1+x^2},\ y_\mu=\rho\frac{2x_\mu}{1+x^2}, \ 
\mu=1,...,n+1\nn 
y^2&=&\rh^2
\eea
We then extend a field $\phi(x)$ into 
$$\phi(\rho,x)\equiv\phi(1,x)\equiv\phi(x)$$
On any scalar field $\phi(y)$, the generators of $SO(n+2)$ are simply
(for generalized treatments along these lines, see for example \cite{ferrara})
\ben
L_{mn}=i(y_m\pa_n-y_n\pa_m)
\een 
with the standard algebra:
\ben
{[}L_{mn},L_{pq}{]}=i\{\del_{np}L_{mq}+\del_{mq}L_{np}-\del_{nq}L_{mp}
-\del_{mp}L_{nq}\} 
\label{soalgebra}
\een
Further we have the vielbeins and their relation to the metric on $S^{n+1}$
(the $y^m\equiv y_m$'s are the flat coordinates, the set 
$(\rho, x^\mu)\equiv (x^0,x^\mu)$ are the 
curvilinear ones).

\bea
e^0_m&=&\frac{\pa\rh}{\pa y^m}=\frac{y_m}{\rh}, \ 
e^\mu_m=\frac{\pa x^\mu}{\pa y^m}\nn
e^m_0&=&\frac{\pa y^m}{\pa\rh}=\frac{y^m}{\rh},\  
e^m_\mu =\frac{\pa y^m}{\pa x^\mu}\nn
y_me^{m\mu}&=&0,\ \ y_me^{m0}=\rho^2,\ \ e^\mu_ne^n_0=0\nn
e^\mu_me^{\nu m}&=&g^{\mu\nu}\nn
e^0_ne^n_0&=&\frac{y_n}{\rh}\cdot\frac{y^n}{\rh}=1  
\eea
We now want to show that 
\ben
L_{mn}\phi(y)L^{mn}\phi(y)=-2\rho^2g^{\mu\nu}\pa_\mu\phi\pa_\nu\phi
\een
It then follows that the dilaton action on $S^{n+1}$ may be formulated as
\bea
S(S^{n+1})&=&-\frac{1}{4}\int d^{n+2}y\del(y^2-1)L_{mn}\phi(y)L^{mn}\phi(y)\nn
&=&\frac{1}{4}\int d^{n+2}y\del(y^2-1)\phi(y)L_{mn}L^{mn}\phi(y)
\eea
(use: $d^{n+2}y\del(y^2-1)=d^{n+1}xd\rho\sqrt{g}\del(\rho -1)$)
and the statement about the connection between the Laplacian and the 
quadratic Casimir
\ben
C_{n+2}\equiv \hf L_{mn}L^{mn}
\een
follows. Hence we work out (the distinction between lower and upper indices 
has no significance)
\bea
L_{mn}\phi L^{mn}\phi&=&-2\Big(y_m\pa_n\phi y^m\pa^n\phi-
y_m\pa_n\phi y^n\pa^m\phi\Big)\nn
&=&-2\left\{ y_m\Big(\frac{\pa x^\mu}{\pa y^n}\frac{\pa}{\pa x^\mu}+
\frac{\pa\rh}{\pa y^n}\frac{\pa}{\pa\rh}\Big)\phi\right.\nn
&&\cdot\left.
\left [y^m\Big(\frac{\pa x^\nu}{\pa y_n}\frac{\pa}{\pa x^\nu}+
\frac{\pa\rh}{\pa y_n}\frac{\pa}{\pa\rh}\Big)\phi-(m\leftrightarrow n)
\right ]\right\}\nn
&=&-2\left\{y_m(e^\mu_n\pa_\mu+e^0_n\pa_\rh)\phi(y)\right.\nn
&&\left.\cdot\left [y^m(e^{\nu n}\pa_\nu+e^{0 n}\pa_\rh)\phi(y)-
(m\leftrightarrow n)\right ]\right\}\nn
&=&-2\left\{y^2(g^{\mu\nu}\pa_\mu\phi\pa_\nu\phi+e^0_ne^{n 0}(\pa_\rh\phi)^2)
-y_me^{0m}\pa_\rh\phi e^0_ny^n\pa_\rh\phi\right\}
\eea
But here we have made the choice that $\pa_\rh\phi\equiv 0$, so we simply get
\ben
L_{mn}\phi L^{mn}\phi=-2\rh^2g^{\mu\nu}\pa_\mu\phi\pa_\nu\phi
\een
as we wanted\footnote{It is easy to verify, that putting instead 
$\phi(\rho,x)=\rho^N\phi(1,x)$ for any $N$, 
would yield exactly the same result.}.

Thus we are let to perform an analysis of the dilaton field on $S^{n+1}$ 
similar to the analysis of scalars on $S^2$ according to spherical harmonics
$Y_\ell^m(\theta,\phi)$. In this latter case the Laplacian is well known to
be identified with (minus) the square of the angular momentum with eigenvalues
$\ell(\ell+1)$ for integer $\ell$. Of course the square of the angular momentum is
just the Casimir of $SO(3)$. We want to arrive at a similar 
understanding in general. 

Let us single out as special coordinate
\ben
u=y_0+iy_{n+1}\equiv Ye^{i\phi}, \ Y,\phi\in\R,\ \ Y=\sqrt{(y^0)^2+(y^{n+1})^2}
\label{Yphi}
\een
On $S^{n+1}$ we have
$$y_0^2+y_{n+1}^2+\sum_1^n y_i^2\equiv Y^2+z^2=1$$
so that we may define
\ben
z^2=\sum_1^n y_i^2=\cos^2\theta, \ Y^2=\sin^2\theta
\label{theta}
\een
In general a representation of $SO(n+2)$ would be characterized by a highest 
weight state, $\ket{\vec{\L}_{n+2}}$ with a certain highest weight, 
$\vec{\L}_{n+2}$ of $SO(n+2)$, a vector in weight space, 
the components of which are eigenvalues
of a mutually commuting set of Cartan generators of the algebra $so(n+2)$. Let 
us take one of these to be 
\ben
H\equiv L_{0,n+1}
\een corresponding to rotations in the complex $u$-plane, and generating an 
$SO(2)$ subgroup. The remaining Cartan generators pertain to an $SO(n)$ 
subgroup commuting with that $SO(2)$. With this choice our weights are then 
automatically labelled by (i) the eigenvalue of $H$ and (ii) by a highest 
weight of that $SO(n)$, i.e. we classify irreducible representations according
to the 
$$SO(2)\times SO(n)$$
subgroup. And we write
\ben
\vec{\L}_{n+2}=(k,\vec{\L}_n)
\een
There are $n$ raising and lowering operators relative to $H$. Indeed, define
\ben
J_i^+\equiv L_{i,n+1}+iL_{i0}
\een
Then work out
\bea
{[}H,J_i^+{]}&=&{[}L_{0,n+1},L_{i,n+1}+iL_{i0}{]}=
{[}L_{0,n+1},L_{i,n+1}{]}+ i {[}L_{0,n+1},L_{i0}{]}\nn
&=&-iL_{0i}+i^2L_{n+1,i}=J_i^+
\eea
Thus we may write
\bea
C_{n+2}&=&\hf L_{mn}L_{mn}=\sum_{1\leq i<j\leq n}L_{ij}L_{ij}+
L_{0i}L_{0i}+L_{n+1,i}L_{n+1,i}+L_{0,n+1}L_{0,n+1}\nn
&=&C_n+(L_{i,n+1}-iL_{i0})(L_{i,n+1}+iL_{i0})-iL_{i,n+1}L_{i0}+iL_{i0}L_{i,n+1}
+H^2\nn
&=&C_n+J_i^-J_i^+-i{[}L_{i,n+1},L_{i0}{]}+H^2\nn
&=&C_n+J_i^-J_i^+ -nL_{n+1,0}+H^2=C_n+J_i^-J_i^+ +nH+H^2
\eea
So acting on a highest weight state, $J^+_i$ will vanish, and we get
\ben
C_{n+1}=C_n+k(k+n)
\een
where $k$ denotes the eigenvalue of the $SO(2)$ generator $H$.

Let us first consider the $(n+2)$-dimensional {\em vector} representation of
$SO(n+2)$. The $L_{mn}$'s are represented by $(n+2)\times (n+2)$ matrices
\ben
(L_{mn})_{ab}=i(\del_{ma}\del_{nb}-\del_{na}\del_{mb})
\een
And $H$ gives eigenvalue 1 for the ``highest weight state''
$$(y_0,y_1,...,y_n,y_{n+1})=\frac{1}{\sqrt{2}}(1,0,...,0,-i)$$
(interpreted as a column).
The raising operators $J_i^+$ give zero on the same ``state'', 
and so do the $SO(n)$ generators 
$L_{ij},\ i,j=1,...,n$. It follows that the highest weight of the vector 
representation is
\ben
\vec{\L}_{n+2}(\mbox{vector})=(1,\vec{0})
\een
and the Casimir has the value above with $k=1$ and $C_n=0$. Next consider the
$k$-fold tensor product of the vector representation with $k$ a positive 
integer. The highest weight is trivially $(k,\vec{0})$ but the representation
is highly reducible. The unique irreducible representation with the same highest
weight is the symmetrized tensor product. This is the one we encountered
in \rf{boundaryfamily}, and we see now that the Casimir of the representation
is given by $k(k+n), k=0,1,2,...$ (generalizing the result $k(k+1)$ for 
$n+2=3$).

We now consider the generalized spherical harmonics of the scalar (dilaton) 
field on $S^{n+1}$. In the case of $SO(3)$ these are the usual spherical 
harmonics $Y_\ell^m(\theta,\phi)$ with $\ell$ integer. They constitute a 
complete
set of scalar functions on $S^2$. They carry irreducible representations of
$SO(3)$ with the highest weight member
\ben
Y_\ell^\ell(\theta,\phi)=N_\ell \sin^\ell\theta e^{i\ell\phi}
\een
The normalization is not interesting here, but is easily evaluated to 
$$N_\ell =\sqrt{\frac{2\ell +1}{4\pi}}\frac{\sqrt{(2\ell)!}}{2^\ell \ell!}$$
All the other spherical harmonics are obtained from this one by applying 
lowering operators.

In general for $S^{n+1}$ and $SO(n+2)$, we may construct a similar highest 
weight scalar function for any fixed positive integer $k$
\ben
Y_k(\hat{y})\equiv N_ku^k=N_k(y_0+iy_{n+1})^k=N_k\sin^k\theta e^{ik\phi}
\label{sonharmonic}
\een 
where we used the definitions in \rf{Yphi} and \rf{theta}. This is a regular 
well defined function on $S^{n+1}$ only for positive integer values of $k$.
Of course the 
similarity with the elementary case is strong (the normalization would depend 
on $n$). It is trivial to verify with the form of the generators we have given,
that this field is indeed a highest weight ``state'' of $SO(n+2)$ with
highest weight
\ben
\vec{\L}_{n+1}=(k,\vec{\L}_n=\vec{0})
\een
just as for the symmetrized $k$-fold tensor representation. Therefore, 
constructing a representation of $SO(n+2)$ by rotating this $Y_k(\hat{y})$ in
all possible ways, or, equivalently by forming all possible linear combinations
of fields obtained from it by applying lowering operators, we shall get a
finite dimensional irreducible representation of $SO(n+2)$, in fact precisely 
the one we met above by considering symmetrized tensor products.

In the present case of $SO(n+2)$ there are of course very many more irreducible 
representations to worry about than for the case of $SO(3)$, and we might 
wonder if the generalized scalar spherical harmonics defined by 
\rf{sonharmonic} (together with all of its multiplet members), would really 
form a {\em complete}
set of functions on $S^{n+1}$ for $n>2$. This will be so, if we
are able to construct arbitrarily good approximations to delta functions with
support at any point on $S^{n+1}$, using (linear combinations of) these 
functions. It is actually intuitively obvious that this should be possible. In 
fact, consider what we may do with the highest weight functions 
$Y_k(\hat{y})$ themselves, directly. For very high values of $k$ it follows 
from \rf{sonharmonic} and the sphere condition
$$y_0^2+y_{n+1}^2=1-\vec{y}^2\leq 1$$
that these functions only have appreciable support in the neighbourhood of the
unit circle
\ben
y_0^2+y_{n+1}^2=1, \ \vec{y}=\vec{0}
\label{circle}
\een
Along this circle the coordinate is $\phi$ and 
$$Y_k\sim e^{ik\phi}$$
essentially a 1-dim. plane wave.
It follows that we may construct arbitrarily good approximations to delta 
functions with support at any point  along the circle \rf{circle}. Since our
representation space also contains all possible rotations of the functions $Y_k$
we may form delta functions with support at any point we please on $S^{n+1}$,
and thus indeed we have found the complete set of generalized spherical 
harmonics. 

The Casimir for this representation is therefore $k(k+n)$ or $k(k+4)$ in the
case of $S^5$. But these are just the squared mass values we needed for the
KK-excitations, \rf{KKdilatonmass}, in order for the Maldacena conjecture to be 
checked in this particular instance.

\section{Breaking SUSY and conformal invariance in the boundary theory. A 
possible new approach to (large $N$) QCD}

The Maldacena conjecture suggests a mathematical equivalence between string/M 
theory in certain backgrounds, and a conformally invariant ordinary quantum 
field theory on ``the boundary''. Field theories, such as the Standard Model
are not conformally invariant: They typically posses a mass gap, there is a 
lightest {\em massive} meson for example. Although the Maldacena conjecture
would seem to throw extremely interesting light on non-perturbative aspects of
quantum field theory, it would therefore also seem that we are 
restricted to theories
rather far removed from reality. However, Witten \cite{witten2} proposed a 
scheme whereby it seems possible to overcome these difficulties. Perhaps the
most interesting case is that of $AdS_7\times S^4$. The boundary theory here is
a certain so called (2,0) exotic 6-dimensional conformally invariant theory 
for which no action seems available \cite{6dim} (see also lectures by 
E. Bergshoeff and by P.C. West, this school). 
By compactifying everything on a 2-torus, $T^2$, however, the 
boundary theory becomes 4-dimensional, and provided the fermions in the theory
are taken anti-periodic around a cycle on the $T^2$, supersymmetry 
and conformal invariance are broken at low energies, and 
the set up provides a novel way of treating (regularized) large $N$ ordinary 
QCD in 4 dimensions (albeit without quarks)! 
For  a long time it has been a dream for theoretical particle physics 
\cite{hooft74} to be able to do something non trivial with that theory.
Glueballs would be stable in that limit
so that perhaps  the theory could be sufficiently tractable to 
furnish an analytic understanding of confinement. Despite intense interest
in this large $N$ limit, however, very little has been achieved in the way of 
concrete results. The Maldacena conjecture combined with Witten's proposal 
seems to introduce a truly novel approach.

Here, at first we shall be more general, following the discussion of
\cite{witten2}, and in the end we shall mostly restrict to the somewhat simpler
case of $AdS_5\times S^5$ for which we introduce a single $T^1=S^1$ 
compactification, thereby rendering a framework for studying large $N$ QCD in 
3 space-time dimensions. At the end we shall indicate some of the first steps
that have to be taken in order to treat also $QCD_4$.

To see how a suitable spin structure can break supersymmetry and conformal 
invariance at low energies, we go to Euclidean time and take it periodic on
a circle of radius $R$, corresponding to an inverse temperature of $2\pi R$. 
A bosonic degree of freedom has to be periodic along this time $t$:
\ben
q(t)=\sum_{n\in\Z}\Big( a_ne^{-int/R}\Big)
\een
There are of course then (KK-like) excitations with masses $n/R$. For fermions,
we have the option of considering a non trivial spin structure. Clearly, if we
also take the fermions periodic, they will be quantized with the same integer
modes as the bosons, and supersymmetry will be preserved. But if we take
fermions to be {\em anti-}periodic around Euclidean time, they have modes
$$(n+\hf)/R$$
In particular, the lowest mode $n=0$ is very different for bosons and for 
fermions. For bosons we have a massless mode, but for fermions the lowest mode
can be considered to decouple for high enough temperatures. So supersymmetry 
is broken. If we investigate the theory with very high frequencies, much 
higher than the temperature, these details are irrelevant and we expect to 
regain a supersymmetric situation. 

The scalar supersymmetric partners of fermions will get masses due to 
renormalization. In the supersymmetric theory the masses are protected from 
being divergent. Thus in the effective low energy theory, we shall have 
divergencies cut off by the ``cut off'' of the effective theory, which is the
temperature. Thus we also expect scalar super partners of fermions to become 
massive and therefore to decouple at low energies. Finally, the effective 
theory will have a non-vanishing beta function, because some of 
the field modes which make the beta function vanish in the full theory 
are absent in the low energy effective theory. Thus, the effective low
energy theory is no longer conformal.

It follows, that apparently we have a scheme for dealing with a realistic 
(QCD like) theory. In that theory, the coupling constant should run at 
high energies to a small value,
since the theory is asymptotically free. The smallest value is the one 
attained at the cut off, the temperature, and it would be given by the fixed
coupling constant of the unbroken theory. Therefore, in order to use this
scheme in a fully realistic way, we should arrange for the coupling constant 
of the boundary theory to be {\em small}. We have previously seen, that the
very simple supergravity approximation is obtained when the coupling constant 
of the boundary theory is {\em large}. Therefore, a straight forward study of
large $N$ QCD in 3 or 4 dimensions based on supergravity, cannot be hoped to be
realistic. It is perhaps similar to a strong coupling analysis of lattice 
gauge theory, known to be in a ``wrong phase''. What we would rather like to do,
would be to study not supergravity, but the full string theory in the 
appropriate background. At large $N$ it should even be enough to study 
{\em classical} string theory - or the tree diagram limit only, in order to have
a realistic framework for large $N$ QCD. This goal has not yet been achieved,
although preliminary proposals have been given \cite{kt98,p98,dpolyakov98}.

Instead, a large number of studies have been performed in the supergravity 
approximation. Such studies can at best be considered to have an exploratory
nature, but it is very instructive to see how several expectations from 
confinement are brought out in a very simple way (see refs. 
\cite{maldacena98,witten2,csaki,hashimoto,ooguri,wittentheta,gkt98,pt98,
olesen,klebanov99}, 
to name but a few).

\subsection{Classical, finite temperature versions of $AdS$} 
According to the previous discussion, we are led to consider 
``finite temperature'' versions of Anti de Sitter spaces. As shown by 
Hawking and Page \cite{hp} and generalized by Witten \cite{witten2} there are
two relevant manifolds denoted $X_1$ and $X_2$ which we must consider.

\subsubsection{The manifold $X_1$}
The first one is the simplest. In the standard case of Poincar\'e isometry, we
may arrange for a finite temperature by taking Euclidean signature, and by 
compactifying the time on $S^1$. That compactification may be thought of as a 
periodic identification of Euclidean time. All bosonic fields would have to be
periodic under: $t\ra t+2\pi R$. This mapping is a translation in time, 
$\tau: t\mapsto \tau(t)$,
and the set of all translations, $\tau^n$, form a group isomorphic to $\Z$. Thus
in the case of $AdS^+_{n+1}$ we may try to do something similar. Consider the 
embedding condition:
\ben
uv-\sum_{i=1}^n x_i^2=b^2
\een
with $AdS_{n+1}^+$ being the branch $u,v >0$. Now introduce a real, positive 
parameter $\l$
(analogous to $R$ and to be related to the temperature), and define the mapping
\ben
u\ra \l^{-1}u,\ \ v\ra \l v,\ \ x_i\ra x_i
\een
This is a mapping, $f_\l$ of $AdS_{n+1}$ onto itself. The set of all 
repeated applications of this mapping (an the inverses) $\{f^n_\l|n\in\Z\}$ 
constitute a group isomorphic to $\Z$. Generalizing the case of Poincar\'e 
isometry we may consider the manifold
$$X_1\equiv AdS^+_{n+1}/\Z$$
So $X_1$ is the set of ``equivalence classes'' where two points $P,P'$ in 
$AdS_{n+1}$ are equivalent if they are related by one of these mappings. We see
that a fundamental domain for $v$ is
\ben
1\leq v/b\leq \l
\een
namely $v/b=1$ and $v/b=\l$ is the same point in $X_1$. These points parametrize
indeed a circle
\bea
v/b&=&\l^{\theta/2\pi}\nn
\theta&=&\frac{2\pi\ln v/b}{\ln\l}
\eea
($\theta =0\Ra v/b=1, \theta =2\pi \Ra v/b =\l\sim 1$). For any $v$ in the 
fundamental domain, we may solve for 
$$u=\Big( b^2+\sum_{n=1}^nx_i^2\Big)/v$$
and use for $X_1$ the coordinates $(x_1,...,x_n)$ together with the angular 
coordinate $\theta$. It follows that topologically 
$$X_1\sim \R^n\times S^1$$
As for the boundary of $X_1$, we know it is obtained by scaling $u,v,x_i$ by
$s\ra\infty$, equivalent to the condition
$$uv-\sum_{i=1}^nx_i^2=0$$
subject to projective equivalence. Thus we may fix a scale so that
\ben
\sum_{i=1}^n\Big( x_i/b\Big)^2=1
\een
defining $S^{n-1}$. We see that the boundary of $X_1$ has topology 
$$\pa X_1\sim S^{n-1}\times S^1$$
These ``spheres'' have two different radii, but only the ratio is relevant 
for the conformal structure. We shall come back to that.

Let us introduce a convenient metric on $X_1$. Write
\bea
r^2&=&\sum_{i=1}^n x_i^2\nn
\sum_{i=1}^ndx_i^2&=&dr^2+r^2d\Om_{n-1}^2
\eea
and define
\ben
t=\ln v/b-\hf\ln \Big(1+(r/b)^2\Big)=\ln\l\cdot\frac{\theta}{2\pi}-
\hf\ln\Big(1+(r/b)^2\Big)
\een
Then work out
\bea
ds^2&=&du dv-\sum dx_i^2\nn
u&=&(b^2+\sum x_i^2)/v\Ra du=-\frac{dv}{v^2}\Big(b^2+r^2\Big) +\frac{2rdr}{v}\nn
dt&=&\frac{dv}{v}-\frac{\frac{r}{b^2}dr}{1+\Big(r/b\Big)^2}
\eea
Then find
\bea
du dv&=&-dt^2(b^2+r^2)+\frac{r^2dr^2}{b^2+r^2}\nn
ds^2&=&dt^2(b^2+r^2)+\frac{b^2dr^2}{b^2+r^2}+r^2d\Om_{n-1}^2,\ \mbox{or}\nn
d\tilde{s}^2&=&ds^2/b^2=dt^2\Big(1+(r/b)^2\Big)+\frac{dr^2}{1+(r/b)^2}
+(r/b)^2d\Om_{n-1}^2
\label{X1metric}
\eea
(we performed as usual the ``mostly minus'' to ``mostly plus'' operation).
This is our final metric on $X_1$.

\subsubsection{The manifold $X_2$}
This is the case where a temperature is introduced by in fact inserting a 
black hole into $AdS_{n+1}$, the Schwarzschild metric generalized to 
$AdS_{n+1}$. Thus we seek a static, spherically symmetric metric of the general
form
\ben
ds^2=A(r)dt^2+B(r)dr^2+r^2d\Om_{n-1}^2
\een
with $r=0$ being ``the position of the black hole''.
Outside the black hole we have Einstein's empty space equation (with a 
cosmological constant)
\ben
R_{\mu\nu}=-\frac{n}{b^2}g_{\mu\nu},\ \ D=n+1
\een
Thus the black hole solution $X_2$ will also be an example of an Einstein space,
but one less symmetric than $AdS_{n+1}$, even though the two are 
``asymptotically similar''.

We find the following non zero Christoffel symbols ($\g_{ij}$ denotes the 
metric on $S^{n-1}$):
\bea
\G^t_{rt}&=&\hf A^{-1}A',\ \G^r_{tt}=-\hf B^{-1}A',\ \G^r_{rr}=\hf B^{-1}B'\nn
\G^r_{ij}&=&-B^{-1}r\g_{ij},\ \G^i_{jk},\ \G^i_{rj}=\frac{1}{r}\del^i_j
\eea
We then find the non vanishing Riemann tensor components contributing to 
$R_{tt}$:
\bea
R^r_{trt}&=&\frac{1}{4}(AB)^{-1}(A')^2-\frac{1}{4}B^{-2}A'B'-\hf(B^{-1}A')'\nn
R^i_{tit}&=&-\frac{n-1}{2r}B^{-1}A'
\eea
and the first Einstein equation
\ben
R_{tt}= \frac{(A')^2}{4AB}-\frac{n-1}{2r}\frac{A'}{B}
 -\frac{A'B'}{4B^2}-\hf\Big(\frac{A'}{B}\Big)'  =-\frac{n}{b^2}A
\een
Further find
\bea
R^t_{rtr}&=&-\hf(A^{-1}A')'-\frac{1}{4}A^{-2}(A')^2+\frac{1}{4}(AB)^{-1}A'B'\nn
R^i_{rir}&=&\frac{n-1}{2r}B^{-1}B'
\eea
and the second Einstein equation
\ben
R_{rr}=\frac{A'B'}{4AB}
+\frac{n-1}{2r}\frac{B'}{B}-\frac{(A')^2}{4A^2}
-\hf\Big(\frac{A'}{A}\Big)'=-\frac{n}{b^2}B
\een
Put
$$B\equiv A^{-1}f;\ \ B'=-A^{-2}A'f+A^{-1}f'$$
Then the two equations become
\bea
-\hf A\left\{f^{-1}\Big(\frac{n-1}{r}A'+A''\Big)+(f^{-1})'\frac{1}{2}A'
\right\}&=& -\frac{n}{b^2}A\nn
-\hf A^{-1}\Big( A''+\frac{n-1}{r}A'\Big)+
f^{-1}f'\Big(\frac{1}{4}A^{-1}A'+\frac{n-1}{2r}\Big)&=&
-\frac{n}{b^2}A^{-1}f
\eea
or
\bea
f^{-1}\Big(\frac{n-1}{r}A'+A''\Big)+(f^{-1})'\frac{1}{2}A'&=&\frac{2n}{b^2}\nn
f^{-1}\Big(\frac{n-1}{r}A'+A''\Big)+(f^{-1})'\Big(
\hf A' +\frac{n-1}{r}A\Big)&=&\frac{2n}{b^2}\nn
\eea
Subtracting, we find 
$$(f^{-1})'=0$$
so $f$ is a constant, and
\ben
A''+\frac{n-1}{r}A'=2n\frac{f}{b^2}
\een
This is solved into
\bea
A&=&\frac{f}{b^2}r^2+\frac{c_1}{r^{n-2}}+c_2\nn
B&=&\frac{f}{A}
\eea
Redefining the scale of $t$ and the meaning of $c_1,c_2$, we may put $f=1$. 
Also,
to get the empty solution for $X_1$ as a special case, we put the new $c_2=1$.
Thus we finally have the Schwarzschild solution in $AdS_{n+1}$ for $X_2$:
\ben
ds^2=\Big(\frac{r^2}{b^2}+1-\frac{w_nM}{r^{n-2}}\Big)dt^2
+\frac{dr^2}{\Big(\frac{r^2}{b^2}+1-\frac{w_nM}{r^{n-2}}\Big)}+r^2d\Om_{n-1}^2
\label{adsblackhole}
\een
We have renamed a constant, putting
\ben
w_n\equiv\frac{16\pi G_N}{(n-1)\Om_{n-1}}
\een
This will turn out to mean that $M$ becomes the ``mass'' of the black hole.

Notice, that this of course generalizes the metric of the standard $b=\infty$, 
4-dimensional 
black hole (Euclidean, $n=3$, $r_g=2G_NM=w_3M$):
$$ds^2=\Big(1-\frac{r_g}{r}\Big) dt^2+\frac{dr^2}{1-\frac{r_g}{r}}+r^2
d\Om_{n-1}^2$$

\subsubsection{Temperature of the black hole}
We expect that a black hole will have a (Beckenstein-Hawking) temperature, and
that of course was our motivation for picking this metric. Let us work 
out what the temperature is. The metric \rf{adsblackhole} is of the form
\ben
ds^2=V(r) dt^2+\frac{dr^2}{V(r)}+r^2d\Om_{n-1}^2
\een
where
\ben
V(r)=\frac{r^2}{b^2}+1-\frac{w_nM}{r^{n-2}}
\label{VX2}
\een
and vanishes at various values of $r$, the largest of which we shall denote, 
$r_+$. This is characteristic of a metric with a horizon: $g_{00}$ vanishes
and $g_{rr}$ has a pole at the horizon, $r=r_+$. Ordinary physical space is 
the region $r\geq r_+$. In our case of a diagonal 
metric, $g^{rr}=1/g_{rr}$. It is rather easy to establish the general formula
\cite{gkt98} for the temperature, $T$ of the black hole in the case of such
a diagonal metric:
\ben
2\pi T=\sqrt{g^{rr}}\frac{d}{dr}\sqrt{g_{00}}|_{r=r_+}
\label{hawkingtemp}
\een
(other equivalent formulas are easily obtained). To see this, notice that the 
horizon property implies that near $r=r_+$
\ben
g_{00}\sim A(r-r_+),\ \ g^{rr}\sim B(r-r_+)
\een
and the metric takes the following form in the vicinity of the horizon
\ben
ds^2\sim \frac{dr^2}{B(r-r_+)}+dt^2 A(r-r_+)+r^2d\Om_{n-1}^2
\label{tempmetric}
\een
This metric has a coordinate singularity at the horizon, which we may remove
by a suitable choice of coordinates, but only if the Euclidean time is periodic
with a particular period, which one then identifies with the inverse 
temperature. In fact \cite{hp} the coordinate singularity may be made analogous 
to a standard 2-dimensional polar coordinate singularity
\ben
ds^2=d\rho^2+\rho^2d\theta^2
\label{polar}
\een
where $\rho$ is the polar distance and $\theta$ is the polar angle. 
At $\rho =0$ 
the metric is singular, but we know very well that the geometry is regular 
{\em provided} the polar angle has period $2\pi$. If the periodicity is anything
else than $2\pi$ there is a genuine geometrical conical singularity.

We may arrange for the metric \rf{tempmetric} to look similar to \rf{polar} 
if we put
\ben
d\rho^2= \frac{dr^2}{B(r-r_+)},\ \frac{d\rho}{dr}=\frac{1}{\sqrt{B(r-r_+)}},
\ \ \rho = \frac{2}{\sqrt{B}}\sqrt{r-r_+}
\een
where we have chosen $\rho$ to vanish at the horizon. Then the first
two terms of \rf{tempmetric} become
\ben
ds^2\sim d\rho^2+\rho^2\frac{AB}{4}dt^2
\een
We see that this metric describes a regular geometry provided the Euclidean time
$t$ is periodic with period
$$\beta=\frac{1}{T}=\frac{4\pi}{\sqrt{AB}}$$
From this \rf{hawkingtemp} is easily obtained. We also see that only 
$r\leq r_+$ is relevant.

In our case, we have near $r=r_+$
\bea
V(r)&=&V'(r_+)\cdot (r-r_+),\ \ 
V'(r_+)=\frac{2r_+}{b^2}+\frac{(n-2)w_nM}{r_+^{n-1}}\nn
0&=&1+\frac{r_+^2}{b^2}-\frac{w_nM}{r_+^{n-2}} \ \Ra \ 
V'(r_+)=\frac{(n-2)b^2+nr_+^2}{r_+b^2}
\eea
It follows that our black hole has inverse temperature
\ben
\beta_0(r_+)=\frac{4\pi r_+b^2}{nr_+^2+(n-2)b^2}
\label{temp0}
\een
Notice that $\b_0(r_+)$ vanishes at $r_+=0$ and for $r_+\ra\infty$. Also it 
attains a maximum (of $4\pi b/\sqrt{n(n-2)}$, at 
$r_+=\sqrt{\frac{n-2}{n}}b$). It follows, that unlike the manifold $X_1$, which 
may be constructed for any temperature, the black hole metric, $X_2$
only exists for ``small'' values of $\b$ or ``large'' values of the temperature.
It will turn out in fact, that $X_1$ dominates the dynamics at low temperatures,
and $X_2$ at high temperatures. We see that to get a high temperature requires
either $r_+\ra 0$ or $r_+\ra \infty$. We shall see below, that the 
thermodynamics is dominated by $r_+\ra\infty$. In that case the $V=0$ equation
for $r_+$ implies that also $M\ra\infty$, indeed that
\bea
0&=&1+\frac{r_+^2}{b^2}-\frac{w_nM}{r_+^{n-2}}  \simeq 
\frac{r_+^2}{b^2}-\frac{w_nM}{r_+^{n-2}}\ \Ra\ r_+=(Mb^2w_N)^{1/n}
\eea
Then approximately in the large mass limit
\ben
ds^2=\Big(\frac{r^2}{b^2}-\frac{w_nM}{r^{n-2}}\Big)dt^2+
\frac{dr^2}{\Big(\frac{r^2}{b^2}-\frac{w_nM}{r^{n-2}}\Big)}+r^2d\Om_{n-1}^2
\een
Remarkably, this form of the $X_2$ metric makes it completely equivalent to the
non-extremal brane solutions \rf{nonexn4}, \rf{nonexn6}, \rf{nonexn3}. The 
relation between the two forms is in all cases $U=r$ and 
$d\Om_{n-1}^2\sim\sum_1^{n-1} d\vex^2$
after a suitable scaling (see below). The relations between the pairs of 
parameters $(L,U_0)$ and $(b,M)$ are in the relevant cases:
\bea
D=10,\  n=4&& L=b,\ \  \ \frac{U_0^4}{L^2}=w_4M\nn
D=11,\  n=6&& L=\hf b,\ \ \frac{U_0^6}{4L^2}=w_6M\nn
D=11,\  n=3&& L=2b,\ \ \frac{4U_0^3}{L^2}=w_3M
\eea
Although the solutions are expressed in terms of two parameters, 
one may in fact bee scaled away (see below). Notice that the $p$-brane 
solutions and the black hole solution have totally different symmetries and 
asymptotic behaviours. Only in the limits considered here (near horizon and 
large mass) do they agree. 

In the same large $M$ limit we have
\ben
\b_0\sim \frac{4\pi b^2}{nr_+}=\frac{4\pi b^2}{n}(w_nb^2M)^{-1/n}
\een
The topology of the solution is
$$X_2\sim S^2\times S^{n-1}$$
since the $(t,r)$ space is topologically similar to the 
2-dimensional plane (described by polar coordinates), which compactifies to 
$S^2$. On this space there are no non-contracitible loops, in contrast to the 
case of $X_1$. 

The topology becomes $S^1\times S^{n-1}$ on the boundary. However, we would 
like to consider limits where the boundary looks more like
$S^1\times \R^{n-1}$, corresponding to the radius of $S^{n-1}$ being 
``much larger'' than the radius of the $S^1$.

Near the boundary, i.e. at $r\ra\infty$ the metric becomes 
\ben
ds^2\sim \frac{r^2}{b^2}dt^2+\frac{b^2}{r^2}dr^2+r^2d\Om_{n-1}^2
\een
At these asymptotically large values of $r$ the metric on $S^1$ is 
$$ \frac{r^2}{b^2}dt^2$$
corresponding to a radius of
$$\frac{r}{b}\cdot\frac{\b_0}{2\pi}$$
On the other hand, the radius of the $S^{n-1}$ 
(the $r^2d\Om_{n-1}^2$ term in the metric) is simply $r$. 
The ratio of the two is therefore
$$\frac{\b_0}{2\pi b}$$
If we want this ratio to be small, we see that we need 
$\b_0$ small.

We now scale as follows
\bea
r&=&\Big(\frac{w_nM}{b^{n-2}}\Big)^{1/n}\rho,\ \ 
t=\Big(\frac{w_nM}{b^{n-2}}\Big)^{-1/n}\tau\nn
V&\sim& \Big(\frac{w_nM}{b^{n-2}}\Big)^{2/n}
\left\{\frac{\rho^2}{b^2}-\frac{b^{n-2}}{\rho^{n-2}}\right\}\ \Ra\nn
ds^2&=&\Big(\frac{\rho^2}{b^2}-\frac{b^{n-2}}{\rho^{n-2}}\Big)d\tau^2
+\Big(\frac{\rho^2}{b^2}-\frac{b^{n-2}}{\rho^{n-2}}\Big)^{-1}d\rho^2+
\Big(\frac{w_nM}{b^{n-2}}\Big)^{2/n}\rho^2d\Om^2
\eea
Notice that now the radius of $S^{n-1}$ is of order $M^{2/n}\ra\infty$, so
that indeed we have managed to replace $S^{n-1}$ effectively by $\R^{n-1}$
as far as coordinates are concerned. The period of $\tau$ is
\bea
\Big(\frac{w_nM}{b^{n-2}}\Big)^{1/n}\b_0&=&\Big(\frac{w_nM}{b^{n-2}}\Big)^{1/n}
\frac{4\pi b^2}{n(w_nb^2M)^{1/n}} \nn
&=&\frac{4\pi b}{n}\equiv \b_1
\eea
and we may write
\ben
ds^2=\Big(\frac{\rho^2}{b^2}-\frac{b^{n-2}}{\rho^{n-2}}\Big)d\tau^2
+\Big(\frac{\rho^2}{b^2}-\frac{b^{n-2}}{\rho^{n-2}}\Big)^{-1}d\rho^2+
\rho^2dx_{n-1}^2
\label{wittenblackhole}
\een
with the boundary being $S^1(\b_1)\times\R^{n-1}$. Notice that in these 
scaled coordinates, the metric is characterized by just one parameter, $b$; 
any explicit reference to $M$ has disappeared.

\subsubsection{Thermodynamics of the black hole/non-extremal brane solution}
In the classical supergravity limit, the CFT partition function should be 
evaluated as
\bea
e^{-I}=e^{-I(X_1)}+e^{-I(X_2)}&=&e^{-I(X_1)}\Big(1+e^{-\Del I}\Big)\nn
&=&e^{-I(X_2)}\Big( e^{\Del I}+1\Big)\nn
\Del I&\equiv&I(X_2)-I(X_1)
\eea
We shall find
\ben
\Del I=\frac{\Om_{n+1}}{4G_N}\cdot\frac{b^2r_+^{n-1}-r_+^{n+1}}{nr_+^2+(n-2)b^2}
\label{deltaI}
\een
We see then that for small $r_+$, $\Del I>0$ and tends to zero when 
$r_+$ tends to zero. For large $r_+$ on the other hand $\Del I<0$. 
Thus we have the following situation
\bea
\mbox{small }\ r_+&:&e^{-I}\sim 2 e^{-I(X_1)}\nn
\mbox{large }\ r_+&:&e^{-I}\sim e^{-I(X_2)}
\eea
For low temperatures, $X_2$ cannot come into play at all and 
$$e^{-I_{CFT}}\sim e^{-I(X_1)}$$
For high temperatures, however, we have both manifolds, and either $r_+\ra 0$ or
$r_+\ra\infty$. In the first case, $I(X_2)\ra I(X_1)$. In the latter case 
(the relevant one as it turns out), $I(X_2)<I(X_1)$:
\bea
I(X_2)&=&I(X_1)+\Del I=I(X_1)+\frac{\Om_{n-1}}{4G_N}
\frac{b^2r_+^{n-1}-r_+^{n+1}}{nr_+^2+(n-2)b^2}\nn
&\sim &I(X_1)-\frac{\Om_{n-1}}{4G_N}\frac{r_+^{n-1}}{n}\ll I(X_1)
\eea
It follows that
\ben
e^{-I(X_1)}+e^{-I(X_2)}|_{r_+\ra 0}\ll e^{-I(X_1)}+e^{-I(X_2)}|_{r_+\ra\infty}
\een
All of that is a consequence of the formula for $\Del I$ \rf{deltaI}. We now
justify that formula.

The relevant part of the action is
\ben
I=-\frac{1}{16\pi G_N}\int d^{n+1}x\sqrt{g}\Big( R+\frac{n(n-1)}{b^2}\Big)
\een 
This is clearly badly divergent. In fact the equations of motion give
\ben
R_{\mu\nu}=-\frac{n}{b^2}g_{\mu\nu}\Ra R=-\frac{n(n+1)}{b^2}
\een
so that
\ben
I=\frac{1}{8\pi G_N}\frac{n}{b^2}\int d^{n+1}\sqrt{g}
=\frac{n}{8\pi G_Nb^2}V_{n+1}
\een
i.e. a volume divergence. We may hope however, to make sense of the 
{\em difference,} $\Delta I$, between $I$ evaluated for the two manifolds 
$X_1$ and $X_2$.
This requires that we arrange for the two manifolds somehow to be 
``asymptotically identical'' in their geometrical respect. Now we have for
both
\ben
ds^2=Vdt^2+V^{-1}dr^2+r^2d\Om^2\Ra \sqrt{g}=r^{n-1}\sqrt{\g}
\een
$\g$ being the determinant of the metric on the sphere $S^{n-1}$. 
$V$ is given by \rf{VX2} for the geometry of $X_2$ and by $V=1+(r/b)^2$
for the geometry of $X_1$, \rf{X1metric}
To regulate
the integrals, we terminate the integral over $r$ at some large $R$:
\bea
\mbox{Volume}(X_1)\equiv V(X_1)&=&\int_0^{\b(X_1)}dt\int_0^R dr r^{n-1}
\Om_{n-1},\ \ 0\leq r\leq R \nn
V(X_2)&=&\int_0^{\b_0}dt\int_{r_+}^R dr r^{n-1}\Om_{n-1},\ \ r_+\leq r\leq R
\eea
So we must choose $\b(X_1)$ such that the geometry at the $R$-surface is the 
same in the two cases, in particular that the circumference of the $S^1$ 
is the same for the two manifolds:
\bea
\mbox{for } X_1&:&\sqrt{1+\frac{R^2}{b^2}}\b(X_1)\nn
\mbox{for } X_2&:&\sqrt{1+\frac{R^2}{b^2}-\frac{w_nM}{R^{n-2}}}\b_0\Ra\nn
V(X_1)&=&\left [\Big(1+\frac{R^2}{b^2}-\frac{w_nM}{R^{n-2}}\Big)/
\Big(1+\frac{R^2}{b^2}\Big)\right ]^{\hf}
\b_0\frac{R^n}{n}\Om_{n-1}\nn
&\simeq&\Big(1-\frac{w_nM}{2R^n}\Big)\frac{R^n\b_0}{n}\Om_{n-1},
\ \ R\ra\infty\nn
V(X_2)&=&\b_0\frac{R^n-r_+^n}{n}\Om_{n-1}\nn
\Del I=I(X_2)-I(X_1)
&\simeq&\frac{\Om_{n-1}\b_0}{8\pi G_N}\left\{\frac{w_nM}{2}-\frac{r_+^n}{b^2}
\right\}
\eea
independent of $R$. But
$$\frac{w_nM}{r_+^{n-2}}=\frac{r_+^2}{b^2}+1$$
so
$$\Del I=\frac{\Om_{n+1}}{4G_N}\frac{b^2r_+^{n-1}-r_+^{n+1}}{nr_+^2+(n-2)b^2}$$
as promised \rf{deltaI}. This completes the discussion.

We now want to interpret $e^{-\Del I}$ as a statistical average $e^{-\b_0 E}$
where
\ben
E=\frac{\pa\Del I}{\pa\b_0}=\frac{\pa\Del I}{\pa r_+}\frac{\pa r_+}{\pa \b_0}
\een
A slightly long but straight forward calculation gives the result
$$E=\frac{(n-1)\Om_{n-1}}{16\pi G_N}\Big(\frac{r_+^n}{b^2}+r_+^{n-2}\Big)=M$$
This in fact justifies the long used notation for $M$.

More interestingly the Beckenstein-Hawking entropy is found from identifying
$\Del I$ with the free energy $F$:
\bea
e^{-F}&=&\sum_{\mbox{``states''}}e^{-\b_0 H}=\br e^{-\b_0 H}\kt\cdot
\Big(\mbox{effective no. of degrees of freedom}\Big)\nn
&=&e^{-\b_0E}e^S
\eea
with $S=\b_0E-\Del I$ the entropy. A very simple calculation now gives
\ben
S=\frac{\Om_{n-1}r_+^{n-1}}{4G_N}=
\frac{\mbox{``area of horizon in $n-1$ dimensions''}}{4G_N}
\een
which of course is the very famous result generalized from the flat case of 
$b\ra\infty$ to the present case of $AdS_{n+1}$. Notice incidentally that 
this kind of calculation does not work in flat space directly since the 
classical action vanishes there.\\[.3cm]
{\bf Exercise:} Verify the expressions given for the mass and entropy of the 
black hole.
 
\subsection{On hadrons and confinement in QCD at large $N$. The case of 
$QCD_3$}
We have emphasized that the supergravity approximation is not satisfactory for
a serious study of large $N$ QCD, even supposing the Maldacena conjecture 
works, and that no unpleasant phenomena corrupt the set up proposed by Witten.
Nevertheless it is very instructive to see how this approximation very 
naturally gives a picture in qualitative agreement with expectations in 
hadronic physics. Thus the area law for Wilson loops,
associated with confinement, comes out and
goes away appropriately at high (physical) temperatures. Several other aspects
have been treated \cite{maldacena98,wittentheta,olesen}. 
Here, however, we shall concentrate on just one of these
aspects, generation of a {\em mass gap} \cite{witten1,witten2,ooguri,csaki,
zyskin,hashimoto,pt98,kjmn,minahan}.

When we use the scheme described above in the case of $AdS_5\times S^5$, 
compactifying Euclidean time, we expect (as argued) in the boundary theory to
obtain an effective (Euclidean) version of $QCD_3$ at energies much below the
temperature cut-off. In the next subsection we shall briefly indicate the 
framework for also dealing with $QCD_4$. In both cases, the hadrons we expect 
to find will be stable glueballs, in particular with a minimal positive 
mass greater than zero: a mass gap. This then is a non-trivial 
confinement effect, not visible in classical YM-theory, nor in perturbation 
theory.

How would we identify such a mass gap? We would need in the boundary theory to 
consider a gauge invariant operator which could create a glueball, the simplest 
example perhaps being
\ben
{\cal O}(\vex)=trF^2(\vex)\equiv\sum_{i,j=1}^N\Big(F_{\mu\nu}\Big)_{ij}
\Big(F^{\mu\nu}\Big)_{ji}(\vex)\ \ \vex\in\R^3
\een
with $(F_{\mu\nu})_{ij}=\sum_{a=1}^{N^2}F^a_{\mu\nu}T^a_{ij}$ being the $U(N)$ 
Yang Mills field strength expressed as an $N\times N$ matrix. We should then 
work out a 2-point function
\ben
\br{\cal O}(\vex){\cal O}(\vec{0})\kt\sim\exp\{-|\vex|m\},\ \ |\vex|\ra\infty
\een
and look for the indicated exponential decay at large distances, dominated by 
the minimal mass, $m>0$, the mass gap. Non leading exponential terms would
correspond to gluon excitations.

To perform the calculation in the {\em bulk theory}, we would have to identify 
the field, $\Phi(y),\ \ y\in$ bulk, to which this operator couples. But we have 
already indicated in sect. 4.5.1 that the expected candidate is the 
dilaton field with
no KK-excitations on $S^5$, the $S^5$ s-wave, or $k =0$ mode \cite{witten2}.
We might then perform a classical supergravity calculation of the action in
the $AdS_5$ background with the black hole and with prescribed boundary 
conditions on the dilaton field.

However, \cite{witten1,witten2} it seems technically simpler to adopt a slightly
different point of view. To see how, begin noticing that (according to
 the 
Maldacena conjecture) the two quantum theories, the bulk theory and the 
boundary theory, are entirely equivalent. They have the same Hilbert space, the 
same operators and correlators, only the physical interpretation of operators
differ drastically in the two theories. First consider the boundary theory. At
$N\ra\infty$, glueballs are free particles. When moving according to plane waves
$$e^{i\vec{k}\cdot\vex}$$
their mass is given by
\ben
m^2=-\vec{k}^2,
\een
(so only imaginary momenta correspond to being on the mass shell, as is usual in
Euclidean space-time).
In the Hilbert space there is an operator describing 
translations in the boundary theory, and $\vk$ is the eigenvalue of that.

In the bulk theory, the very same operators occur, just with different 
interpretation. We need to think of the Hilbert space of the dilaton theory.
First recall the situation for a free scalar in flat space-time. It satisfies the
KG equation
$$(\pa^2 - m^2)\phi=0$$
To build the Hilbert space (in this case the Fock space) one finds ``modes'',
solutions of the KG equation, such as $e^{ipx}$, and build field operators as
sums over modes, each mode being multiplied by a creation or annihilation 
operator. States in the Hilbert space are spanned by multi-particle states
created by the creation operators. Actually, although we often consider plane
waves, we really need to put them in a quantization volume, or better in fact,
consider wave packets, {\em square integrable modes} for which
$$\int_{\mbox{space}}(\mbox{mode})^2 <\infty$$
We wish to imitate this procedure in the present case of a modified $AdS_5$ 
background with a black hole in it, the manifold $X_2$. Thus we must
\begin{enumerate}
\item Formulate the equation of motion on $X_2$,
\item find the square integrable modes,
\item interpret those in terms of possible $m^2(\mbox{glueball})$ values.
\end{enumerate}
We follow here the treatments in \cite{witten2,csaki,zyskin,ooguri}. In the case
of $AdS_5\times S^5$ we have
\ben
b^2=\ell_s^2\sqrt{4\pi g_sN}=\ell_s^2\sqrt{g_{YM}^2N}
\een
Putting $n=4$ in the black hole metric \rf{wittenblackhole} we find
\ben
\frac{ds^2}{\ell_s^2\sqrt{4\pi g_sN}}=
\Big(\frac{\rho^2}{b^4}-\frac{1}{\rho^2}\Big)d\tau^2   
+\frac{d\rho^2}{\Big(\rho^2 -\frac{b^4}{\rho^2}\Big)}
+\frac{\rho^2}{b^2}dx_3^2+d\Om_5^2
\een
(with a {\em unit radius} $S^5$).
Now put $\tau=b^2\tilde{\tau}$ and get (scaling also $x_i$)
\ben
\frac{ds^2}{b^2}=\Big(\rho^2-\frac{b^4}{\rho^2}\Big)d\tilde{\tau}^2
+\frac{d\rho^2}{\Big(\rho^2-\frac{b^4}{\rho^2}\Big)}
+\rho^2 dx_3^2+d\Om_5^2
\label{ooguriblackhole}
\een
Here the period of $\tau$ is $\b_1=4\pi b/n=\pi b$ and the period of 
$\tilde{\tau}$ is 
\ben
\b(\tilde{\tau})\equiv 2\pi R(\tilde{\tau})=\b_1/b^2=\pi/b
\een
while the horizon is still at $\rho = b$.

Now, the dilaton is massless in 10 dimensions, and the $k =0$ mode on $S^5$ 
which we should consider, is still massless in 5 dimensions. This mode does
not depend on coordinates of $S^5$ and it satisfies the equation of motion on
$X_2$
\ben
\pa_\mu\left [\sqrt{g}\pa_\nu\Phi g^{\mu\nu}\right ]=0
\een
We now wish to look for modes that are (i) square integrable on $X_2$, and 
(ii) correspond to a definite momentum in the boundary theory. 
Thus we consider the ansatz
\ben
\Phi(\rho,\vex)=f(\rho)e^{i\vk\cdot\vex}
\een
Also our Euclidean time $\tilde{\tau}$ is compactified corresponding to a 
``high'' temperature, and we do not want the mode to depend on  $\tilde{\tau}$
in the ``low energy approximation''. Clearly we may then identify 
$\vk^2=-m^2$ as the glueball mass in the boundary theory. Now we have
\ben
g^{00}=\Big(\rho^2-\frac{b^4}{\rho^2}\Big)^{-1}, \ 
g^{\rho\rho}=\Big(\rho^2-\frac{b^4}{\rho^2}\Big), \ g^{ij}=\del^{ij}\rho^{-2}, 
\ \sqrt{g}=\rho^3
\een
The equation of motion then becomes
\bea
\pa_\rho\left [ \rho^3\pa_\rho\Phi \Big(\rho^2-\frac{b^4}{\rho^2}\Big)
\right ]+\pa_i\left [ \rho^3\pa_i\Phi \rho^{-2}\right ]&=&0\ \ \mbox{or}\nn
\pa_\rho\left [ \Big(\rho^5 -b^4\rho\Big)f'(\rho)\right ]-k^2\rho 
f(\rho)&=&0\ \mbox{or}\nn
\rho^{-1}\frac{d}{d\rho}\left [\rho\Big(\rho^4-b^4\Big)f'(\rho)\right ]
&=&-m^2 f(\rho)
\eea
Put 
$$x=\rho^2,\ \frac{d}{d\rho}=\frac{dx}{d\rho}\frac{d}{dx}=2\rho\frac{d}{dx}$$
and obtain
\ben
4x(x^2-b^4)\frac{d^2f}{dx^2}+4(3x^2-b^4)\frac{df}{dx}-k^2f=0
\label{diffeqf}
\een
and of course we may scale $b$ away and replace it by $1$. This ordinary
differential equation is the equation of motion. It has solutions for any 
values of $\vk^2$, but now we must understand the additional information
coming from boundary conditions and square integrability. These will imply that
for generic $\vk^2$ there is no acceptable solution, only for a 
particular {\em spectrum} 
of {\em strictly positive} $-\vk^2$ values do such solutions exist.

\subsubsection{Boundary condition}
Putting $b=1$ the metric \rf{ooguriblackhole} becomes (we write $\tau$ for 
$\tilde{\tau}$ in the following):
\ben
ds^2\sim\Big(x-\frac{1}{x}\Big)^{-1}d\rho^2+\Big(x-\frac{1}{x}\Big)d\tau^2+...
=\frac{dx^2}{4(x^2-1)}+\Big(x-\frac{1}{x}\Big)d\tau^2+...
\een
As before there is a coordinate singularity at the horizon $x=1$ which we wish
to cast into the form of a 2-dimensional polar coordinate singularity
\bea
dz^2&=&\frac{dx^2}{4(x^2-1)}\ \ (z \ \mbox{is polar radius})\nn
\frac{dz}{dx}&=&\frac{1}{2\sqrt{x^2-1}},\ \ z=\hf\cosh^{-1}x,\ \ x=\cosh 2z
\eea
with $z=0$ at the horizon $x=1$, and we find
\ben
\Big(x-\frac{1}{x}\Big)=\frac{\sinh^2 2z}{\cosh 2z}\simeq 4z^2 \ \mbox{near } z=0
\een
and
$$ds^2\simeq dz^2+4z^2d\tau^2+...$$
near the horizon, showing that we have ``polar-like'' coordinates with polar
distance $\propto z$. The function $f(\rho)$ above is then a function of the
``polar distance'' only, not of the ``angle'' $\tau$. The proper boundary 
condition for such a function to be smooth at the origin is therefore
$$\frac{df}{dz}=0$$
But
$$\frac{df}{dz}=\frac{dx}{dz}\frac{df}{dx}
=2\sinh 2z\frac{df}{dx}\simeq 4z\frac{df}{dx}$$
near $z\simeq 0$. So we merely want to ensure that $f$ is {\em regular} at 
$x=1$.

\subsubsection{Square integrability}
We want to demand that
$$\int \sqrt{g}d\rho |f(\rho)|^2<\infty$$
But
$$\sqrt{g}d\rho=\rho^3 d\rho=\rho^2 \rho d\rho=\hf xdx$$
Hence we should demand, that if $f$ is inverse power bounded, 
by $f\sim {\cal O}(x^{-a})$, then $a>1$.

\subsubsection{Determination of the spectrum}
The differential equation \rf{diffeqf} divided by $4x(x^2-1)$ becomes
\ben
y''+\Big(\frac{1}{x}+\frac{1}{x-1}+\frac{1}{x+1}\Big)y'-\frac{p}{x(x^2-1)}y=0
\een
$p\equiv\vk^2/4=-m^2/4$, and $y\equiv f$. This homogeneous, linear, 2. order,
ordinary differential equation has a 2-dimensional space of solutions for
any value of the mass parameter, $p$. We follow now the treatment of 
\cite{zyskin}.
These solutions may be expressed as linear 
combinations of any 2 linearly independent solutions. Generically these would 
be analytic functions of $x$ with singularities at $x=0,1,\infty$. 
Therefore these solutions cannot be represented by series expansions, convergent
throughout the physical region $1\leq x <\infty$. Instead, however, we may
consider expansions convergent either in 
$$I(\infty)\equiv\{x\in\C |1<|x|<\infty\}$$
or in 
$$I(1)\equiv\{x\in\C ||x-1|<1\}$$
respectively. Thus for the case of $I(\infty)$ we use the ans\"atze
\bea
y_1^\infty(x)&=&\frac{1}{x^2}+\sum_{n=1}^\infty a_n^\infty\frac{1}{x^{n+2}}\nn
y_2^\infty(x)&=&\frac{p^2}{2} y_1^\infty(x)\log x +\sum_{n=1}^\infty
b_n^\infty\frac{1}{x^n}
\eea
whereas for the case of $I(1)$ we may use
\bea
y_1^1(x)&=&1+\sum_{n=1}^\infty a_n^1 (x-1)^n\nn
y_2^1(x)&=&y_1^1(x)\log(x-1)+\sum_{n=1}^\infty b_n^1(x-1)^n
\eea
Inserting these expansions into the differential equation, one finds recursion 
relations for the unknown coefficients with simple unique solutions.

Any solution of the differential equation may then be represented, {\em either}
as a linear combination of $y_1^\infty(x)$ and  $y_2^\infty(x)$, {\em or}
a linear combination of $y_1^1(x)$ and  $y_2^1(x)$. The series expansions only 
converge in $I(\infty)$ or $I(1)$ respectively, but the solutions may be 
uniquely analytically continued to the entire complex plane (with
the exception of the singularities at $x=0,1$). In the overlap (containing
$1<x<2$), they
may furthermore be directly compared. It is now clear that $y_1^\infty(x)$
but not $y_2^\infty(x)$ satisfies the square integrability condition. Also,
only $y_1^1(x)$, but not $y_2^1(x)$ satisfies the boundary condition at $x=1$.

We may therefore assert, that an acceptable solution is expressed in the
$y_i^\infty$ basis simply as
$$c\cdot y_1^\infty(x)$$
and such a solution may be built for any $p$. However, when analytically 
continued to $I(1)$, it will generically be a expressed as a linear combination
$$\a_1 y_1^1(x)+\a_2y_2^1(x)$$
with $\a_2\neq 0$, and therefore be unacceptable from the point of view of the
behaviour near $x=1$. Similarly, we may assert, that any acceptable solution is
expressed in the $y_i^1$ basis simply as
$$c'\cdot y_1^1(x)$$
and of course also such a solution may be built for any $p$, but when 
analytically continued to $I(\infty)$ it will generically be expressed as a 
linear combination 
$$\b_1 y_1^\infty(x)+\b_2y_2^\infty(x)$$
with $\b_2\neq 0$, and therefore be unacceptable from the point of view of 
square integrability.

It follows, that if we restrict ourselves to the overlap $1<x<2$, we are able
to require the condition
\ben
y_1^\infty(x)\propto y_1^1(x)
\label{condition}
\een
and this condition is only satisfied for certain discrete values of $p$. These
represent the sought for spectrum of mass values. More concretely, we may pick a
value such as $x=x_0=3/2\in I(1)\cap I(\infty)$, and require the existence of
a common constant, $c$ such that
\bea
y_1^\infty(x_0)&=&c y_1^1(x_0)\nn
(y_1^\infty)'(x_0)&=&c(y_1^1)'(x_0)
\label{wronski1}
\eea
This will ensure that \rf{condition} is satisfied throughout, since we are 
dealing with a 2-dimensional set of solutions. These conditions 
are of course just equivalent to the Wronski condition
\ben
\left |\begin{array}{rr}y_1^\infty(x_0)& y_1^1(x_0)\\
(y_1^\infty)'(x_0)&(y_1^1)'(x_0)\end{array}\right |=0
\label{wronski2}
\een
Here we should think of $y_i^a(x_0)$ and $(y_i^a)'(x_0)$ ($a=\infty,1$)
as functions of
$p$ (for fixed $x_0$). Indeed the coefficients $a_n^a$ are determined uniquely
as polynomials in $p$ of order $n$. 
Truncating the series expansions at some high
cut-off, \rf{wronski2} becomes the condition for zeros of a certain high order 
polynomial in $p$.
As the summation cut-off is taken higher and higher, the numerically 
smaller zeros of the polynomial rather quickly converge, whereas the larger 
zeros take more terms to stabilize.

Concretely and by way of illustration, it is easy from the differential equation
to establish the following recursion relations (using the notation
$a_{n<0}\equiv 0, a_0^\infty=1=a_0^1$):
\bea
a_{n+1}^\infty&=&\frac{1}{(n+2)^2-1}\Big(a_{n-1}^\infty (n+1)^2+pa_n^\infty\Big)
\nn
a^1_{n+1}&=&\frac{1}{2(n+1)^2}\Big(a^1_{n-1}(n^2-1)+(3n(n+1)-p)a_n^1\Big)
\eea
Thus
$$a_1^\infty=\frac{p}{3},\ a_2^\infty=\frac{p^2+12}{24}, ...$$
$$a_1^1=-\frac{p}{2},\ a_2^1=\frac{p(p-6)}{16},...$$
The glueball masses thus determined may be shown to pertain to $J^{PC}=0^{++}$
states. Numerically one finds (in units of $1/b$) the following strictly 
positive mass values
$$11.6,\ 34.5,\ 69.0, \ 114.9,..$$
(close, but not equal to the values, $6n(n+1)$ \cite{csaki,zyskin}).
Similar studies have been made for other $J^{PC}$ quantum numbers. Comparisons
with results from lattice calculations indicate reasonable agreement for the 
mass {\em ratios.} 

\subsection{On QCD in 4 dimensions}
Witten \cite{witten2} has also explained how to obtain 4-dimensional QCD based
on the $AdS_7\times S^4$ compactification of 11-dimensional M-theory with an 
$AdS_7$ scale \rf{L74}
$$b=2\ell_{11}(\pi N)^{1/3}$$
In that case the boundary theory is the conformally invariant 6-dimensional so 
called (2,0) theory\cite{6dim}.
Excitations of this 5-brane cannot be understood in terms of open strings 
ending on it, but in fact in terms of 2-branes ending on it. Upon 
compactification on a circle of radius $R_1$, we obtain type IIA string theory,
and for small $R_1$, IIA string theory  in the perturbative regime. 
There the 2-branes of M-theory
turn into strings and we {\em can} use string theory to describe excitations
of that once compactified theory. Now the effective boundary theory is 
5-dimensional. We still have to compactify once more, on a circle of radius 
$R_2$ before we obtain the desired 4-dimensional theory. In this second step
we must take fermions anti periodic around the second $S^1$ in order to break
supersymmetry and conformal invariance. It is seen that we assumed it made 
sense to perform the compactification in two steps, corresponding to 
$R_1\ll R_2$. 

In the first step, we take SUSY preserving boundary conditions for the fermions,
and we get a string theory 4-brane with an effective 5-dimensional SYM theory
with YM coupling \rf{YMcoupling}
\ben
g_5^2=8\pi^2\ell_s g_s=8\pi^2 R_1^{(10)}
\label{g5}
\een
where the last step is the standard identification between M-theory and IIA 
string theory upon compactification on a circle of 10-dimensional radius 
$R_1^{(10)}$ \cite{wittenvarious}. 
In the second compactification on $S^1(R_2)$, we may write
$$\int d^5x\frac{1}{g_5^2}\ra \int d^4x\frac{2\pi R_2}{g_5^2}\Ra 
\frac{1}{g_4^2}=\frac{2\pi R_2}{g_5^2}$$
So we get for the 4-dimensional YM coupling
\ben
\frac{g_{YM}^2}{4\pi}=\frac{R_1}{R_2}
\label{g4}
\een
At low energies this effective D3 brane theory is large $N$ QCD with a running
coupling constant, which at large energies becomes smaller and smaller 
(asymptotic freedom), until the theory changes over into a 5-dimensional one 
at the ``cut-off'' $R_2$ (in string units). Thus the YM coupling \rf{g4} is the
value of $\a_s$ at that cut-off, and ideally we want that to be small in 
accord with the requirement $R_1\ll R_2$. If the second compactification also 
preserves SUSY, we get the $N=4$ superconformal theory, but if we take fermions
anti periodic around $S^1(R_2)$ we expect to obtain ordinary large $N$ QCD 
(with no quarks).

Let $\eta\equiv \frac{g_{YM}^2N}{4\pi}$ denote the 't Hooft coupling, so
\ben
R_1=\frac{\eta R_2}{N}
\label{R1etaR2}
\een
with fixed  (preferably small) $\eta$ at $N\ra\infty$. Let us in fact begin
with the second compactification first. The $AdS_7$ metric with a black hole
is given by \rf{wittenblackhole}
\ben
ds^2=\Big(\frac{\rho^2}{b^2}-\frac{b^4}{\rho^4}\Big)d\tau^2+
\frac{d\rho^2}{\Big(\frac{\rho^2}{b^2}-\frac{b^4}{\rho^4}\Big)}
+\rho^2 dx_5^2
\een and $b=2\ell_{11}(\pi N)^{1/3}$. Here $\tau$ has period ($n=6$)
$$\b_1=\frac{4\pi b}{n}=\frac{2\pi b}{3}=\ell_{11}\frac{4\pi}{3}(\pi N)^{1/3}$$
Then put
$$\tau=\frac{b}{3}\theta,\ \ \rho=b\l$$
with $\theta$ an angular variable $\in{[}0,2\pi{]}$. Then (with a trivial
rescaling of the $x_i$)
\ben
ds^2=\Big(\l^2-\frac{1}{\l^4}\Big)\frac{1}{9}b^2d\theta^2+
\frac{b^2d\l^2}{ \Big(\l^2-\frac{1}{\l^4}\Big)} + \frac{b^2\l^2}{9}dx_5^2
\een
This metric of course is in 11-dimensional units. But we should also compactify 
on a circle of radius $R_1$, say use the coordinate $x_5$ for that, with
that radius being $\eta/N$ times the radius of the first compactification 
\rf{R1etaR2}, at least at the boundary, i.e. for very large values of $\l$. For
large $\l$, the circle with coordinate $\theta$ (the $R_2$ circle) has metric
$$\frac{1}{9}\l^2 b^2d\theta^2$$
and therefore radius $b\l/3$. For large $\l$ the $x_5$ coordinate has metric
$$\frac{1}{9}b^2\l^2dx_5^2$$
(before, $dx_5^2$ meant the flat metric in 5 dimensions, now it refers 
just to the 
5'th coordinate) so to compactify with the correct radius, we want to write
$$x_5=\frac{\eta}{N}\psi$$
with $\psi$ an angular variable. The full 11-dimensional metric, including the
$S^4$ part is now
\bea
ds^2&=&\frac{b^2}{9}\Big(\l^2-\frac{1}{\l^4}\Big)d\theta^2 + 
\frac{b^2\l^2\eta^2}{9N^2}d\psi^2\nn
&&+\frac{b^2d\l^2}{ \Big(\l^2-\frac{1}{\l^4}\Big)} + \frac{b^2\l^2}{9}
\sum_1^4 dx_i^2 +\frac{b^2}{4}d\Om_4^2
\eea
It is the circle parametrized by $\psi$ which gives us IIA compactification.
It is seen to have a radius of
$$\frac{1}{3}b\l\frac{\eta}{N}$$
in 11-dimensional units. The general rule to go between 11-dimensional and 
10-dimensional units is \cite{wittenvarious}
\bea
L^{(11)}&=&g_s^{-1/3}L^{(10)};\ \ R_1^{(11)}=g_s^{2/3}\ell_{11},\ \ 
R_1^{(10)}=g_s\ell_s\nn
ds^2_{(11)}&=&g_s^{-2/3}ds^2_{(10)}+(R_1^{11})^2d\psi^2\Ra ds^2_{(10)}=
\frac{R_1^{(11)}}{\ell_{11}}ds^2_{(11)}- (x_5-\mbox{part})\nn
g_s&=&e^{\phi},\ \ \phi=\mbox{ dilaton field}
\eea
So the final 10-dimensional metric in the string frame becomes independent 
of $N$ (!)
\bea
3\frac{ds^2_{(10)}}{8\pi\eta\ell^3}&=&\frac{1}{9}\Big(\l^3-\frac{1}{\l^3}\Big)
d\theta^2+\frac{\l^2d\l^2}{\Big(\l^3-\frac{1}{\l^3}\Big)}\nn
&&+\frac{1}{9}\l^3\sum_1^4 dx_i^2+\frac{1}{4}\l d\Om_4^2
\label{witten10d74}
\eea
This is the form given in \cite{witten2}.
From the radius of the circle parametrized by $\psi$ we find
$$\frac{b\l\eta}{3N}=\frac{2\pi^{1/3}\l\eta\ell_{11}}{3N^{2/3}}=
g_s^{2/3}\ell_{11}$$
and
\ben
g_s=e^{\phi}=\frac{1}{N}\sqrt{\pi\Big(\frac{2\l\eta}{3}\Big)^3}
\een
So this time we see that we have a non-trivial dilaton field depending on the
coordinate $\l$. This fact, actually makes the analysis considerably more 
complicated. We must set up fluctuation equations for dilatons around the
background provided by the above solution \cite{csaki}. 
But a number of subtleties exist
\cite{hashimoto,gkt98,pt98} which we do not want to discuss here.

It is instructive to derive an equivalent version of the above, 
starting from the non-extremal near horizon limit of the 11-dimensional 
5-brane solution \rf{nonexn6} in 11 dimensional units
\ben
ds^2=\frac{U^2}{4L^2}\left\{\Big(1-\Big(\frac{U_0}{U}\Big)^6\Big)dt^2+d\vex^2
\right\}+4L^2\frac{dU^2}{U^2\Big(1-\Big(\frac{U_0}{U}\Big)^6\Big)}
+L^2d\Om_4^2
\een
with $L^2=\ell_{11}^2(\pi N)^{2/3}$. Now put 
$$L^2r=\frac{U^2}{4L^2}$$
and recast this into
\ben
\frac{ds^2}{L^2}=r\left\{\Big(1-\Big(\frac{r_0}{r}\Big)^3\Big)dt^2+d\vex^2
\right\} +\frac{dr^2}{r^2\Big(1-\Big(\frac{r_0}{r}\Big)^3\Big)}
+d\Om_4^2
\een
From the discussion above we have seen that when compactifying on a circle to
get to the 10-dimensional IIA metric, we must in fact write 
\ben
ds^2_{(11)}=\Big(R_1^{(11)}\Big)^2d\psi^2+g_s^{-2/3}ds^2_{(10)}=
\ell_{11}^2e^{4\phi/3}d\psi^2+e^{-2\phi/3}ds^2_{(10)}
\een
Thus we find (in suitable units)
\bea
r&=&e^{4\phi/3}\Ra e^\phi=r^{3/4}\nn
ds^2_{(10)}&=&r^{3/2}\left\{\Big(1-\Big(\frac{r_0}{r}\Big)^3\Big)dt^2
+\frac{dr^2}{r^3-r_0^3}+\sum_1^4 dx_i^2\right\}+r^{\hf} d\Om_4^2
\label{stringD4}
\eea
This is the form given in \cite{hashimoto,gkt98}. The transformation 
$$r\propto \l^2$$
makes this form in agreement with \rf{witten10d74}.
It is instructive to see that it also agrees with the 
non-extremal D4-brane solution.
From \rf{branesolution} we find with 
$$D=10,\ p=4,\ d=5,\ a=-\hf,\ \Delta =16$$
\bea
H&=&1+\Big(\frac{h}{r}\Big)^3\sim\Big(\frac{h}{r}\Big)^3\ \ 
\mbox{in the near horizon approximation}\nn
f&=&1-\Big(\frac{r_0}{r}\Big)^3
\eea
Then (in un units of $h$)
\bea
ds^2_{(10)}(\mbox{Einstein})&=&r^{9/8}\left\{
\Big(1-\Big(\frac{r_0}{r}\Big)^3\Big)dt^2 +\sum_1^4 dx_i^2\right\}\nn
&&+r^{-15/8}\left\{\frac{dr^2}{1-\Big(\frac{r_0}{r}\Big)^3}+r^2 d\Om_4^2
\right\}\nn
e^{\phi}&=&r^{3/4}
\eea
The dilaton agrees with above, but the metric being in the Einstein frame does 
not yet. However, we may put it in the appropriate string frame using 
\rf{einstein2string}, by multiplying the Einstein frame metric by 
$$e^{\hf\phi}=r^{3/8}$$
This precisely reproduces \rf{stringD4}.

In the supergravity approximation we should now set up fluctuation equations
of motion for dilatons and other fields in the backgrounds given. However,
\cite{hashimoto} it becomes important to diagonalize these fluctuations 
appropriately in a non trivial way, since the dilaton background itself is non
trivial. We shall not, however, pursue these finer points. 
\\[.3cm]
{\bf Acknowledgement:} I have benefited greatly from discussions with P. Di 
Vecchia, T. Harmark, R. Marotta, N. Obers and R. Szabo. I thank the 
organizers of the two schools: ``Quantum aspects of gauge theories, 
supersymmetry and unification'', School in Leuven, January 18-23, 1999; and 
``Nordic Course on Duality in String and Field 
Theories'', 4-14 August, 1998, Nordita, Copenhagen, 
for the invitation to present these lectures.

\end{document}